\documentclass[journal=jctcce,manuscript=article]{achemso}

\usepackage{amssymb}
\usepackage{amsmath}
\usepackage{graphicx}
\usepackage{mathtools}
\usepackage{bm}
\usepackage[dvipsnames,svgnames]{xcolor}
\usepackage{hyperref}
\usepackage{placeins}
\usepackage{cleveref}
\usepackage{xr}
\usepackage[normalem]{ulem}

\newcommand{\revision}[1]{#1}
\newcommand{\revisionremoved}[1]{}
\newcommand{\revisionhideable}[1]{}

\title{Coupling all-electron full-potential density functional theory with grid-based continuum embeddings}

\author{Jakob Filser}
\email{jakobfilser@boisestate.edu}
\affiliation{Boise State University, Boise, ID}
\alsoaffiliation{Fritz Haber Institute of the Max Planck Society, Berlin, Germany}
\author{Edan Bainglass}
\affiliation{PSI Center for Scientific Computing, Theory and Data, Villigen PSI, Switzerland}
\author{Karsten Reuter}
\affiliation{Fritz Haber Institute of the Max Planck Society, Berlin, Germany}
\author{Oliviero Andreussi}
\affiliation{Boise State University, Boise, ID}

\date{\today}

\begin{document}

\maketitle

\begin{abstract}

Recent advances in continuum embedding models have enabled the incorporation of solvent and electrolyte effects into density functional theory (DFT) simulations of material surfaces, significantly benefiting electrochemistry, catalysis, and other applications. To extend the simulation of diverse systems and properties, the implementation of continuum embedding models into the Environ library adopts a modular programming paradigm, offering a flexible interface for communication with various DFT programs. The speed and scalability of the current implementation rely on a smooth definition of the key physical properties of the atomistic system, in particular of its electronic density. This has hindered the coupling of Environ with all-electron simulation packages, as the sharp electron density peaks near atomic nuclei are difficult to represent on regular grids. In this work, we introduce a novel smoothing scheme that transforms atom-centered electron densities into a regular grid representation while preserving the accuracy of electrostatic calculations. This approach enables a minimal and generic interface, facilitating seamless interoperability between Environ and all-electron DFT programs. We demonstrate this development through the coupling of Environ with the FHI-aims package and present benchmark simulations that validate the proposed method.

\end{abstract}

\section{Introduction}

Continuum embedding models are commonly used to incorporate solvation effects in molecular simulations.\cite{andreussi2019,tomasi2005} These models treat the solvent as a structureless dielectric continuum interacting with a solute system while the latter is treated on an atomistic level. The popularity of these methods arises from their high efficiency. For typical applications where the solute is treated on an electronic structure level of theory, continuum embedding causes only a small to moderate computational overhead.\cite{filser2022} By contrast, simulating individual solvent molecules explicitly would not only increase the system size of the electronic structure calculation but also require adequate thermodynamic sampling of the solvent molecules' phase space, thus increasing the computational cost by multiple orders of magnitude.\cite{steinmann2016}

A variety of continuum embedding models, also referred to as implicit solvation models, have been proposed in the literature. \cite{tomasi2005} Historically, these were typically implemented with one particular host program in mind.\cite{marenich2009,andreussi2012,kiran2014} However, with the modular programming paradigm getting increasingly adopted in computational science, continuum embedding packages are now available which can be incorporated as libraries into different host programs or used as standalone software.

These implementations often rely on fast Fourier transforms (FFT) for their numerical solvers.\cite{andreussi2012,kiran2014} This allows for a straightforward coupling with \textit{density functional theory with plane waves and pseudopotentials} (DFT-PWP) host codes.\cite{giannozzi2017} FFT requires data to be represented on a regular grid, i.e., a grid with even spacing in the directions of all three lattice vectors. In contrast, all-electron full-potential DFT (henceforth shortened to all-electron DFT) relies on overlapping atom-centered grids to resolve the potential near the nuclei.\cite{blum2009} In this region, all-electron DFT correctly predicts a steep variation of the charge density with a characteristic cusp, which cannot be adequately represented on a regular grid. This leads to inconsistent results for any practically feasible grid spacing and precludes coupling regular grid based continuum embedding with all-electron DFT in the minimalist fashion which is possible with DFT-PWP. This hinders interoperability despite the underlying physical models of continuum embedding and all-electron DFT being compatible in most cases.


To bridge this gap, we present here an interface between the all-electron DFT program FHI-aims\cite{blum2009,havu2009,yu2018,abbott2025} and the regular grid based continuum embedding package Environ.\cite{andreussi2012} The interface is kept generic in the sense that communication is, for the most part, restricted to grid information and such physical data which is practically always needed, namely the electron density, molecular geometry, potential, energy, and forces. Such an interface makes only minimal assumptions on either side about the methods which are used by the other side. This allows the user to freely choose these methods as long as the underlying physical models are compatible. This interoperability will immediately include methods developed in the future without requiring any adaptations to the interface. At the center of the interface lies a smoothing scheme which we use to represent the all-electron density on the regular grid in a way that accurately preserves atomic multipole moments, leading to consistent results at a feasible grid spacing.

\section{Theoretical background}

The physical background and computational details of the methods available in FHI-aims and Environ have been described in other publications. We restrict ourselves here to reiterating only those parts of the formalism which are relevant for the communication interface that is the subject of the present work.

\subsection{All-electron full-potential density functional theory}

Regular grid based software typically store the electron density in a straightforward value-per-grid-point manner. By contrast, FHI-aims uses a partitioning scheme to store the density as atomic multipole components which are functions of only one radial coordinate. In a first step, the full density $\rho^\text{el}$ of all electrons in the system is separated into a free atom contribution and a $\textit{delta}$ density $\delta\rho$
\begin{equation}
    \rho^\text{el}(\mathbf{r}) = \sum_{at} \rho^\text{free}_{at} (r_{at}) + \delta \rho(\mathbf{r})
\end{equation}
where the sum runs over all atoms $at$ in the system and $\rho^\text{free}_{at}$ is the density of a single, free (i.e., in vacuum) atom of the same species (i.e., nuclear charge) as $at$. Throughout this work, we use the shorthand notations $r_{at}$, $\Omega_{at}$ to denote the radial and angular components of the atom-centered position $\mathbf{r}-\mathbf{R}_{at}$ in spherical coordinates, where $\mathbf{R}_{at}$ is the nuclear position of $at$. $\delta\rho$ is then further partitioned into atomic contributions according to
\begin{equation}
    \delta \rho_{at}(\mathbf{r}) = p_{at}(\mathbf{r})\delta \rho(\mathbf{r})
\end{equation}
The exact form of the atomic partitioning functions $p_{at}$ is not relevant for the present work. Suffice it to say that they fulfill $\sum_{at}p_{at}(\mathbf{r})=1$ at every point $\mathbf{r}$ where $\rho^\text{el}$ or any of its components are non-zero such that $\sum_{at} \delta \rho_{at} = \delta \rho$, and that each of them is centered around its nucleus $\mathbf{R}_{at}$ such that it decays to zero at some finite distance from $\mathbf{R}_{at}$. The atomic contributions are then finally partitioned into multipoles
\begin{equation}\label{eq:integrate_MP}
    \delta \widetilde{\rho}_{at,lm}(r) = \oint\limits_{|\mathbf{r}-\mathbf{R}_{at}|=r} d\Omega_{at}\, \delta \rho_{at}(\mathbf{r}) Y_{lm}(\Omega_{at})
\end{equation}
where $Y_{lm}$ are the real-valued linear combinations of the spherical harmonics. Truncating the expansion at a finite order $l_\text{max}$, a \textit{multipole density} can be reconstructed which reproduces the original $\rho^\text{el}$ to a good approximation.
\begin{equation}\label{eq:rho_MP}
    \rho^\text{MP}(\mathbf{r}) = \sum_{at} \rho^\text{free}_{at} (r_{at}) + \sum_{at}\sum_{l=0}^{l_\text{max}}\sum_{m=-l}^{l} \delta \widetilde{\rho}_{at,lm}(r_{at}) Y_{lm}(\Omega_{at})
\end{equation}
Although it slightly differs from $\rho^\text{el}$, we refer to $\rho^\text{MP}$ as the \textit{original} electron density throughout this article to distinguish it from the smoothened electron density which we introduce in \cref{sec:smoothing}. A few things are worth pointing out about this formalism. In \cref{eq:rho_MP}, the analytical form of $Y_{lm}$ is well-known, and the two remaining types of functions $\rho^\text{free}_{at}$ and $\delta \widetilde{\rho}_{at,lm}$ depend only on a single, radial coordinate rather than the full vector $\mathbf{r}$. They can, thus, be tabulated on a dense grid without requiring much memory space. The evaluation at an arbitrary radius uses cubic spline interpolation between the tabulated values. $\rho^\text{free}_{at}$ and $\delta \widetilde{\rho}_{at,lm}$ are tabulated on different types of grids. Without going into their detailed definitions, we note that both types of grids get denser closer to the nucleus. This allows the characteristic cusp of the electron density at the nucleus to be resolved with high accuracy, enabling an all-electron full-potential formalism which does not rely on pseudopotentials. While the evaluation of \cref{eq:rho_MP} at an arbitrary point $\mathbf{r}$ requires only data stored on 1-dimensional grids, the evaluation of \cref{eq:integrate_MP} at one given $r$ still requires $\delta \rho_{at}(\mathbf{r})$ on a 2-dimensional angular grid centered around $\mathbf{R}_{at}$. Consequently, evaluating \cref{eq:integrate_MP} at all grid values of $r$ at which we want to tabulate $\delta \widetilde{\rho}_{at,lm}$ ultimately does require an atom-centered 3-dimensional grid consisting of radial shells of angular grids. The full grid in FHI-aims which corresponds to the regular grids used in DFT-PWP thus consists of overlapping atom-centered grids.

\subsection{Solving the Poisson equation}\label{sec:Poisson}

The electrostatic potential $\Phi$ generated by a charge density $\rho$ in the absence of any other charge carriers or external fields is defined by the Poisson equation (PE)
\begin{equation}\label{eq:Poisson}
    \nabla^2 \Phi(\mathbf{r}) = - 4\pi\rho(\mathbf{r})
\end{equation}
A unique solution (in some cases up to a constant offset) is defined by appropriate boundary conditions, e.g. $\lim_{r\to\infty}\Phi=0$ or periodicity at the boundary of a unit cell. The density $\rho$ here is more general than the electron density discussed in the previous section and may, for example, include nuclei.

Solving \cref{eq:Poisson} is an integral part of electronic structure theory and consequently has to be implemented in any DFT software package either directly or through external libraries. Different practical approaches have been developed. In FHI-aims, the linearity of the $\nabla^2$ operator is exploited to decompose the electronic contribution to the potential into individual $v^\text{free}_{at}$ and $\delta\widetilde{v}_{at,lm}$ in complete analogy to \cref{eq:rho_MP}. These components are obtained from $\delta\widetilde{\rho}_{at,lm}$
\begin{equation}\label{eq:Poisson_aims2}
    \delta\widetilde{v}_{at,lm}(r^\prime) = \int_0^{r^\prime} dr\, r^2 g_l(r,r^\prime) \delta\widetilde{\rho}_{at,lm}(r) + \int_{r^\prime}^\infty dr\, r^2 g_l(r^\prime,r) \delta\widetilde{\rho}_{at,lm}(r)
\end{equation}
using the symmetry dependent radial Green's function $g_l(r_1,r_2) = 4\pi / (2l+1) \,\cdot\, r_1^l / r_2^{l+1}$ of the $\nabla^2$ operator.

By contrast, starting from $\rho(\mathbf{r})$, regular grid based codes typically use FFT to obtain $\hat{\rho}(\mathbf{g})$. In reciprocal space, the Laplace operator $\nabla^2$ is transformed into a simple multiplication with $|\mathbf{g}|^2$, which can easily be inverted to obtain $\hat{\Phi}(\mathbf{g})$ and, through inverse FFT, $\Phi(\mathbf{r})$.\cite{andreussi2012}

\subsection{The generalized Poisson and Poisson-Boltzmann equations}\label{sec:GPE_PBE}

The situation becomes more complicated compared to \cref{eq:Poisson} when investigating systems in which charges are present which we do not want to explicitly include in $\rho$. In this case, we need to make assumptions about the behavior of the remaining charges and their impact on $\Phi$. For example, consider some explicit charge distribution $\rho$ which is immersed in a local, linear, and isotropic dielectric medium. We may not want to include the polarization charge of the medium in $\rho$. Instead, we can describe it by its relative dielectric permittivity $\varepsilon$ in the generalized Poisson equation (GPE)\cite{andreussi2019}
\begin{equation}\label{eq:GPE}
    \nabla\left(\varepsilon(\mathbf{r})\nabla\Phi(\mathbf{r})\right) = - 4\pi\rho(\mathbf{r})
\end{equation}
If, additionally, an electrolyte is dissolved in the solvent medium and we do not want to treat the electrolyte charge explicitly as part of $\rho$ either, then we can treat each ionic species $ion$ as a separate source term $\rho_{ion}$
\begin{equation}\label{eq:PBE1}
    \nabla\left(\varepsilon(\mathbf{r})\nabla\Phi(\mathbf{r})\right) = - 4\pi\left(\rho(\mathbf{r})+\sum_{ion}\rho_{ion}(\mathbf{r})\right)
\end{equation}
which follows the Poisson-Boltzmann equation (PBE)\cite{nattino2018}
\begin{equation}\label{eq:PBE2}
    \rho_{ion}(\mathbf{r}) = z_{ion}\,\rho_{\infty, {ion}}\gamma(\mathbf{r}) \,\exp\left(-\frac{z_{ion}\Phi(\mathbf{r})}{k_\text{B} T}\right)
\end{equation}
with the charge $z_{ion}$ and bulk concentration $\rho_{\infty,{ion}}$ of species $ion$, the Boltzmann constant $k_\text{B}$, the temperature $T$, and a function $\gamma$ which is $=1$ in regions of space that are fully accessible to the electrolyte and $=0$ in regions that the electrolyte cannot enter, with some transition between the regions.

\subsection{Continuum embedding}

The modifications of PE described in the previous section are generally accurate only on a macroscopic scale. Nonetheless, the idea of using them as an approximation on an atomic scale has led to the development of continuum embedding models. When simulating molecules or surfaces in solution, this allows us to treat only the solute (and possibly a small number of solvent molecules) on an atomistic level while representing (most of) the solvent as a dielectric continuum.\cite{andreussi2019,tomasi2005} In the present work, we focus on the continuum embedding software package Environ.

Different variations of the two basic models described in the previous section are implemented in Environ, such as a modification of PBE due to finite ion size and a linearized version of it\cite{nattino2018}, as well as Stern layer corrections.\cite{dabo2010,ringe2016} The simpler planar countercharge, Gouy-Chapman-Stern,\cite{nattino2018} and Mott-Schottky\cite{campbell2019} electrolyte models for surfaces are also available, as well as further modifications of the potential such as periodic boundary corrections\cite{andreussi2014}, confining potentials\cite{nattino2019}, and external charges.

What these models have in common is that they cannot be solved directly using the formalism in \cref{eq:Poisson_aims2}. This is because the symmetry dependent radial Green's function $g_l(r_1,r_2)$ is derived from a multipole decomposition of the three-dimensional Green's function of the $\nabla^2$ operator.\cite{jacksonElectrodynamics} No general and straightforward decomposition of this kind exists for the $\nabla(\varepsilon(\mathbf{r})\nabla)$ operator with arbitrary $\varepsilon(\mathbf{r})$. Furthermore, \cref{eq:PBE1,eq:PBE2} are interdependent and generally require iterative solver algorithms.

In addition to implementing these solvers, Environ handles a range of additional tasks. The permittivity $\varepsilon(\mathbf{r})$ and the function $\gamma(\mathbf{r})$ in PBE need to be defined. To this end, the region occupied by the solute, called the solvation \textit{cavity}, is introduced. It is assumed to be inaccessible to solvent molecules. It is defined by a cavity function $s(\mathbf{r})$ which is $=1$ for $\mathbf{r}$ inside the cavity and $=0$ in the solvent. This allows for a definition $\varepsilon(\mathbf{r})=\varepsilon(s(\mathbf{r}))$ such that $\varepsilon(s=1)=1$ and $\varepsilon(s=0)=\varepsilon_\text{bulk}$ with the permittivity $\varepsilon_\text{bulk}$ of the bulk solvent. Similarly, $\gamma(\mathbf{r})=\gamma(s(\mathbf{r}))$ with $\gamma(s=1)=0$ and $\gamma(s=0)=1$.\cite{andreussi2019,nattino2018} The transition of $\varepsilon(s)$ and $\gamma(s)$ as functions of $s$ between these extremes is continuous and monotonic. Throughout the present work, we use the term `cavity' in a strict sense only for the region in space where $s\cong 1$. The transition region with $0<s\ll 1$ is considered to be outside of the cavity. We will use the shorthand terms \textit{inside}/\textit{outside} for `inside/outside of the cavity' henceforth.

Environ provides various options to define the cavity. The self-consistent continuum solvation (SCCS) model\cite{andreussi2012} defines $s$ as a function of the electron density
\begin{equation}\label{eq:s_sccs}
    s_\text{SCCS}(\mathbf{r}) = \begin{cases}
        0, & \rho^\text{el}(\mathbf{r}) \leq \rho_\text{min} \\
        t(\ln(\rho^\text{el}(\mathbf{r}))), & \rho_\text{min} < \rho^\text{el}(\mathbf{r}) < \rho_\text{max} \\
        1, & \rho^\text{el}(\mathbf{r}) \geq \rho_\text{max}
    \end{cases}
\end{equation}
where $t$ is a function which smoothly and monotonically switches from 0 to 1, and $\rho_\text{min}$ and $\rho_\text{max}$ are model parameters. The soft-sphere continuum solvation (SSCS) model\cite{fisicaro2017} defines the cavity based on the scaled van-der-Waals radii $r^\text{vdW}_{at}$ of the atoms
\begin{equation}
    s_\text{SSCS}(\mathbf{r}) = 1 - \prod_{at} \frac{1}{2}\left(1 +\text{erf}\left(\frac{r_{at}-\alpha r^\text{vdW}_{at}}{\Delta}\right) \right)
\end{equation}
where the scaling factor $\alpha$ and the softness $\Delta$ are global model parameters which, in the base version of SSCS, do not depend on atom $at$. However, an additional electric flux dependent prefactor can be introduced into the $\alpha r^\text{vdW}_{at}$ term to define a field-aware interface.\cite{truscott2019} It is clear that $s_\text{SSCS}$ is never analytically equal to 1, which is why we defined the cavity by $s\cong 1$ rather than $s=1$ above. Both SCCS and SSCS can be modified to take the finite size of the solvent molecules into account in a solvent-aware cavity definition.\cite{andreussi2019_2} Finally, $s$ can be defined as a sphere (cluster-like system geometry), cylinder (system periodic in 1 dimension), or slab (system periodic in 2 dimensions)
\begin{equation}\label{eq:s_system}
    s_\text{system}(\mathbf{r}) = \frac{1}{2}\text{erfc}\left(\frac{\sqrt{\sum_i(\mathbf{r}_i-\mathbf{r}_{0,i})^2}-W_\text{system}}{\Delta}\right)
\end{equation}
where $\text{erfc}(x)=1-\text{erf}(x)$, the softness $\Delta$ and width $W_\text{system}$ are parameters, the sum runs over all non-periodic dimensions of the system, $\mathbf{r}_i$ is the cartesian coordinate of $\mathbf{r}$ in dimension $i$, and $\mathbf{r}_{0,i}$ are the respective coordinates of the system center $\mathbf{r}_0$. Additionally, Environ provides an option to include one or multiple additional dielectric regions of similar shape to \cref{eq:s_system} with individual dimensionality, softness, width, and center, originally introduced to model ice surfaces with molecular adsorbates.\cite{bononi2020}

To account for effects such as surface tension or dispersion and Pauli repulsion interactions between solute and solvent, continuum embedding models commonly include some non-electrostatic energy contribution. In Environ, this takes the empirical form of an effective pressure and an effective surface tension which are applied to the cavity volume and surface area, which are calculated as the space integrals of $s$ and $|\nabla s|$, respectively.\cite{andreussi2019} Furthermore, the continuum embedding software needs to compute the solvent potential which enters the Kohn-Sham (KS) operator in the host DFT program. Apart from the solvent contribution to the electrostatic potential, this may also include functional derivatives due to the electron-density-dependent cavity definition in SCCS and any derived terms such as the cavity surface area. Finally, the force contributions from the solvent model are computed.

To solve GPE, PBE, or their variations in practice, iterative algorithms are employed which at each step solve the regular PE~(\ref{eq:Poisson}) with the source term (right hand side) modified in such a way that after convergence, the resulting $\Phi$ fulfills the specified electrostatic problem for the original $\rho$. Environ closely follows the formalism of its original host DFT-PWP code\cite{giannozzi2017} using an FFT based Poisson solver. This makes it necessary that the charge density and all derived quantities such as $\varepsilon$ and $\gamma$ are stored on a regular grid. All operations performed on them or necessary to compute them are implemented in a way that assumes this structure.

It would, in principle, be possible to recreate these iterative algorithms in FHI-aims using \cref{eq:Poisson_aims2} for the `inner' Poisson solver which is called at each iteration step. Indeed, FHI-aims currently features implementations of the conductor-like screening model (COSMO)\cite{klamt1993}, Stern layer modified Poisson-Boltzmann (SMPB),\cite{ringe2016} and multipole expansion (MPE)\cite{sinstein2017,filser2022,stishenko2024} methods. While these specialized implementations yield optimized computational performance\cite{filser2022}, their limited scope is a considerable disadvantage compared to the wide range of functionality implemented in a general purpose continuum embedding library like Environ.

Reproducing all of Environ's already existing methods individually in FHI-aims would lead to a difficult-to-maintain, inflexible software extension which would require constant updates as novel embedding methods are developed. Conversely, one might consider introducing additional levels of abstraction into Environ to enable it to operate on data stored on different kinds of grids. However, apart from having to rewrite all existing objects and operations in an overlapping atom-centered grid framework, such an abstraction would also impose that same duplicate effort on future developments in this package.

Instead, here we describe a generic procedure to convert charge densities between the grids of FHI-aims and Environ. Following the modular programming paradigm, this minimal, easy-to-maintain interface immediately makes any methods implemented in either code (currently or in the future) interoperable as long as the underlying physical models are compatible. This is possible because, despite its complexity, continuum embedding generally needs to communicate only a very limited set of data with the host DFT program, as described in the next section.

\section{The interface}

\subsection{Communication}

Continuum embedding interacts with DFT by modifying the electrostatic potential in the KS operator. This modification is applied at each self-consistent field (SCF) step using the electron density at that step, thereby allowing for full self-consistency of the embedded DFT system. After convergence, energy and force contributions are computed, also allowing for geometry relaxation in full consistency with the continuum model. Generally, the DFT program does not require any further input from the embedding program. Conversely, the continuum embedding package Environ is designed to operate on only a minimal set of input data from its host DFT program, namely the electron density, the nuclear positions and charges, and the unit cell.

\begin{figure}[!tb]
    \centering
    \includegraphics[width=0.95\linewidth]{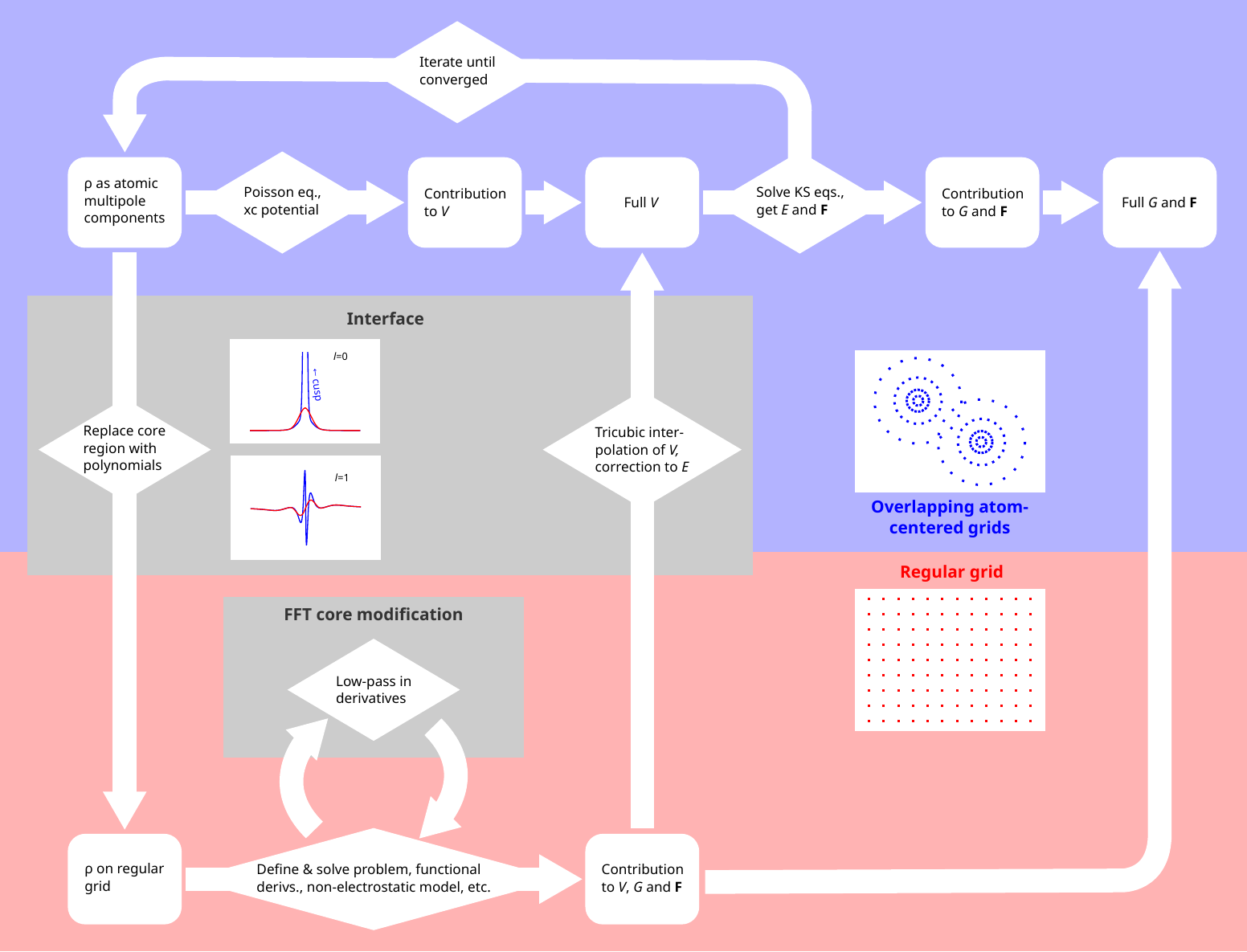}
    \caption{Communication workflow between FHI-aims (blue/top) and Environ (red/bottom). At each SCF step, the components of $\rho^\text{MP}$ are smoothened in a way that conserves the long range multipole moments of the core regions, cf. \cref{sec:smoothing}. Shown as inserts are examples of original $l=0$ and $l=1$ components (blue) and their smoothened counterparts (red). The smoothened density is evaluated on Environ's regular grid. Environ defines the cavity and solves the specified electrostatic problem. It computes any additional contributions and corrections specified by the user. To avoid artifacts in $\nabla s$, Environ's FFT core subroutines are modified to include a low-pass filter in the derivatives, cf.~\cref{sec:lowpass}. The solvent contributions to the effective potential $V_\text{eff}^\text{solv}$ in the KS operator, the free energy in solution $G^\text{solv}$, and the forces $\mathbf{F}^\text{solv}_{at}$ (symbols shortened in figure for clarity) are communicated to FHI-aims. The potential is interpolated back onto the overlapping atom-centered grids and added to the vacuum contribution computed by FHI-aims. An additional energy correction is computed, as described in \cref{sec:full_E}. The KS equations are solved, the density is updated, and the process is repeated to self-consistency.}
    \label{fig:workflow}
\end{figure}

The general workflow of the resulting interface between FHI-aims and Environ is outlined in \cref{fig:workflow}. Within this framework, both FHI-aims and Environ are free to use any methods to compute their contributions to potential, energy, and forces as specified by the user in the separate input files for both programs. At the center of the interface stands a smoothing scheme described in \cref{sec:smoothing}. This scheme allows us to convert the electron density from the grid of FHI-aims to that of Environ without losing important information contained in the core region where an atom-centered grid is considerably more fine-grained than a regular grid. Furthermore, we include a low-pass filter in the FFT based derivatives as described in \cref{sec:lowpass} to avoid high frequency artifacts in the smoothened density. We note in passing that in contrast to the electron density and its electrostatic potential, Environ already has methods implemented to smear out the nuclei and their potentials, which will therefore not be further discussed here.

Specifically, the quantities listed below are computed and communicated at certain steps of the simulation. This procedure assumes that FHI-aims and Environ are parallelized using the same number of processes. In both programs, each process handles a part of the respective program's grid and the corresponding data. However, the regions in space covered by one process in Environ and by the process with the same process ID in FHI-aims generally do not coincide, owing to the different nature of the grids and the corresponding partitioning schemes.
\begin{itemize}
    \item During (re-)initialization, the unit cell and nuclear positions and charges are sent from FHI-aims to Environ. Environ returns smoothing information ($r^\text{cut}_{at}$ (SSCS) or $\rho_\text{max}$ (SCCS), see \cref{sec:cutoff}) to FHI-aims.
    \item At the beginning of the potential update, Environ sends its grid points and meta-data to FHI-aims. FHI-aims computes the smoothened electron density on those grid points and sends it to Environ. Parallelization is trivial at this point. While each FHI-aims process handles only a part of the 3-dimensional grids, the 1-dimensional tabulated multipole components of every atom are synchronized across all processes. Each FHI-aims process simply computes the smoothened density on the grid points of the Environ process with the same process ID. Environ computes potential, energy, and, if requested, force components. Energy (and force) are sent to FHI-aims. Each Environ process writes the potential on its part of the grid to a separate temporary file.
    \item During the potential update, FHI-aims interpolates the potential contribution from Environ back onto its own grid using tricubic interpolation and adds it to its own potential contribution. The latter is computed in the same way that FHI-aims would if no continuum embedding was used. Each FHI-aims process loads the Environ potential only from those temporary files which it needs for its part of the grid as determined using the stored grid meta-data. A double counting energy correction is integrated on-the-fly as described in \cref{sec:full_E}.
\end{itemize}

\subsection{Multipole-conserving smoothing}\label{sec:smoothing}

While at first it may seem straightforward to evaluate \cref{eq:rho_MP} on a regular grid and pass this data to Environ, this would lead to inconsistencies. In all-electron DFT, a substantial amount of charge is localized in a narrow cusp around the core which is smaller than the volume element corresponding to one regular grid point. Therefore, when integrating the total charge or computing the electrostatic potential, the result would depend on where exactly or whether at all a grid point happens to lie in the cusp region.

A solution to this problem becomes apparent as we realize that when combining continuum embedding with DFT, we do not actually need the full potential of the former. The DFT program can compute the nuclear and electronic potentials as usual and requires only the solvent contribution from the embedding program. For the electrostatic part of the potential, this contribution is simply defined as
\begin{equation}\label{eq:solv_pot_def}
    \Delta \Phi^\text{solv} = \Phi^\text{solv} - \Phi^\text{vac}
\end{equation}
where $\Phi^\text{vac}$ is the solution of PE~(\ref{eq:Poisson}) in vacuum and $\Phi^\text{solv}$ is the solution of the electrostatic problem defined by the user, e.g. \cref{eq:GPE} or \cref{eq:PBE1,eq:PBE2}. `Vacuum' in this context refers to the energy and potential without any solvent contributions but not necessarily of the vacuum ground state. The ground state electron density in solution or, in fact, any instantaneous electron density may be used in \cref{eq:solv_pot_def}, as long as the same density is used for both terms. The different electrostatic models can generally be described as variations of \cref{eq:PBE1} distinguished by the definition of $\varepsilon$ on the left hand side and by the source terms in addition to $\rho$ on the right hand side. However, all of these models locally reduce to \cref{eq:Poisson} inside the cavity, leading to
\begin{equation}\label{eq:laplace_delta_phi}
    \nabla^2 \Delta \Phi^\text{solv}(\mathbf{r}) = 0 \quad \forall \mathbf{r}: s(\mathbf{r})=1
\end{equation}
In other words, the solvent potential $\Delta \Phi^\text{solv}$ \textit{inside} does not locally depend on $\rho$, and it is the harmonic continuation of the \textit{outside} part of $\Delta \Phi^\text{solv}$. This leads to the idea of using a smoothened charge density $\Bar{\rho}$ to compute smoothened potentials $\Bar{\Phi}^\text{solv}$, $\Bar{\Phi}^\text{vac}$, and ultimately $\Delta \Bar{\Phi}^\text{solv}$ in complete analogy to \cref{eq:solv_pot_def}. As long as $\Bar{\Phi}^\text{solv}$ and $\Bar{\Phi}^\text{vac}$ are equal to their non-smoothened counterparts \textit{outside}, $\Delta \Bar{\Phi}^\text{solv}$ will be equal to $\Delta \Phi^\text{solv}$ both \textit{inside} and \textit{outside} because \cref{eq:laplace_delta_phi} together with the value and derivatives at the cavity boundary uniquely defines the behavior \textit{inside} and it does not locally depend on $\rho$, $\Bar{\rho}$, or any other quantity. The full electrostatic potential in solution can then be obtained as
\begin{equation}\label{eq:Phi_aims_plus_env}
    \Phi^\text{solv} = \Phi^\text{vac} + \Delta \Bar{\Phi}^\text{solv}
\end{equation}
where $\Phi^\text{vac}$ can be computed in the way that the DFT program normally would. Here and in the following, the bar $\Bar{\quad}$ indicates quantities which are computed using the smoothened charge density $\Bar{\rho}$. 

It is sufficient for $\Bar{\rho}$ to be $=\rho$ \textit{outside} and conserve the long range atomic multipole moments of the portion of $\rho$ that is \textit{inside}. If this is the case, then $\Bar{\Phi}^\text{vac}$ reproduces $\Phi^\text{vac}$ \textit{outside} as will be shown below, cf.~\cref{eq:point_multipole}. $\nabla^2\Phi^\text{solv}$ is generally a semi-local function of $\rho$ and of $\Phi^\text{solv}$ itself, implying that it has to be found iteratively. Using $\Phi^\text{vac}$ as an initial guess for $\Phi^\text{solv}$, this means that $\Bar{\Phi}^\text{solv}$ reproduces $\Phi^\text{solv}$ \textit{outside} if $\Bar{\Phi}^\text{vac}=\Phi^\text{vac}$ and $\Bar{\rho}=\rho$, both \textit{outside}. This is described in more detail in \cref*{si-sec:elstat_pot_equal} of the supplementary information (SI) of the present article.

Furthermore, if $\Bar{\rho}$ is $=\rho$ \textit{outside} and smoothly increases moving \textit{inside} from the cavity boundary, then $s$ will be unaltered in SCCS. In field-aware SSCS, the scaling factors depend on the electric flux through the exposed surfaces of the individual soft spheres which at the point of determining the factors are not yet scaled by these same factors. These scaling factors and, consequently, $s$ are unaltered if $\nabla\Bar{\Phi}^\text{solv}=\nabla\Phi^\text{solv}$ at the cavity surface. We reiterate that the cavity surface in Environ is a smeared out region defined by $|\nabla s|\neq 0$ which we consider to be \textit{outside} in this work.

\begin{figure}[!htb]
    \centering
    \includegraphics[width=0.95\linewidth]{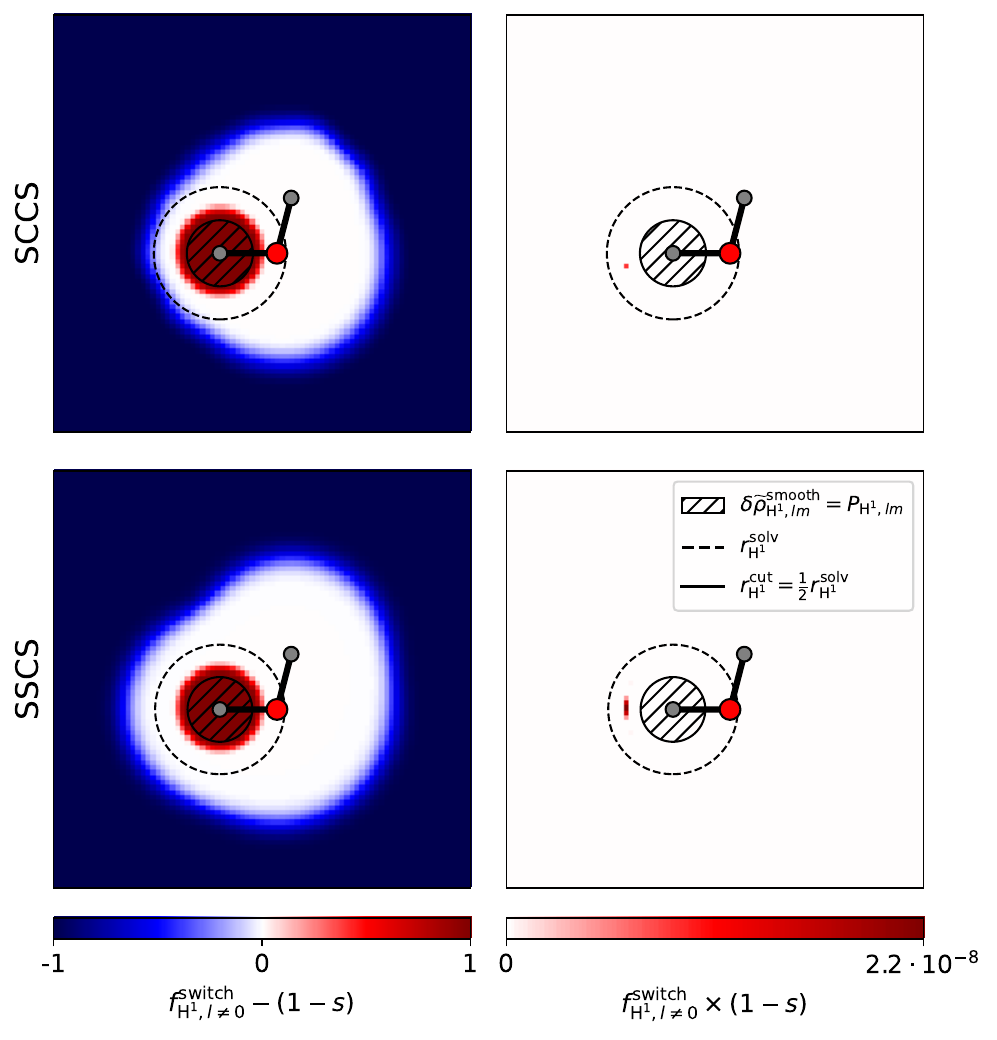}
    \caption{Water molecule in implicit solvent. Difference (left) and overlap (right) between $f^\text{switch}_{\text{H}^1,l\neq0}$ of one hydrogen atom H$^1$ and inverted cavity function $1-s$ of entire molecule for SCCS (top) and SSCS (bottom), cut through the molecular plane. $s$ computed from converged $\rho^\text{el}$ for SCCS. Contrast in right images chosen to capture maximum value across both images. Radius $r^\text{solv}_{\text{H}^1}$ of dashed circle estimates shortest distance of H$^1$ from any point \textit{outside}. Inside of a sphere with half of that radius (hatched area), $\delta \widetilde{\rho}_{\text{H}^1,lm}(r)$ is replaced by a polynomial $P_{\text{H}^1,lm}(r)$. In transition region of $f^\text{switch}_{\text{H}^1,l\neq0}$, original $\delta \widetilde{\rho}_{\text{H}^1,lm}(r)$ is used but integration errors in this region are still partially compensated in inner (hatched) region. For $l=0$, integration errors over entire $\mathbb{R}^3$ are compensated. Corresponding plots for the O atom, different cavity definitions, and ionic systems shown in \cref*{si-sec:separation_smoothing_solvent} in the SI.}
    \label{fig:regions}
\end{figure}

These considerations allow us to modify the multipole components $\delta \widetilde{\rho}_{at,lm}(r_{at})$ within a certain cutoff radius $r_{at}^\text{cut}$ around the nuclei which lies safely \textit{inside}, as illustrated in \cref{fig:regions}. The choice of this radius is discussed in \cref{sec:cutoff}. The modified multipole components should reproduce the long range potential of the unmodified $\delta \widetilde{\rho}_{at,lm}$ outside $r_{at}^\text{cut}$. To this end, we replace the multipole components with 6th order polynomials inside $r_{at}^\text{cut}$
\begin{equation}
    \delta \widetilde{\rho}^\text{smooth}_{at,lm}(r)=\begin{cases}
        P_{at,lm}(r), & r < r_{at}^\text{cut} \\
        \delta \widetilde{\rho}_{at,lm}(r), & r\geq r_{at}^\text{cut}
    \end{cases}
\end{equation}
We choose the polynomial coefficients to fulfill two types of conditions. First, the value and first and second derivatives of $\widetilde{\rho}^\text{smooth}_{at,lm}(r_{at})Y_{lm}(\Omega_{at})$ are continuous at $r_{at}=0$ and at $r_{at}=r_{at}^\text{cut}$. Continuity in the angular directions is fulfilled by construction, leaving only continuity in $r$ to be considered. This leaves us with 6 continuity conditions so far. Second, the polynomial reproduces the correct long range multipole moment of the multipole component it replaces. The electrostatic potential generated by $\delta \widetilde{\rho}^\text{smooth}_{at,lm}(r)$ at any $r^\prime>r_{at}^\text{cut}$ is exactly equal to that generated by $\widetilde{\rho}_{at,lm}(r)$ if
\begin{equation}\label{eq:multipole_condition_analytic}
    \int\limits_0^{r_{at}^\text{cut}} dr\, r^{l+2} P_{at,lm}(r) = \int\limits_0^{r_{at}^\text{cut}} dr\, r^{l+2}\delta \widetilde{\rho}_{at,lm}(r)
\end{equation}
is fulfilled because the potential generated at $r^\prime$ by a multipole charge distribution confined to $r<r^\prime$ can be expressed exactly using a point multipole. This can be seen by inserting $g_l$ in \cref{eq:Poisson_aims2} and separating the $r^\prime$ dependent term out of the first integral
\begin{equation}\label{eq:point_multipole}
    \delta\widetilde{v}_{at,lm}(r^\prime) = \frac{1}{(r^\prime)^{l+1}}\int_0^{r^\prime} dr\, r^{l+2} \delta\widetilde{\rho}_{at,lm}(r) + \int_{r^\prime}^\infty dr\, r^2 g_l(r^\prime,r) \delta\widetilde{\rho}_{at,lm}(r)
\end{equation}
When replacing $\delta\widetilde{\rho}_{at,lm}(r)$ with $\delta \widetilde{\rho}^\text{smooth}_{at,lm}(r)$, the first integral in the above expression is unchanged if \cref{eq:multipole_condition_analytic} is fulfilled and $r^\prime>r_{at}^\text{cut}$. In the second integral, the integrand itself is unchanged if $r^\prime>r_{at}^\text{cut}$.

This leaves us with a total of 7 conditions, hence the use of 6th order polynomials. For the low order multipoles with $l\leq 2$, some of the continuity conditions at the origin are fulfilled by construction and do not need to be enforced, as shown in \cref*{si-sec:polynomial_order} in the SI. A 4th order polynomial is thus sufficient for $l=0$, as is 5th order for $l=1$ and $l=2$. The conditions are first enforced computing the left hand side of \cref{eq:multipole_condition_analytic} analytically and the right hand side numerically on the radial grid of FHI-aims. $\rho^\text{free}_{at}$ is smoothened to produce $\rho^\text{free, smooth}_{at}$ in almost the exact same manner, effectively treating it like $\delta \widetilde{\rho}_{at,00}$. The only difference is that FHI-aims uses different $r$ grids to store the free atom and multipole densities.

In a second step, defining the smoothened atomic multipole density
\begin{equation}\label{eq:rho_MP_smooth}
    \rho^\text{MP, smooth}_{at}(\mathbf{r}) = \rho^\text{free, smooth}_{at}(r_{at})+\sum_{l^\prime=0}^{l^\prime_\text{max}}\sum_{m^\prime=-l^\prime}^{l^\prime} \delta \widetilde{\rho}_{at,l^\prime m^\prime}^\text{smooth}(r_{at}) Y_{l^\prime m^\prime}(\Omega_{at})
\end{equation}
we check the similar condition
\begin{equation}\label{eq:multipole_condition_numeric}
\begin{split}
    \int\limits_{\mathbb{R}^3} d\mathbf{r}\, (r_{at})^{l}&\, \rho^\text{MP, smooth}_{at}(\mathbf{r})\, Y_{lm}(\Omega_{at})\, f^\text{switch}_{at,l}(\xi_{at}(r_{at})) \\
    &\stackrel{?}{=} \delta_{l0}\sqrt{\frac{1}{4\pi}}Z_{at}+ \int\limits_0^{\infty} dr\, r^{l+2}\,\delta\widetilde{\rho}_{at,lm}(r) \,f^\text{switch}_{at,l}(\xi_{at}(r))
\end{split}
\end{equation}
evaluating the left hand side on the regular grid and, again, the integral on the right hand side on the radial grid. 
$\delta_{l0}$ is the Kronecker delta and $Z_{at}$ is the nuclear charge. The switching function $f^\text{switch}_{at,l}$ for $l\neq 0$ smoothly switches from 1 inside $r^\text{cut}_{at}$ to 0 outside of it and is discussed in \cref{sec:switch}. For $l=0$, $f^\text{switch}_{at,l=0}=1$ everywhere to conserve the total charge exactly. \Cref{eq:multipole_condition_numeric} checks the apparent atomic multipole moments when the smoothened atomic multipole density is evaluated on the regular grid, and it compares them to the exact atomic multipole moments obtained from $\delta\widetilde{\rho}_{at,lm}$. The residual error in \cref{eq:multipole_condition_numeric} is computed, and the polynomial coefficients are adjusted to correct for it while still conserving continuity conditions.

The adjustment of the polynomial coefficients due to \cref{eq:multipole_condition_numeric} is made using once again analytical integrals for the polynomials, which leaves room for residual errors when evaluated on the regular grid. To reproduce the atomic multipole moments exactly, the correction scheme needs to be iterated to self-consistency. In principle, this could be done at each step of the encompassing SCF procedure. In practice, we find the following scheme which requires only one density evaluation per SCF step to be sufficient: Keep record of the residual errors in \cref{eq:multipole_condition_numeric} cumulatively over SCF steps. At each SCF step, use \cref{eq:multipole_condition_analytic} to get polynomial coefficients. Using the cumulative errors from the previous SCF steps, add a correction to the coefficients. Compute $\Bar{\rho}^\text{el}=\sum_{at}\rho^\text{MP, smooth}_{at}$ on the regular grid and pass it to Environ. On the fly, check the new residual errors in \cref{eq:multipole_condition_numeric}. Add these new residual errors to the stored cumulative errors. Before the first SCF step, assume that $\sum_{at}\rho^\text{free}_{at}$ is the electron density, get $\rho^\text{free, smooth}_{at}$ using \cref{eq:multipole_condition_analytic}, set all $\delta \widetilde{\rho}_{at,lm}=0$, and compute \cref{eq:multipole_condition_numeric} to get the stored residual errors for the first SCF step. This update scheme effectively converges the residual error as the encompassing SCF procedure converges. In fact, the residual errors are typically small already in the initial SCF step and remain small throughout the procedure.

\subsection{Choice of cutoff radii}\label{sec:cutoff}

We choose
\begin{equation}\label{eq:default_rcut}
    r_{at}^\text{cut} = \frac{1}{2} r_{at}^\text{solv}
\end{equation}
where $r_{at}^\text{solv}$ is an estimate of the shortest distance of nucleus $at$ from any point \textit{outside}. In practice, $r_{at}^\text{cut}$ is slightly adjusted to conincide with the halfway point (in parameter space, cf.~ref.~\citenum{blum2009}: parameter $i$ in $r(i)$, sec.~3.1., and parameter $s$ in eq.~(18)) between two shells of the radial grid of FHI-aims. For SCCS, we choose $r_{at}^\text{solv}$ such that
\begin{equation}\label{eq:r_solv_SCCS}
    \rho^\text{free}_{at}(r_{at}^\text{solv}) = \rho_\text{max}
\end{equation}
As the electron density relaxes to the molecular geometry and the solvent potential, $s_\text{SCCS}$ can become $<1$ at points with shorter distances from $\mathbf{R}_{at}$ than $r_{at}^\text{solv}$ especially for cations and atoms with positive partial charges. Nonetheless, $r_{at}^\text{cut}$ stays contained \textit{inside} by setting it to half of $r_{at}^\text{solv}$, as illustrated in \cref{fig:regions} as well as \cref*{si-sec:separation_smoothing_solvent} in the SI. For SSCS, we choose
\begin{equation}\label{eq:r_solv_SSCS}
    r_{at}^\text{solv} = \alpha r^\text{vdW}_{at} - 2\Delta
\end{equation}
If the solute consists of only one atom, then $s_\text{SSCS}(r_{at}^\text{solv})=1/2\cdot\text{erfc}(-2)=0.998\cong 1$. Any additional atoms will push the solvent boundary away from nucleus $at$, never towards it. Although $1-s_\text{SSCS}$ is never analytically zero, we find that it is numerically zero inside $r_{at}^\text{cut}$, cf.~\cref{fig:regions} as well as \cref*{si-sec:separation_smoothing_solvent} in the SI. Field-aware and solvent-aware modifications of $s$ as well as additional dielectric regions are not taken into account when computing $r_{at}^\text{solv}$. For the `system' cavity \cref{eq:s_system}, we use \cref{eq:r_solv_SCCS} with default SCCS parameters. Alternatively, the user may specify $r_{at}^\text{cut}$ for each species manually regardless of cavity definition.

\subsection{Switching function for integration error check}\label{sec:switch}

Including $f^\text{switch}_{at,l}$ in \cref{eq:multipole_condition_numeric} is necessary because small errors at large $r$ (far ouside $r^\text{cut}_{at}$) have a strong impact on the multipole moment, whereas small corrections at small $r$ (inside $r^\text{cut}_{at}$) only have a much smaller impact. This is due to the factor $(r_{at})^l$ or $r^{l+2}$, respectively, in the integrand and is accordingly more pronounced for higher $l$. This would require large correction terms for small errors which, paradoxically, would lead exactly to those steep variations of the density in the core region which we want to avoid and, consequently, integration errors on the regular grid. Moreover, if integration errors do occur \textit{outside}, then corrections \textit{inside} will not fully recover the correct $\Delta \Bar{\Phi}^\text{solv}$ \textit{outside}, cf. the integration boundaries in \cref{eq:point_multipole}. If a numerical error occurs in the first integral, the correct potential at $r^\prime$ can still be recovered by adjusting the integrand in the same integral. However, if an error occurs in the second integral, no single adjustment in the first integral will generally recover the correct potential for all possible $r^\prime$. The appropriate remedy for such errors is a tighter integration grid. In this spirit, $f^\text{switch}_{at,l}$ also ensures that we only attempt to correct those errors which we actually can correct. The only exception to these considerations is $l=0$ for which $f^\text{switch}_{at,0}=1$ because we want the total charge of the entire system to be conserved exactly.

The reason why we do not sharply switch $f^\text{switch}_{at,l}$ from $1$ to $0$ at $r^\text{cut}_{at}$ is because it would lead to inconsistencies when, due to a small geometry change, this radius moves across a point on the regular grid. The density at that grid point would then sharply switch from being fully included in the left hand side integral in \cref{eq:multipole_condition_numeric} to being fully excluded. A smeared out switching function allows for smoothly varying results with small geometry changes. The switching function is defined such that it is continuous and continuously differentiable to second order.
\begin{equation}
    f^\text{switch}_{at,l}(\xi) = \begin{cases}
        1, & \xi \leq 0 \quad \text{or} \quad l=0 \\
        1-10\xi^3+15\xi^4-6\xi^5, & 0<\xi<1 \quad \text{and} \quad l\neq0 \\
        0, & \xi \geq 1 \quad \text{and} \quad l\neq0
    \end{cases}
\end{equation}
with a shifted and scaled radius
\begin{equation}
    \xi_{at}(r) = \frac{r-r_{at}^\text{cut}}{2d_\text{diag}}
\end{equation}
with
\begin{equation}\label{eq:d_diag}
    d_\text{diag} = \sqrt{3}\max_i d_i
\end{equation}
where $d_i$ is the grid spacing along lattice vector $i$. In the case of an orthorhombic cell and equal spacing of the regular grid along all three lattice vectors, the volume element corresponding to one grid point is a cube and $d_\text{diag}$ is the length of its space diagonal. Assuming such cubic volume elements, we consider a case where a grid point $\mathbf{r}$ exists such that $\mathbf{r}-\mathbf{R}_{at}$ is pointing along the space diagonal and $\xi_{at}(r_{at})=0$, i.e., the sphere with radius $r_{at}^\text{cut}$ around $\mathbf{R}_{at}$ cuts exactly through $\mathbf{r}$. In such a case, $f^\text{switch}_{at,l}$ will be $=1$ at gridpoint $\mathbf{r}$, it will be $=1/2$ at the diagonally next neighboring grid point, and it will be $=0$ at the diagonally second neighbor.


\subsection{Low-pass filter}\label{sec:lowpass}

In addition to the smoothing which happens before the electron density is passed to Environ as described in the previous sections, we introduce a second modification in Environ itself. This modification is a low-pass filter in the FFT based core subroutines which compute the derivatives of a function on the regular grid. 
%
While our smoothing scheme makes $\rho^\text{el}$ representable on the real space regular grid, high frequency artifacts may occur in the discrete Fourier transform. 
The $\nabla$ operator transforms into a multiplication with the frequency vector $i\mathbf{g}$ in Fourier space. The imaginary unit $i$ in the prefactor corresponds to a phase shift which makes oscillations visible in $\nabla \rho^\text{el}$ which are hidden in $\rho^\text{el}$ because they pass through zero at the grid points before the phase shift. In addition, artifacts at a high frequency are amplified in the gradient by that same frequency.


\begin{figure}[!htb]
    \centering
    \includegraphics[width=0.95\linewidth]{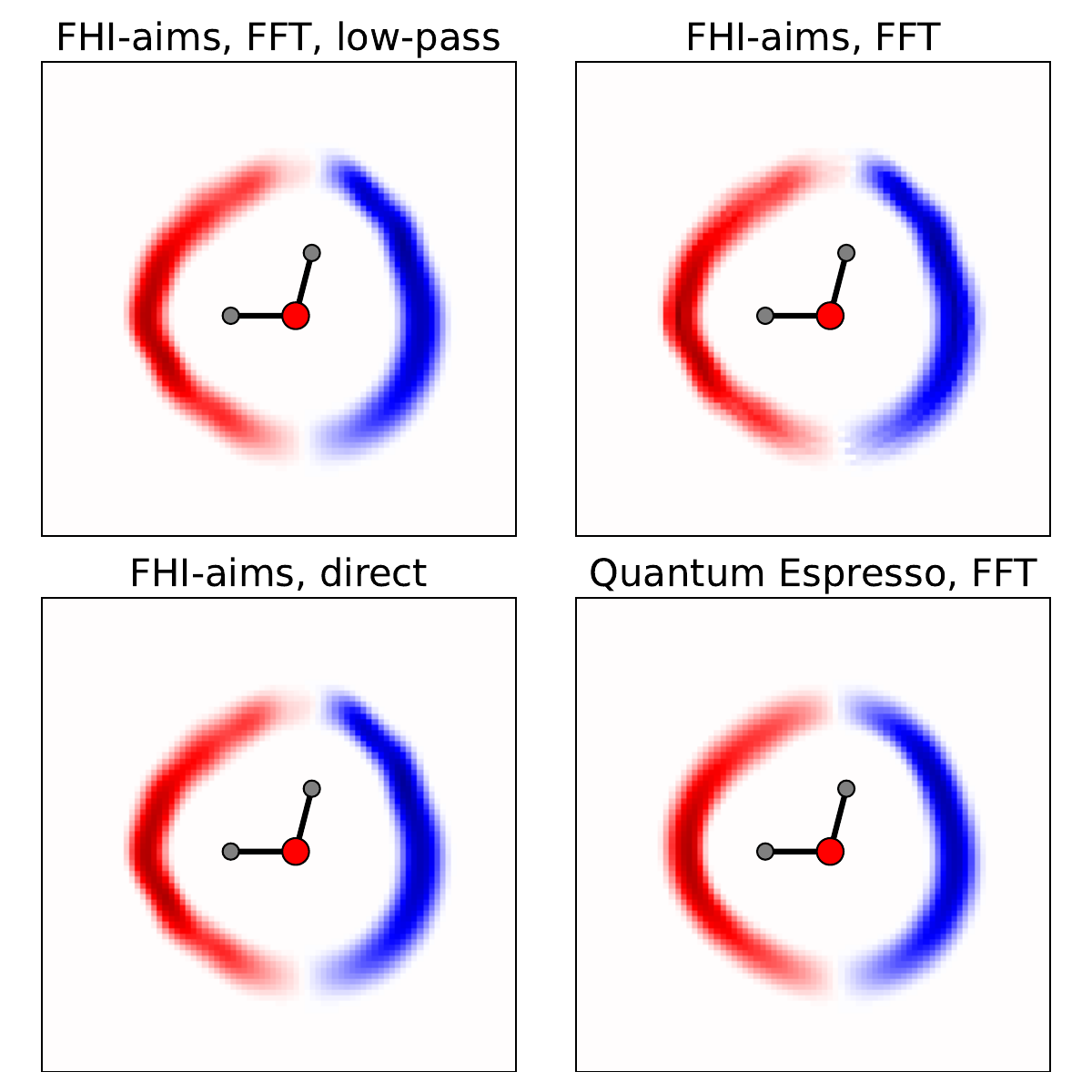}
    \caption{Gradient component $\partial s / \partial x$ for the same system as in \cref{fig:regions}, SCCS. Red and blue show positive and negative values, respectively, at an arbitrary scale. $\nabla s$ computed from $\nabla \Bar{\rho}^\text{el}$ via chain rule after SCF convergence. Titles refer to computation of $\nabla\Bar{\rho}^\text{el}$. Top left: FFT with low-pass. Top right: FFT without low-pass. Bottom left: computed in FHI-aims using the analytical derivative of \cref{eq:rho_MP}. Filtered FFT version correctly reproduces this reference. Bottom right: for reference, FFT without low-pass, the DFT program being Quantum Espresso\cite{giannozzi2017} instead of FHI-aims. While the overall less smooth shape of the cavity in FHI-aims is a result of its atom-centered basis set and is consistently present, cf. also \cref{fig:regions}, the ripples in the unfiltered FFT derivative are an artifact that is absent in the analytical derivatives and can be removed by a low-pass filter. Corresponding plots for $\nabla^2 s$ and differences of filtered to unfiltered and direct gradients shown in \cref*{si-fig:gradients_diff,si-fig:laplacians} in the SI.}
    \label{fig:gradients}
\end{figure}

For our purposes, this mainly affects the computation of the atomic forces, as well as the gradient, Laplacian, and Hessian of $s$ in SCCS. The latter terms enter the KS potential (cf.~\cref{sec:full_E}) and are themselves obtained from the derivatives of $\Bar{\rho}^\text{el}$ via the chain rule. Artifacts and instabilities in the derivatives of $s$ therefore directly translate to the energy, KS potential and forces. It would, in principle, be possible to compute the derivatives of $\Bar{\rho}^\text{el}$ in FHI-aims using the analytical derivatives of \cref{eq:rho_MP}. 
To keep communication minimal, we choose to address both issues in Environ's core FFT subroutines instead. For consistency, this affects all FFT based derivatives,
not just when applied to $\Bar{\rho}^\text{el}$. We replace these four subroutines (gradient, Laplacian, Hessian, and forces) with modified versions which include a low-pass filter. $\mathbf{g}$ space in Environ is confined to a sphere with
\begin{equation}
    |\mathbf{g}|^2 \leq |\mathbf{g}_\text{cut}|^2
\end{equation}
We choose a filter
\begin{equation}
    \text{filter}(\mathbf{g}) = \frac{1}{2}\text{erfc}\left(p_1 \frac{|\mathbf{g}|^2}{|\mathbf{g}_\text{cut}|^2}-p_2 \right)
\end{equation}
with two tunable parameters $p_1 > p_2 > 0$. As illustrated in \cref{fig:gradients}, this effectively removes artifacts in $\nabla s$. If inaccuracies arise due to actual data (as opposed to artifacts) being filtered, these errors can be systematically reduced by choosing a denser grid which corresponds to a higher $|\mathbf{g}_\text{cut}|$.

\subsection{Full energy, force and potential expressions}\label{sec:full_E}

The full free energy in solution $G^\text{solv}$, the force $\mathbf{F}^\text{solv}_{at}$ acting on atom $at$, and the effective potential $V_\text{eff}^\text{solv}$ which enters the KS operator are obtained by summing up the respective vacuum contributions computed by FHI-aims and the solvent contributions computed by Environ.
\begin{subequations}\label{eq:sum_all}
\begin{align}
    G^\text{solv} &= E^\text{vac} + \Delta \Bar{G}^\text{solv} \label{eq:sum_E}\\
    \mathbf{F}^\text{solv}_{at} &= \mathbf{F}^\text{vac}_{at} + \Delta\Bar{\mathbf{F}}^\text{solv}_{at} \label{eq:sum_F} \\
    V_\text{eff}^\text{solv} &= V_\text{eff}^\text{vac} + \Delta \Bar{V}_\text{eff}^\text{solv} \label{eq:sum_V}
\end{align}
\end{subequations}
$G^\text{solv}$ is formally a Gibbs free energy since simplifying the solvent degrees of freedom into a dielectric model formally amounts to an (approximate) integration over these degrees of freedom and we generally consider the pressure to be a constant property of the model solvent. We reiterate that the `vacuum' contributions are calculated using the pure DFT Hamiltonian without embedding but on the ground state electron density in solution. In this context, $\Delta \Bar{G}^\text{solv}$ is \emph{not} the experimentally accessible free energy of solvation which would be the difference between the free energies of the respective ground state densities.

The exact definition of $G^\text{solv}$ as a functional of $\rho^\text{el}$ depends on the specific solvent model.\cite{andreussi2019} We refrain from stating all possible energy, potential, and force expressions here and refer to the respective original publications. We do not need to concern ourselves with this because Environ provides the functionality to compute $\Delta \Bar{G}^\text{solv}=G^\text{solv}-E^\text{vac}$ for any implemented $G^\text{solv}$. Similarly, $\Delta \Bar{V}_\text{eff}^\text{solv}$ is computed such that the energy functional $G^\text{solv}$ is minimized when using $V_\text{eff}^\text{solv}$ according to \cref{eq:sum_V} in the KS operator. The forces can be decomposed in a similar manner. It is worth noting that the inclusion of $\Delta \Bar{V}_\text{eff}^\text{solv}$ in the KS operator automatically yields a corresponding contribution to the Pulay forces, i.e., the correction to the Hellmann-Feynman forces which arises from the dependence of the atom centered electronic basis functions on the nuclear positions..\cite{blum2009,pulay1969}

The vacuum and solvent parts of \cref{eq:sum_F,eq:sum_V} as well as $\Delta \Bar{G}^\text{solv}$ can be computed straightforwardly without any further modification to either program beyond the already described electron density smoothing, low-pass filter in the FFT derivatives, and potential interpolation. Only $E^\text{vac}$ requires one additional term compared to a calculation without embedding. 
\begin{equation}\label{eq:E_vac}
\begin{split}
    E^\text{vac} &= \sum_i f_i \epsilon_i - \int d\mathbf{r}\, \rho^\text{el}(\mathbf{r})v_\text{xc}(\mathbf{r}) + E_\text{xc}[\rho^\text{el}] \\
    &\quad -\frac{1}{2} \int d\mathbf{r}\, \rho^\text{MP}(\mathbf{r})\Phi^\text{vac}(\mathbf{r}) + \frac{1}{2} \sum_{at} -Z_{at}\Phi^\text{vac}(\mathbf{R}_{at}) - \int d\mathbf{r}\, \rho^\text{MP}(\mathbf{r})\Delta \Bar{V}_\text{eff}^\text{solv}(\mathbf{r})
\end{split}
\end{equation}
The first five terms are how FHI-aims normally defines the total energy. In order, these terms are the sum of KS eigenvalues $\epsilon_i$ weighted by occupation numbers $f_i$, the removal of the interaction of the electron density with the exchange-correlation (xc) potential $v_\text{xc}$, the xc energy $E_\text{xc}$, the removal of the double-counted electron-electron interaction, and the nuclear-nuclear interaction. It is implied here that for any given $at$, $\Phi^\text{vac}(\mathbf{R}_{at})$ does not include the singularity of $at$ itself. The formulation with the fourth and fifth term containing the total potential $\Phi^\text{vac}$ instead of, respectively, the electronic and nuclear potentials is obtained by subtracting and adding half of the electron-nuclear interaction. The additional sixth term removes the interaction of the electrons with the solvent contribution to the KS potential. The actual interaction energy is contained in $\Delta \Bar{G}^\text{solv}$, similarly to how the interaction with $v_\text{xc}$ is replaced with the actual $E_\text{xc}$. The integration of this additional term is performed in the same loop as the interpolation of $\Delta \Bar{V}_\text{eff}^\text{solv}$ onto the atom-centered grids. Note that the double counting correction for the electron-electron interaction (second integral above) is computed using the multipole density $\rho^\text{MP}$ (cf.~\cref{eq:rho_MP}) rather than the exact electron density $\rho^\text{el}$. This is a deliberate choice in FHI-aims and leads to a faster convergence of the total energy with multipole expansion order.\cite{blum2009} We find that using $\rho^\text{MP}$ also in the solvent potential correction (last integral) similarly leads to faster convergence with multipole order, as shown in \cref{sec:conv_lmax} A more detailed derivation of \cref{eq:E_vac} is given in \cref*{si-sec:E_vac} in the SI.

It can be shown that the exact solvent contributions in \cref{eq:sum_all} calculated with a smoothened electron density and smeared cores are equal to the same quantities for the original electron density and nuclear point charges. In practice, Environ can compute a numerical solution only for the smoothened density, which we can use without further modification due to this equality. The equality follows from the fact that the solvent contribution to the electrostatic potential can be represented by an additional \textit{polarization} charge density which is $=0$ in regions of space where the potentials of the original and smoothened explicit density differ, i.e., \textit{inside}. Similarly, any terms beyond classical electrostatics can be reduced to semilocal expressions which vanish \textit{inside} by construction. A detailed discussion of each of these terms is given \cref*{si-sec:equivalence} in the SI.

\section{Results and discussion}

\subsection{Computational details}\label{sec:comp_det}

All calculations were carried out with modified versions based on FHI-aims version 241018 and Environ version 3.1. \revision{except for the CO on Pt(111) system in \cref{sec:consisteny_and_surfaces} and all calculations in \cref{sec:no_smoothing_limit,sec:overhead} for which FHI-aims version 250822 and Environ version 3.1.1 were used}. Unless mentioned otherwise, for FHI-aims we used the Perdew-Burke-Ernzerhof exchange correlation functional,\cite{perdew1996} \textit{tight} species defaults (2020 variant),\cite{blum2009} no spin-polarization, atomwise scalar zeroth-order regular approximation (ZORA) for relativistic eﬀects,\cite{blum2009} and a Gaussian broadening scheme with a width of 0.01 eV for the KS states.\cite{fu1983} Unless mentioned otherwise, for Environ we used a parabolic correction for periodic boundary conditions\cite{dabo2008,dabo2008_erratum}, a value of $\varepsilon_\text{bulk}=78.3$ for the bulk permittivity of water, and, respectively, the fitg03+$\beta$ parameters for SCCS\cite{andreussi2012} or the UFF\cite{rappe1992} parameters for neutral molecules for SSCS with the exception of the systems containing Pt, for which we used the Bondi\cite{bondi1964} parameters to avoid the extremely small radius of Pt in the UFF parameterization.\cite{fisicaro2017} In any case where two parameterizations were available, one with a non-zero effective pressure and one without but otherwise equivalent, we used the former option. We used Environ's fixed-point iterative solver for the electrostatic problem throughout, as we found the conjugate-gradient solver to yield inconsistent forces in preliminary calculations.

The water molecule, NaF dimer, F$^-$ anion, and Na$^+$ cation used in \cref{fig:regions,fig:gradients}, the following sections, and various examples in \cref*{si-sec:add_fig} in the SI were placed in $15\,\text{\AA}\times 15\,\text{\AA}\times 15\,\text{\AA}$ unit cells. The PtCO trimer was placed in a $20\,\text{\AA}\times 20\,\text{\AA}\times 20\,\text{\AA}$ cell. Only the Gamma point of $\mathbf{k}$ space was sampled. The bounding boxes of the figures do not represent the unit cell.

\subsection{Energy and force convergence with grid density, low-pass filter}\label{sec:conv_ecut}

\begin{figure}[!tb]
    \centering
    \includegraphics[width=0.95\linewidth]{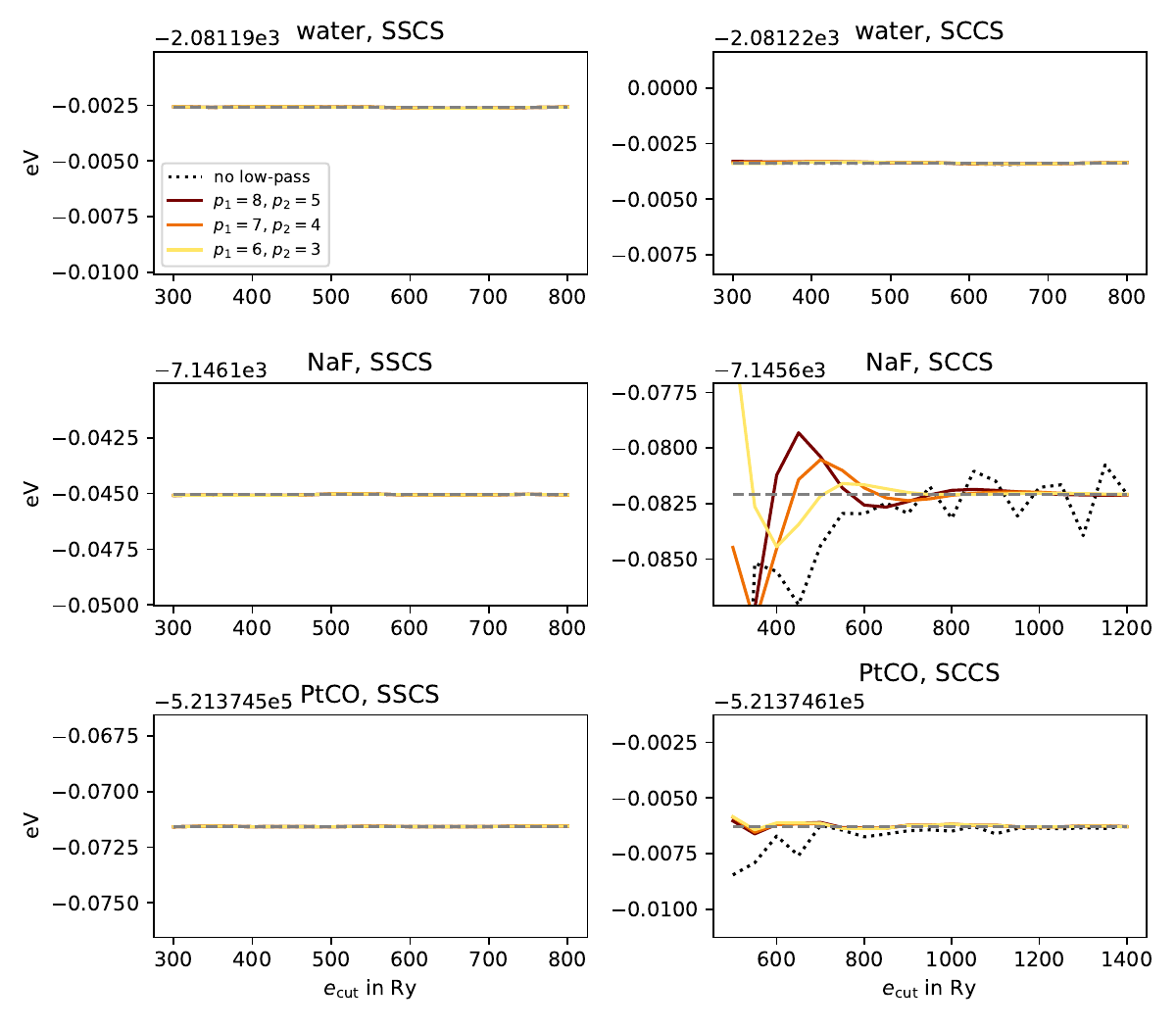}
    \caption{Convergence of total energy with Environ grid density parameter $e_\text{cut}$ for different test systems and solvent models. Dashed lines only as visual guideline.}
    \label{fig:E_ecut_tot_all}
\end{figure}

\begin{figure}[!tb]
    \centering
    \includegraphics[width=0.95\linewidth]{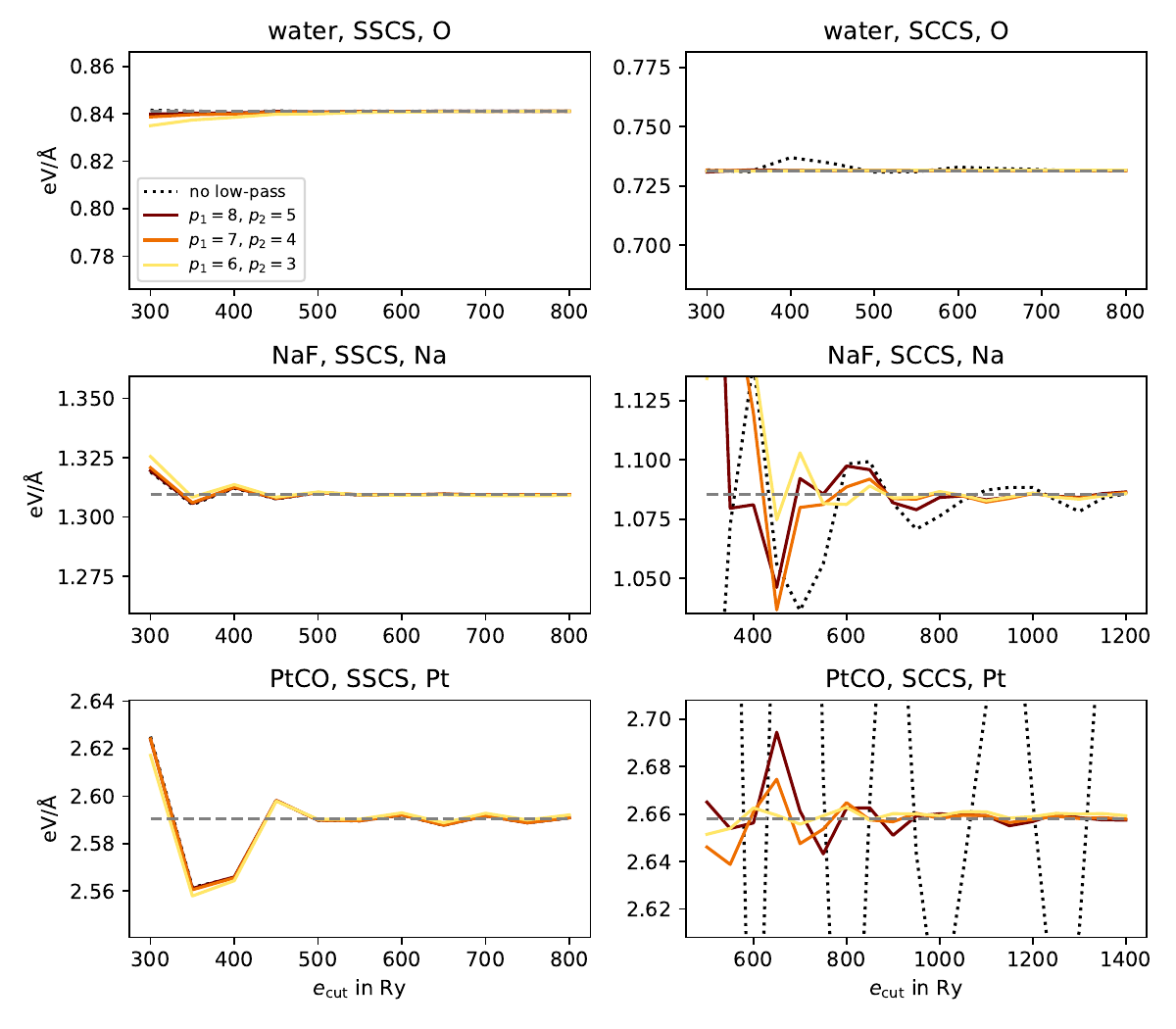}
    \caption{Convergence of total forces of selected atoms in different test systems and solvent models with Environ grid density parameter $e_\text{cut}$. Dashed lines only as visual guideline.}
    \label{fig:F_ecut_tot_all}
\end{figure}

We tested the convergence of the total energy and forces and their respective components with increasing Environ grid density for different low-pass filter parameters. The three test systems were a water molecule as a simple base example, a NaF dimer at 2~{\AA} separation as an extreme case of a polarized system, and a linear PtCO trimer with bond lengths of 1.85~{\AA} (Pt-C) and 1.1~{\AA} (C-O) in anticipation of the CO on Pt(111) example discussed in \cref{sec:consisteny_and_surfaces}. We tested the parameterizations of SSCS and SCCS described in \cref{sec:comp_det}. In Environ, the grid density is controlled by the $e_\text{cut}$ parameter where higher values correspond to denser grids. We sampled $e_\text{cut}$ in steps of 50~Ry over a range from 300~Ry to 800~Ry. For PtCO in SCCS, we started from 500~Ry instead, as some of the initial low $e_\text{cut}$ calculations did not result in SCF convergence. For NaF in SCCS, we extended the $e_\text{cut}$ range to 1200~Ry and for PtCO in SCCS to 1400~Ry, as results up to 800~Ry were not entirely conclusive. For the low-pass filter, we reduced the number of free parameters from 2 to 1 by fixing $p_2=p_1-3$ (implied throughout this section), which corresponds to $\text{filter}(\mathbf{g}_\text{cut})=1/2\cdot\text{erfc}(3)\approx 10^{-5}$. We tested filters with $p_1=6, 7, 8$, as well as no filter. The primary results are shown in \cref{fig:E_ecut_tot_all,fig:F_ecut_tot_all}. Complete results including the forces on all atoms as well as all individual energy and force components are shown in \cref*{si-sec:conv_E_ecut_sscs,si-sec:conv_F_ecut_sscs,si-sec:conv_E_ecut_sccs,si-sec:conv_F_ecut_sccs} in the SI. For better comparison, all plots throughout this and the following section are shown on the same energy or force scale, respectively, unless mentioned otherwise.

In SSCS, the total energies and their components did not exhibit any noticeable dependencies on $e_\text{cut}$ for NaF and water. For PtCO, minor changes in the eigenvalues and the Hartree and solvent potentials occur which do, however, cancel in the total and kinetic energy. The forces on the Pt and Na atoms exhibit small fluctuations originating from the Pulay force term which converge around $e_\text{cut}=500\,\text{Ry}$, with some small but noticeable residual fluctuations $<10\,\text{meV}/\text{\AA}$ on the Pt atom. The forces on the other atoms appear to be already converged at $e_\text{cut}=300\,\text{Ry}$. The low-pass filters do not have any impact on the energies, as they do not enter the energy and potential in SSCS. In the forces, they do not appear to reduce fluctuations. On the contrary, the most aggressive filter ($p_1=6$) actively introduces an error for low $e_\text{cut}$. We conclude that converged energies and forces in SSCS can be obtained simply by choosing high enough $e_\text{cut}$. For all SSCS calculations in the following, we use no low-pass filter and $e_\text{cut}=500\,\text{Ry}$.

In SCCS, the kinetic, xc, and total energies appear to converge around $e_\text{cut}=900\,\text{Ry}$ for NaF and significantly faster for water and PtCO. In PtCO, a systematic error cancellation between the Hartree and solvent potentials becomes apparent, with neither of the terms individually converging in the sampled range of $e_\text{cut}$. While residual fluctuations are observed also in the eigenvalues and, to a lesser extent, the kinetic energy of PtCO, the total energy does converge. The forces of the water and CO molecules do not exhibit any noticeable fluctuations as long as any low-pass filter is applied. The total forces appear to converge around $e_\text{cut}=700\,\text{Ry}$ for the NaF dimer and around $e_\text{cut}=900\,\text{Ry}$ for the Pt atom, with the residual fluctuations on the metal atoms Na and Pt significantly higher than on any of the nonmentals. Again, the individual force contributions on the Pt atom do not converge in the sampled range of $e_\text{cut}$. Nonetheless, total force convergence to below $10\,\text{meV}/\text{\AA}$ accuracy is possible for all tested systems with the more aggressive low-pass filters ($p_1=6$, $p_1=7$). Even the most conservative of the tested filters ($p_1=8$) significantly reduces energy and force fluctuations. Without a filter, we observe massive fluctuations in the energies and forces. In the converged range of $e_\text{cut}$, the grid dependent fluctuations of the energy and force components are much more pronounced than the differences between different low-pass filters as long as any filter is used. However, in the total forces on the Pt atom, a subtle systematic error originating from the Environ force contribution can be observed with the most aggressive filter ($p_1=6$) compared to the other two filters. We conclude that the use of a low-pass filter is necessary to ensure consistent results between different grid densities in SCCS. Too aggressive filters can lead to systematic errors. For more conservative filters, a safe regime exists where the exact choice of parameters has little impact when used with a sufficiently dense grid. $p_1=7$ appears to produce slightly faster and more stable convergence than $p_1=8$. For all SCCS calculations in the following, we use $p_1=7$ and $e_\text{cut}=1000\,\text{Ry}$ as a safe choice.

\subsection{Energy and force convergence with multipole order}\label{sec:conv_lmax}

\begin{figure}[!tb]
    \centering
    \includegraphics[width=0.95\linewidth]{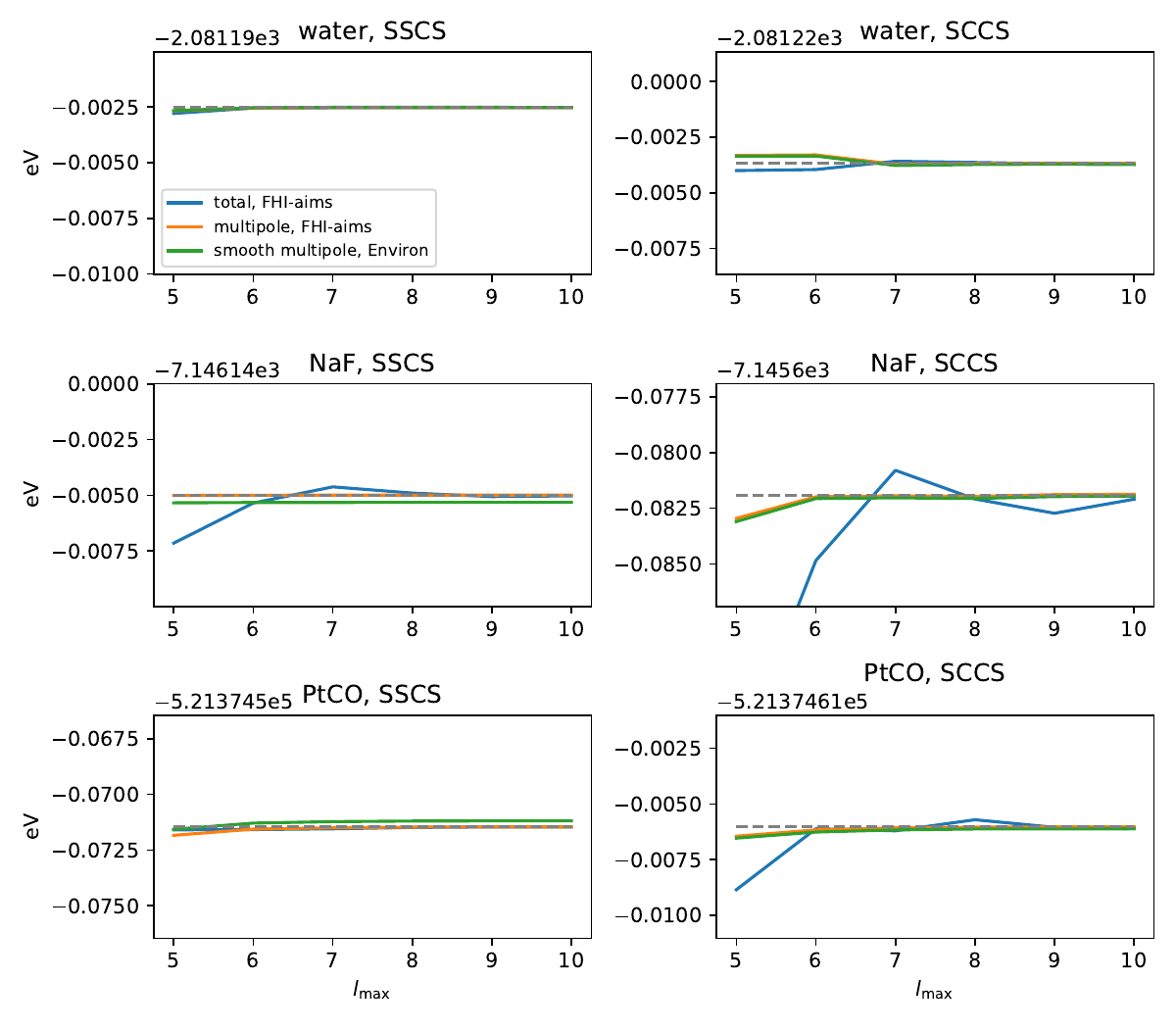}
    \caption{Convergence of total energy with multipole expansion order $l_\text{max}$ for different test systems, solvent models, and schemes for the solvent-electron double counting correction. Dashed lines only as visual guideline.}
    \label{fig:E_lmax_tot_all}
\end{figure}

\begin{figure}[!tb]
    \centering
    \includegraphics[width=0.95\linewidth]{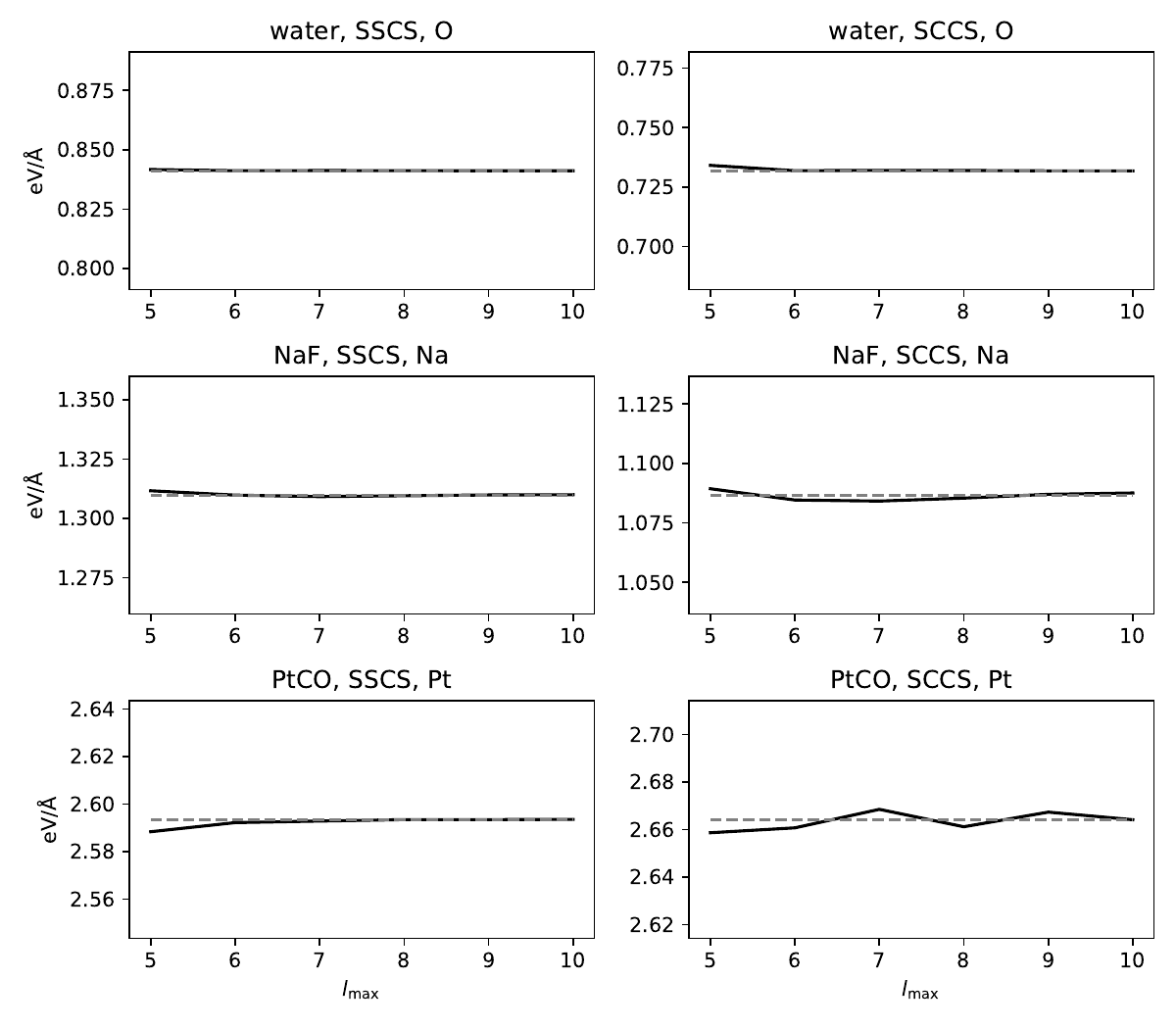}
    \caption{Convergence of total forces of selected atoms in different test systems and solvent models with multipole expansion order $l_\text{max}$. Dashed lines only as visual guideline.}
    \label{fig:F_lmax_tot_all}
\end{figure}

Using the such determined grid and low-pass filter settings for Environ, we tested energy and force convergence with the multipole order $l_\text{max}$ in \cref{eq:rho_MP}. To ensure an appropriate representation of the higher order multipoles on the angular grids of FHI-aims, we used the \textit{really tight} species defaults for this series of tests. For the tested elements, these defaults use the same electronic basis functions and radial grids as \textit{tight}, but denser angular grids and a higher multipole order $l_\text{max}=8$ instead of $l_\text{max}=6$, the latter being the \textit{tight} default value. We varied $l_\text{max}$ in the range from 5 to 10 and used the \textit{really tight} basis sets and grids otherwise unmodified. We also use this opportunity to test three possible ways of computing the solvent-electron double counting correction, i.e., the last term in \cref{eq:E_vac}: Using $\rho^\text{MP}$ and performing the integration in FHI-aims (as written in \cref{eq:E_vac}), using the exact $\rho^\text{el}$ instead of its multipole-decomposed counterpart and performing the integration in FHI-aims, and using the smoothened multipole density and performing the integration on Environ's regular grid. The primary results are shown in \cref{fig:E_lmax_tot_all,fig:F_lmax_tot_all}. Complete results including the forces on all atoms as well as all individual energy and force components are shown in \cref*{si-sec:conv_E_lmax_sscs,si-sec:conv_F_lmax_sscs,si-sec:conv_E_lmax_sccs,si-sec:conv_F_lmax_sccs} in the SI.

In SSCS, total energies and forces are converged at $l_\text{max}=6$ for all tested systems to good accuracy. Only the forces on the C atom in PtCO show small fluctuations which converge at $l_\text{max}=8$. These fluctuations are, however, of the same order of magnitude $<10\,\text{meV}/\text{\AA}$ as the $e_\text{cut}$ dependent fluctuations on the Pt reported in the previous section and can thus be considered converged within the overall accuracy of our method. By contrast, the individual energy and force contributions exhibit a clear dependence on $l_\text{max}$. This demonstrates that Environ can be seamlessly integrated into the intrinsic correction and error cancellation schemes of FHI-aims. In the following, we use the \textit{tight} species defaults for all SSCS calculations.

In SCCS, the energy of water and the forces in NaF and PtCO are not converged at $l_\text{max}=6$. At $l_\text{max}=8$, almost all total energies and forces are converged. At this value, only the total forces on the Na and Pt atoms still exhibit some small error, of the same order of magnitude as the residual fluctuations with increasing grid density for these atoms ($<10\,\text{meV}/\text{\AA}$). Energies and forces become unstable for $l_\text{max}>10$ using the \textit{really tight} angular grids, precluding any definitive statement about the convergence of the forces on these two atoms. However, increasing $l_\text{max}$ from 8 to 10, the residual net force on the center of mass of the total systems, which should be zero, reduced from $5.2\,\text{meV}/\text{\AA}$ to $2.4\,\text{meV}/\text{\AA}$ for PtCO and remained constant at $1.5\,\text{meV}/\text{\AA}$ for NaF. Of course, the interpretability of these few data points is limited. It appears intuitively reasonable that SCCS, in which the cavity definition depends on the electron density, requires a more precise representation of this density compared to SSCS, which has no such dependency. In the following, we use the \textit{really tight} species defaults for all SCCS calculations.

Across both solvent models it becomes apparent that of the three possible ways to compute the solvent-electron double counting correction, using $\rho^\text{el}$ and performing the integration in FHI-aims does not lead to reliable energy convergence. This is particularly pronounced for NaF. The other two options both yield fast and reliable convergence to approximately the same value, with a minor difference $<1\,\text{meV}$ for NaF and PtCO in SSCS. We opt to use $\rho^\text{MP}$ and perform the integration in FHI-aims as it allows us to use Environ's existing functions for computing $\Delta \Bar{G}^\text{solv}$ without modification.



\subsection{Consistency between energies and forces, surfaces}\label{sec:consisteny_and_surfaces}

Using the such determined settings, we confirmed the consistency between energies and forces by checking the residual force on the center of mass of each system, which should be zero. Additionally, we displaced one atom in each system by 0.005~{\AA} away from the central (water, PtCO) / other (NaF) atom to compute finite difference numerical forces and compared them to the analytical forces. The results are shown in \cref{tab:forces}.

\begin{table}[!htb]
\begin{tabular}{ll|l|lll}
system & model & residual total & atom & numerical & analytical \\
\hline
water & SSCS & 0.84 & H & 558.2 & 559.1 \\
NaF & SSCS & 1.3 & F & 1290.5 & 1290.3 \\
PtCO & SSCS & 4.3 & O & 6842.2 & 6846.1 \\
CO@Pt(111) & SSCS & \revision{10.0} & O & \revision{3063.6 (3072.5)} & \revision{3086.0} \\
\hline
water & SCCS & 0.80 & H & 505.1 & 504.7 \\
NaF & SCCS & 1.5 & F & 1065.4 & 1068.2 \\
PtCO & SCCS & 5.2 & O & 6853.4 & 6857.6 \\
CO@Pt(111) & SCCS & \revision{4.8} & O & \revision{2841.7 (2850.8)} & \revision{2864.0}
\end{tabular}
    \caption{Forces in meV/{\AA} for different systems in SSCS and SCCS. Center column: Residual total force (Euklidean norm) on center of mass of system (should be zero). Right columns: Forces along bond axis (unsigned) acting on specific atoms, computed numerically from total energy change on small displacement, and computed analytically (mean value between undisplaced and displaced geometry). Numerical forces in brackets are derived from the extrapolated energies at zero broadening of occupation numbers for CO@Pt(111).}
    \label{tab:forces}
\end{table}

We also test the applicability of our method in the context of computational surface science by including in the test systems a CO molecule (bond length 1.12~{\AA}) adsorbed at the top site of a Pt(111) surface (Pt-C distance 1.85~{\AA}). The Pt surface was represented by a slab consisting of 3 layers of $4\times 4$ Pt atoms in an orthorhombic cell with cell parameters 11.10~{\AA} and 9.61~{\AA} in plane, corresponding to a nearest neighbor distance of 2.77~{\AA}. In the direction perpendicular to the surface, the cell had an extent of 27.50~{\AA}. We used the 2 dimensional version of Environ's parabolic periodic boundary correction. We increased the broadening width for the occupation numbers to 0.1~eV to allow for convergence of the electronic structure of the metal substrate. \revisionremoved{We only sampled the Gamma point of $\mathbf{k}$ space.}\revision{We used a $\mathbf{k}$ grid of $15\times 15\times 1$ for which we found energies to be converged to 1~meV accuracy for the undisplaced geometry in both SSCS and SCCS, cf.~\cref*{si-sec:kgrid} in the SI.}

We consistently find the residual force on the center of mass to be \revisionremoved{$<10\,\text{meV}/\text{\AA}$ when normalized by the number of atoms in the system}\revision{$\leq 10\,\text{meV}/\text{\AA}$}. The same accuracy is also found in the agreement between numerical and analytical forces for the isolated systems. For CO on Pt(111), a larger disagreement is found. This is likely due to the non-zero broadening of the occupation numbers in the metal part of the system. If the derivatives of the occupation numbers are not explicitly accounted for, the analytical forces no longer correspond to the total energy.\cite{blum2009} FHI-aims also outputs the extrapolated energy at zero broadening. Computing the numerical forces from these extrapolated energies reduces the disagreement with the analytical forces to \revisionremoved{$\approx$}\revision{$<$} $ 20\,\text{meV}/\text{\AA}$. Implementing explicit derivatives of the occupation numbers taking the solvent into account is beyond the scope of this article and left for future work.

\subsection{\revision{Behavior in the no-smoothing limit} \label{sec:no_smoothing_limit}}

\revision{So far, we have only shown that for a given set of $r^\text{cut}_{at}$ there exists a free energy and corresponding forces to which a calculation converges with sufficiently tight numerical settings. However, we have not yet empirically shown that this is, indeed, the correct free energy of the original electron density in the chosen solvent model. The theory of the equivalence between original and smoothened quantities is discussed in \cref*{si-sec:equivalence} in the SI, but only in terms of exact analytical solutions. It is worth verifying how accurately this equivalence holds for the practically obtainable numerical solutions.}

\revision{An exact reference using the original electron density is elusive because the cusps of this density near the nuclei cannot be resolved on any regular grid with a feasible grid density due to the discontinuity of the derivatives at the cusp. Conversely, implementing Environ's models in an overlapping atom-centered grid formalism lies beyond the scope of this article and is, in fact, exactly what we want to avoid.}

\revision{However, we can make use of the fact that the original electron density is recovered in the limit of small $r^\text{cut}_{at}$
\begin{equation}
    \lim_{r^\text{cut}_{at}\to 0^+} \delta \widetilde{\rho}^\text{smooth}_{at,lm} = \delta \widetilde{\rho}_{at,lm}
\end{equation}
and equivalently for $\rho^\text{free}_{at}$. Although this limit cannot be reached, we can get arbitrarily close to it by choosing arbitrarily dense regular grids.}

\revision{To verify that the energy of in the no-smoothing limit is the same as with the choice of $r^\text{cut}_{at}$ described in \cref{sec:cutoff}, we systematically scale these radii by a factor $f_\text{scale}$
\begin{equation}\label{eq:rcut_scaled}
    r^\text{cut}_{at} = \frac{1}{2}f_\text{scale} r^\text{solv}_{at}
\end{equation}
where $f_\text{scale}=1$ reproduces the default \cref{eq:default_rcut}.}

\revision{On this occasion, we also demonstrate how errors emerge once smoothing and solvent regions overlap. Starting at $f_\text{scale}=3$, we decreased the scaling factor in steps of 0.1 until the SCF procedure no longer converged. The same factor was applied to all atoms in a given system. We repeated this process for different $e_\text{cut}$, starting at the previously determined default values and increasing it in steps of 500~Ry up to 3500~Ry for water and NaF and 2000~Ry for PtCO.}

\begin{figure}[!tb]
    \centering
    \includegraphics[width=0.95\linewidth]{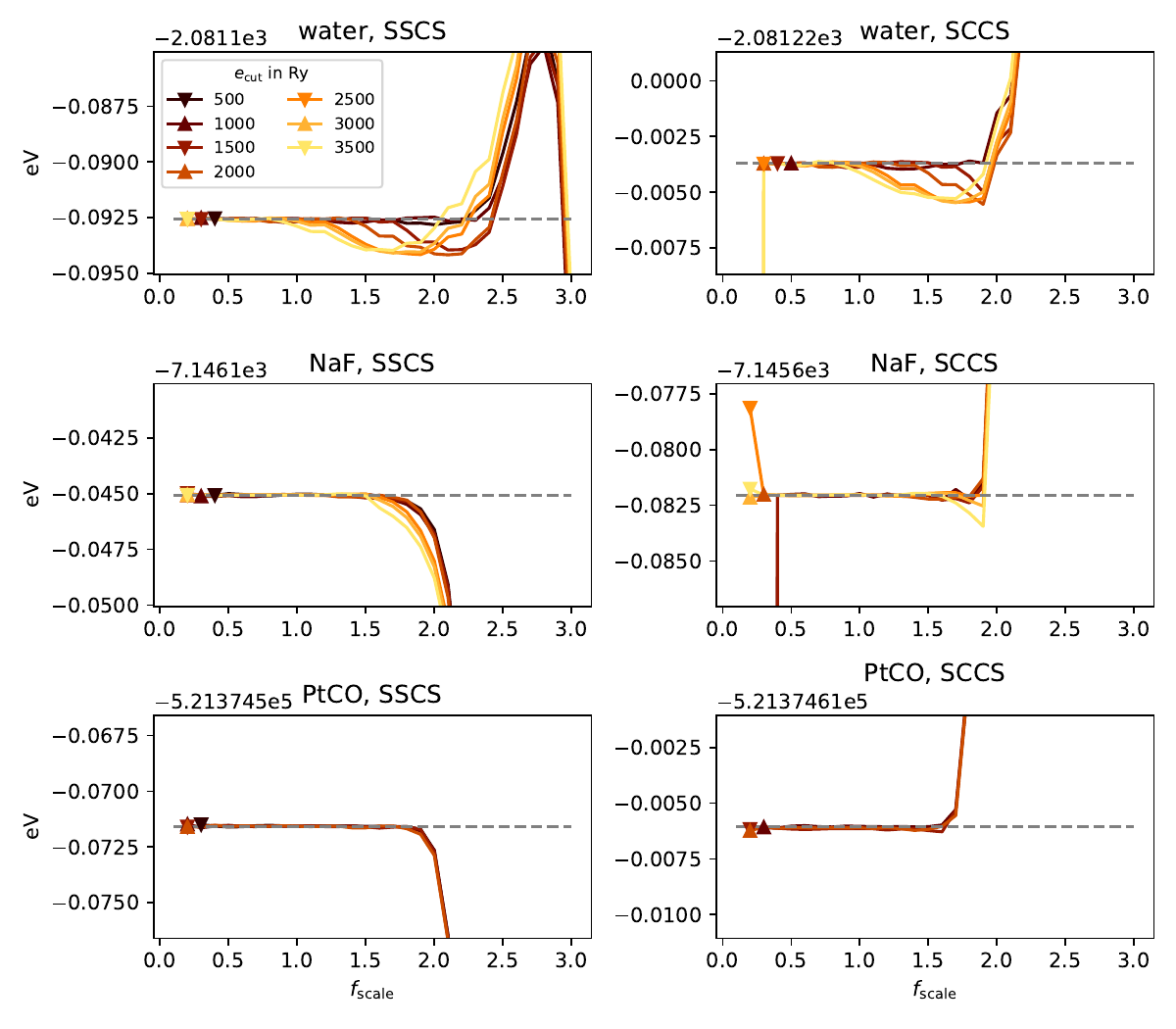}
    \caption{\revision{Convergence of total energy with scaling factor for $r^\text{cut}_{at}$ for different test systems, solvent models, and Environ grid densities as defined by $e_\text{cut}$. Markers indicate the lowest $f_\text{scale}$ at which SCF convergence was still achieved. Dashed lines only as visual guideline.}}
    \label{fig:E_rcut_tot_all}
\end{figure}

\revision{The results are shown in \cref{fig:E_rcut_tot_all}. Across all test systems, we consistently observe a stable regime up to around $f_\text{scale}=1$. Around $f_\text{scale}=2$, a steep deviation of the energy from this stable regime is found. This is expected as at this value, $r^\text{cut}_{at}$ is equal to the estimated solvation radius $r^\text{solv}_{at}$, cf.~\cref{eq:rcut_scaled}. For larger $r^\text{cut}_{at}$, the smoothing region overlaps noticeably with the \textit{outside} region and the assumptions made in \cref{sec:smoothing} no longer hold. Remarkably, for water both in SSCS and SCCS the $f_\text{scale}$ value at which the energy starts to deviate from the converged value is smaller for denser grids. Nonetheless, at $f_\text{scale}=1$, the error was $<1\,\text{meV}$ for all tested grid densities.}

\revision{In the direction of smaller smoothing regions, stable SCF convergence can be achieved down to $f_\text{scale}$ between 0.3 and 0.5 with the default grid densities. Smaller $f_\text{scale}$ down to 0.2 (except for water in SCCS, for which only 0.3 was possible with the tested grid densities) can be reached but require a steeply increasing Environ grid density. In most cases, the SCF procedure simply does not converge below a certain scaling factor. For water and NaF in SCCS, however, the lowest possible $f_\text{scale}$ for some $e_\text{cut}$ resulted in outliers instead. All of these outliers are associated with additional SCF steps compared to the stable $f_\text{scale}$ region. Furthermore, in NaF it can be seen how increasing the grid density further can remedy these outliers. At $f_\text{scale}=0.3$, $e_\text{cut}$ values of 1000~Ry and 1500~Ry resulted in outliers, but $e_\text{cut}=2000$ recovers the stable energy. Similarly, $f_\text{scale}=0.2$ and $e_\text{cut}=2500\,\text{Ry}$ also results in an outlier which is recovered at higher $e_\text{cut}$. These outliers can thus be attributed to the SCF just narrowly converging to a spurious state. The outliers for the two highest $e_\text{cut}$ and $f_\text{scale}=0.2$ in water in SCCS are not recovered within the sampled $e_\text{cut}$ range. However, the considerable number of additional SCF steps (45 for $e_\text{cut}=3000\,\text{Ry}$ and 22 for 3500~Ry compared to 10 for $f_\text{scale}=0.3$) also suggests a spurious state which could presumably be recovered with an even denser grid.}

\revision{Apart from these outliers, our results suggest that the no-smoothing limit of the total energy is already reached at $f_\text{scale}=1$ for the default grid densities. We thus find \cref{eq:default_rcut} to be a reasonable choice which avoids both spurious states for very small smoothing regions and overlap with the \textit{outside} region for very large smoothing regions.}

\subsection{\revision{Computational overhead}\label{sec:overhead}}

\revision{The relative computational overhead of adding solvent effects via Environ into an FHI-aims calculation depends strongly on the methods used in both programs. The computational cost of different electronic structure methods famously differs by orders of magnitude. In Environ, we have shown that different models require different grid densities. Furthermore, the Poisson-Boltzmann model for electrolytes, cf.~\cref{eq:PBE1,eq:PBE2}, requires a nested iterative solver scheme which takes more computational time than the generalized Poisson model for pure dielectrics, cf.~\cref{eq:GPE}. It follows from these wide varieties within both parts of the simulation that either part may be the computational bottleneck, depending on the simulated system and the chosen models.}

\revision{Nonetheless, a very rough estimate of the computational overhead is possible based on the dimensionality of the system. In Environ, the grid is uniform throughout the cell. Environ's computational cost is, thus, dictated primarily by the cell size but not by the atoms inside it. By contrast, the grids in FHI-aims are attached to atoms. Therefore, its computational cost is dictated by the number (and species) of atoms but not by the cell size. This means that empty space in the cell increases the cost of Environ but not of FHI-aims. Empty space typically occurs in combination with periodic boundary corrections, as a buffer between the system and the cell boundary. Accordingly, we find the relative overhead on isolated systems where empty space is added in three directions to be significantly higher than on slab calculations where empty space is added only in one direction. Adding to this difference is the fact that for slab calculations in FHI-aims, $\mathbf{k}$ space integration is necessary while in Environ, this is not the case because only charge densities and electrostatic potentials are considered, but no wavefunctions.}

\revision{\Cref{tab:times} shows the total calculation time, average time for one SCF step, and number of SCF steps for our test systems in SSCS, SCCS, and without continuum embedding. We note that some part of the overhead for SCCS is caused not by Environ itself but rather by the use of \textit{really tight} instead of \textit{tight} species defaults in FHI-aims. As expected from the considerations above, for each solvent model the overheads on one SCF step were all in a similar order of magnitude across all systems, 9~-~18~s for SSCS and 22~-~106~s for SCCS. This is because the simulation cells all had similar sizes, between 2933 and 8000 \AA$^3$. By contrast, FHI-aims can take full advantage of the non-periodicity of the isolated systems, resulting in one SCF step of the water molecule being almost four orders of magnitude faster than for the CO@Pt(111) slab. Additionally, with FHI-aims being an all-electron code, heavy elements like Pt are intrinsically more computationally expensive than lighter elements.}

\begin{table}[!htb]
\revision{\begin{tabular}{ll|lll}
system &   & vacuum & SSCS & SCCS  \\
\hline
 & avg. step & 0.034 & 9.2 & 26 \\
water & total & 0.64 & 88 & 273 \\
& steps & 20 & 10 & 11 \\
\hline
 & avg. step & 0.044 & 11 & 22 \\
NaF & total & 2.7 & 132 & 227 \\
& steps & 24 & 12 & 11 \\
\hline
 & avg. step & 0.056 & 18 & 57 \\
PtCO & total & 1.5 & 198 & 670 \\
& steps & 26 & 12 & 13 \\
\hline
 & avg. step & 248 & 265 & 354 \\
CO@Pt(111) & total & 12729 & 22157 & 29489 \\
& steps & 50 & 82 & 82
\end{tabular}}
    \caption{\revision{Wall clock time in s for average SCF step except first and last, total calculation time, and number of SCF steps for different systems and solvent models as well as without solvent. All calculations were performed on 36 Intel(R) Xeon(R) CPU E5-2695 v4 @ 2.10GHz CPU cores. Both FHI-aims and Environ were built with Intel oneAPI compilers and libraries, version 2022.0.2. For the vacuum calculation of CO@Pt(111), periodic boundary conditions in $z$ direction were removed using the dipole correction implemented in FHI-aims. The same $15\times 15\times 1$ $\mathbf{k}$ point grid was used as in the calculations with Environ. For the isolated systems, vacuum calculations were performed without periodic boundary conditions. All vacuum calculations used \textit{tight} species defaults.}}
    \label{tab:times}
\end{table}

\revision{In summary, the relative overhead on a single SCF step is small to moderate for the CO@Pt(111) test system. The total calculation time is still noticeably increased because more SCF steps were necessary with continuum embedding than without. This may possibly be remedied by optimizing the mixing parameters. However, for the isolated systems, Environ is clearly the bottleneck of the calculation. The intrinsic continuum embedding models of FHI-aims which can take advantage of non-periodic systems are significantly faster for these systems, albeit more limited in scope. For example, the MPE model was shown to not be the bottleneck of the calculation even for small isolated systems\cite{filser2022}, but it lacks most of Environ's features like electrolyte models.}

\section{Conclusion and outlook}

We have presented an interface between the all-electron full-potential DFT program FHI-aims and the regular grid based continuum embedding package Environ. FHI-aims passes a modified electron density to Environ which returns contributions to the KS potential, energy, and forces, allowing for integration of Environ into the SCF loop of FHI-aims. To overcome the complication of representing the narrow cusps of the electron density around the nuclei on an evenly spaced grid, we have introduced a smoothing scheme which conserves the charge density and electrostatic potential outside the solvation cavity, resulting in the correct solvent potential. Furthermore, we have included a low-pass filter in the FFT based derivatives in Environ to suppress high frequency artifacts in the electron density passed from FHI-aims.

The resulting method allows for reliable convergence of energies to $<1\,\text{meV}$ and forces to $<10\,\text{meV}/\text{\AA}$ accuracy. For SSCS, the \textit{tight} species defaults of FHI-aims are sufficient and no low-pass filter is necessary. For SCCS, the higher multipole order of the \textit{really tight} defaults is required as well as a low-pass filter, for which we found parameters $p_1=7$, $p_2=4$ to be a safe choice. Even higher multipole orders and accordingly denser angular grids than in the \textit{really tight} defaults might possibly be necessary in some cases. The required grid density in Environ depends strongly on the system and the solvent model. For all systems which were tested in this work, a parameter of $e_\text{cut}=500\,\text{Ry}$ appeared to be a safe choice for SSCS and 1000~Ry for SCCS. However, we stress here that the chosen applications intentionally cover challenging systems, and much smaller values may be sufficient in many cases. As in most electronic structure applications, testing convergence of energies and forces with $e_\text{cut}$ for each specific system is recommended.

The presented work showcases a robust scheme for coupling simulation methods with very different numerical domains. Given the accuracy and transferability of the proposed methodology, we expect the proposed scheme to have a positive impact on the further development of modular libraries for electronic structure and multiscale/multiphysics calculations. Other forms of environment/electrostatic embedding, in particular density embedding, subsystem DFT, and polarizable QM/MM  models, may benefit from the possibility of combining grid-based solvers with accurate full-electron calculators that rely on atom-centered numerical domains.  

\section*{Acknowledgments}

We thank Volker Blum for his useful suggestions on potential representation on different grids. We acknowledge funding from the NSF Cybertraining program, award {\#}2321102 and from the NSF CAREER award {\#}2306929.

\section*{Data availability}

Data sharing is not applicable to this article as no new data were created or analyzed in this study.

\newpage
\bibliography{refs}

\end{document}


\maketitle

\section[Equivalence of certain solvent terms derived from original and smoothened density]{Equivalence of certain solvent terms derived from original and smoothened density}\label{sec:equivalence}

In this section, we discuss the solvent contributions to the potential and their interaction with the solute. We consider the exact solution of the electrostatic problem and non-electrostatic model for the original explicit charge density $\rho$ and its smoothened counterpart $\Bar{\rho}$. Most equations are expressed in terms of $\Bar{\rho}$ and quantities derived from it since this is what we compute in practice. However, for the exact solution, the same equations hold for $\rho$ and its derived quantities. The numerical solution on the regular grid is a good approximation to this (unknown) exact solution for $\Bar{\rho}$ but not for $\rho$. We will show that the exact solvent potential is equal for both densities, allowing us to use the numerical solution for the smoothened density and treat it as if it was the solution for the original density.

Throughout this and the following sections, the superscript `vac' is used to denote any quantities resulting from the Poisson equation without any embedding. Note that such quantities are still derived from a charge density which may be the converged result of an embedded calculation rather than a vacuum calculation or, in fact, any instantaneous charge density. The superscript `solv' indicates any quantities resulting from an embedded calculation. The bar $\Bar{\quad}$ is used to denote smoothened electron densities as described in the main article, smeared nuclear charge densities according to Environ's\cite{andreussi2012} intrinsic procedures, as well as any total or other charge densities, potentials, energies, and forces derived from the former two quantities. The corresponding quantities without the bar and the term \textit{original} refer to the non-smoothened electron density from FHI-aims\cite{blum2009} and nuclear point charges as well as any derived quantities. Unless mentioned otherwise, any equations presented here refer to the (unknown) exact solutions of the respective models rather than the numeric solutions on the regular (or any other) grid.

\subsection{Electrostatic potential}\label{sec:elstat_pot_equal}

Environ implements a variety of electrostatic problems, not just the generalized Poisson and Poisson-Boltzmann models.\cite{dabo2010,ringe2016,nattino2018,andreussi2019} However, all of these models are defined in a way which allows $\Bar{\Phi}^\text{solv}$ to be computed as the solution of the Poisson equation with an additional polarization source term. In the original literature on Environ, the dielectric polarization and the electrolyte charge are considered separately. Such a distinction is not necessary for the derivation here. Instead, we summarize both into $\Bar{\rho}^\text{pol}$.
\begin{equation}\label{eq:Poisson_add_source}
    \nabla^2 \Bar{\Phi}^\text{solv}(\mathbf{r}) = - 4\pi\left(\Bar{\rho}(\mathbf{r}) + \Bar{\rho}^\text{pol}(\mathbf{r})\right)
\end{equation}
The linearity of the $\nabla^2$ operator allows us to separate
\begin{subequations}\label{eq:pot_pol}
\begin{align}
    \Bar{\Phi}^\text{solv} &= \Bar{\Phi}^\text{vac} + \Bar{\Phi}^\text{pol} \\
    \nabla^2 \Bar{\Phi}^\text{pol}(\mathbf{r}) &= - 4\pi\Bar{\rho}^\text{pol}(\mathbf{r}) \label{eq:Poisson_pol}
\end{align}
\end{subequations}
Here, we have renamed the quantity which is called $\Delta\Bar{\Phi}^\text{solv}$ in the main article to $\Bar{\Phi}^\text{pol}$ to make the relation to $\Bar{\rho}^\text{pol}$ more apparent and avoid confusion with $\Bar{\Phi}^\text{solv}$ in the following. We can use \cref*{main-eq:Phi_aims_plus_env} in the main article to obtain the non-smoothened potential in solution as long as
\begin{equation}\label{eq:delta_phi_equal}
    \Bar{\Phi}^\text{pol} = \Phi^\text{pol}
\end{equation}
The iterative solver finds $\Bar{\rho}^\text{pol}$ such that $\Bar{\Phi}^\text{solv}$ solves the specified problem for the explicit source term $\Bar{\rho}$. All implemented electrostatic problems allow for $\Bar{\rho}^\text{pol}$ to be computed as a semi-local function
\begin{equation}\label{eq:iter1}
    \Bar{\rho}_{(i)}^\text{pol}(\mathbf{r}) = \Bar{\rho}^\text{pol}\left(\Bar{\Phi}_{(i)}^\text{solv}(\mathbf{r}),\nabla \Bar{\Phi}_{(i)}^\text{solv}(\mathbf{r}), \Bar{\rho}(\mathbf{r}), \nabla \Bar{\rho}(\mathbf{r}), s(\mathbf{r}), \nabla s(\mathbf{r}) \right)
\end{equation}
at each iteration step $i$, using $\Bar{\Phi}_{(i)}^\text{solv}$ at that step which is computed from the polarization at the previous step
\begin{equation}\label{eq:iter2}
    \nabla^2 \Bar{\Phi}_{(i)}^\text{pol}(\mathbf{r}) = - 4\pi\Bar{\rho}_{(i-1)}^\text{pol}(\mathbf{r})
\end{equation}
For all implemented models, we can generally say that $\Bar{\rho}^\text{pol}=0$ everywhere \textit{inside}. This means that at any given iteration step, the polarization density for the original and smoothened explicit density are equal
\begin{equation}\label{eq:pol_identical}
\begin{split}
    &\Bar{\rho}^\text{pol}\left(\Bar{\Phi}_{(i)}^\text{solv}(\mathbf{r}),\nabla \Bar{\Phi}_{(i)}^\text{solv}(\mathbf{r}), \Bar{\rho}(\mathbf{r}), \nabla \Bar{\rho}(\mathbf{r}), s(\mathbf{r}), \nabla s(\mathbf{r})\right) \\
    = &\rho^\text{pol}\left(\Phi_{(i)}^\text{solv}(\mathbf{r}),\nabla \Phi_{(i)}^\text{solv}(\mathbf{r}), \rho(\mathbf{r}), \nabla \rho(\mathbf{r}), s(\mathbf{r}), \nabla s(\mathbf{r})\right)
\end{split}
\end{equation}
under the condition that
\begin{subequations}
\begin{align}
    \Bar{\rho}(\mathbf{r}) & = \rho(\mathbf{r}) \label{eq:rho_identical} \\
    \Bar{\Phi}_{(i)}^\text{solv}(\mathbf{r}) = \Bar{\Phi}^\text{vac}(\mathbf{r}) + \Bar{\Phi}_{(i)}^\text{pol}(\mathbf{r}) &= \Phi^\text{vac}(\mathbf{r}) + \Phi_{(i)}^\text{pol}(\mathbf{r}) = \Phi_{(i)}^\text{solv}(\mathbf{r}) \label{eq:pot_identical}
\end{align}
\end{subequations}
for all $\mathbf{r}$ \textit{outside}. The equality of $s$ and $\Bar{s}$ is discussed in \cref*{main-sec:smoothing} the main article. \textit{Inside}, \cref{eq:pol_identical} is trivially fulfilled by being $=0$. \Cref{eq:rho_identical} is always fulfilled \textit{outside} because we defined our smoothing scheme to only manipulate the density \textit{inside}. By conserving the long range multipole moments of the smoothened regions, $\Bar{\Phi}^\text{vac}(\mathbf{r})=\Phi^\text{vac}(\mathbf{r})$ \textit{outside}. By simply choosing
\begin{equation}\label{eq:init_0}
    \Bar{\Phi}_{(1)}^\text{pol}(\mathbf{r}) = \Phi_{(1)}^\text{pol}(\mathbf{r}) = 0
\end{equation}
as an initial guess, we find that \cref{eq:pot_identical} is fulfilled \textit{outside} for $i=1$. This means that \cref{eq:pol_identical} is fulfilled in this step and, according to \cref{eq:iter1,eq:iter2}, \cref{eq:pot_identical} remains fulfilled in the next step etc., eventually converging to the same result 
\cref{eq:delta_phi_equal}. It does not matter which initial guess for $\Bar{\Phi}_{(1)}^\text{pol}$ is actually used as long as the iteration converges to the same result as when starting from \cref{eq:init_0}.

\subsection{Free energy}\label{sec:free_energy}

The exact definition of the free energy functional differs between models.\cite{andreussi2019} Nonetheless, we can make some general statements about the equivalence of the solvent contribution to the free energy for the original and smoothened density. 

The electrostatic interaction between the explicit charge and the solvent potential is a term that occurs as a part of $\Delta \Bar{G}^\text{solv}$ in all of these models. It can equivalently expressed as the interaction between the potential of the explicit charge, i.e., the vacuum potential, and the polarization density.
\begin{equation}\label{eq:G_elstat_rephrase}
    \int d\mathbf{r}\, \Bar{\rho}(\mathbf{r})\Bar{\Phi}^\text{pol}(\mathbf{r}) = \int d\mathbf{r}\, \Bar{\Phi}^\text{vac}(\mathbf{r})\Bar{\rho}^\text{pol}(\mathbf{r})
\end{equation}
As established in the previous section, $\Bar{\rho}^\text{pol}=\rho^\text{pol}=0$ everywhere \textit{inside} for all models implemented in Environ and $\Bar{\Phi}^\text{vac}=\Phi^\text{vac}$ and $\Bar{\rho}^\text{pol}=\rho^\text{pol}$ everywhere \textit{outside}. This means that the integrand in the right hand side expression of the above equation vanishes everywhere \textit{inside} both for the smoothened and original potential and density, and it is equal to its non-smoothened counterpart everywhere \textit{outside}. The electrostatic solute-solvent interactions for the original and smoothened density are, thus, equal.

Beyond these electrostatic interactions, the total free energy may contain a confining potential,\cite{nattino2019} the electrostatic energy, chemical potential and entropy of an electrolyte, or a non-electrostatic contribution in the form of an effective pressure and an effective surface tension\cite{andreussi2019} applied to the cavity volume and surface area. All of these terms are computed as volume integrals. In all of them, the integrand depends locally only on quantities which we have already established to be equivalent between the smoothened and non-smoothened case everywhere \textit{outside}. Specifically, these are $s$ and $\Bar{\rho}^\text{el}$ for the confining potential, $s$ and $\Bar{\Phi}^\text{solv}$ for the electrolyte terms, $s$ for the volume, and $|\nabla s|$ for the surface area. Furthermore, all integrands vanish \textit{inside} by construction except for the cavity volume. However, the integrand \textit{inside} is simply $=s=1$ in this case, both in the smoothened and the non-smoothened case. Therefore, $\Delta \Bar{G}^\text{solv}=\Delta G^\text{solv}$.

\subsection{Kohn-Sham potential}

The KS potential is given by the functional derivative
\begin{equation}\label{eq:V_eff_def}
    V_\text{eff}=\frac{\delta E_\text{pot}}{\delta\rho_\text{el}}
\end{equation}
where the potential energy $E_\text{pot}=E-T$ is simply the energy without the KS kinetic energy $T$, with $E$ being the energy expression that is minimized by using $V_\text{eff}$ in the KS equations. In our context, this can be either $E=E^\text{vac}$ in vacuum or $E=G^\text{solv}$ in solution. In the former case, the potential is given by
\begin{equation}\label{eq:V_eff_vac}
    V_\text{eff}^\text{vac} = \Phi^\text{vac} + v_\text{xc}
\end{equation}
with the functional derivative of the exchange-correlation (xc) energy $v_\text{xc}=\delta E_\text{xc} / \delta \rho^\text{el}$. With continuum embedding, it is given by
\begin{equation}\label{eq:V_eff_solv}
    V_\text{eff}^\text{solv} = \Phi^\text{vac} + v_\text{xc} + \Delta \Bar{V}_\text{eff}^\text{solv}
\end{equation}
$\Delta \Bar{V}_\text{eff}^\text{solv}$ contains $\Bar{\Phi}^\text{pol}$. Additionally, it may contain any functional derivatives due to an electron density dependent cavity definition in SCCS\cite{andreussi2012} including non-electrostatic terms as well as the confining potential if the respective methods are used.

$T$ depends only on $\rho^\text{el}$ and is computed in FHI-aims, without any smoothing. It is, therefore, fully contained in $E^\text{vac}$. Furthermore, $V_\text{eff}^\text{vac}$ is actually the functional derivative of $E^\text{vac}$, see also \cref{sec:E_vac}.
\begin{subequations}
\begin{align}
    G^\text{solv}&=  E^\text{vac}_\text{pot} + T + \Delta \Bar{G}^\text{solv} \\
     V_\text{eff}^\text{solv} &= V_\text{eff}^\text{vac} + \frac{\delta \Delta \Bar{G}^\text{solv}}{\delta\rho_\text{el}}
\end{align}
\end{subequations}
This allows us to directly relate $\Delta \Bar{V}_\text{eff}^\text{solv}$ to $\Delta \Bar{G}^\text{solv}$
\begin{equation}
    \Delta \Bar{V}^\text{solv}_\text{eff}=\frac{\delta \Delta \Bar{G}^\text{solv}}{\delta\rho_\text{el}}
\end{equation}
We have already established that $\Delta \Bar{G}^\text{solv}=\Delta G^\text{solv}$. This is true for any electron density, not just ground- or other eigenstates. Therefore, $\delta \Delta \Bar{G}^\text{solv}=\delta \Delta G^\text{solv}$ regardless of the chosen perturbation, and consequently $\Delta \Bar{V}^\text{solv}_\text{eff}=\Delta V^\text{solv}_\text{eff}$.

\subsection{Forces}

The electrostatic forces in Environ can equivalently be expressed through the solute charge density and solvent potential or vice versa
\begin{equation}
    -\int d\mathbf{r}\, \Bar{\Phi}^\text{pol}(\mathbf{r})\nabla_{at}\Bar{\rho}(\mathbf{r}) = -\int d\mathbf{r}\,\Bar{\rho}^\text{pol}(\mathbf{r})\nabla_{at}\Bar{\Phi}^\text{vac}(\mathbf{r})
\end{equation}
$\nabla_{at}$ is the gradient with respect to $\mathbf{R}_{at}$. It can be seen in the right hand side formulation that, in analogy to \cref{eq:G_elstat_rephrase}, the integrand vanishes \textit{inside} and is equal to its original counterpart \textit{outside}.

Additional terms may emerge due to an explicit dependency of $s$ on the nuclear coordinates. Similarly to the additional free energy contributions discussed in \cref{sec:free_energy}, these are evaluated as volume integrals the integrands of which vanish $\textit{inside}$ and locally depend only on smoothened quantities which are equal to their original counterparts \textit{outside}. Therefore, $\Delta\Bar{\mathbf{F}}^\text{solv}_{at}=\Delta\mathbf{F}^\text{solv}_{at}$.

\section[Continuity of multipole components and their derivatives at the origin]{Continuity of multipole components and their \\derivatives at the origin}\label{sec:polynomial_order}

The smoothened multipole components follow the general form
\begin{equation}\label{eq:f_times_Y}
    \delta \widetilde{\rho}^\text{smooth}_{at,lm}(\mathbf{r}) = f(r)Y_{lm}(\Omega)
\end{equation}
with a polynomial
\begin{equation}\label{eq:polynomial}
    f(r) = a_0 + a_1 r + a_2 r^2 + \hdots + a_{n_\text{max}}r^{n_\text{max}}
\end{equation}
Throughout this section, we use the shorthand notation $f(r)$ for the polynomial that is called $P_{at,lm}(r)$ in the main article, do not explicitly denote $0\leq r<r^\text{cut}_{at}$, and drop the subscript $at$ in the spherical coordinates $r,\Omega$ for reasons of clarity. It is implied that $f(r)$ depends on $at$, $l$ and $m$, that $r$ lies inside the cutoff radius, and that the coordinate system is shifted to have atom $at$ at the origin. The gradient of \cref{eq:f_times_Y} is given by
\begin{equation}\label{eq:gradient_Y}
    \nabla \big( f(r) Y_{lm}(\Omega)\big) = \frac{f(r)}{r}r\nabla Y_{lm}(\Omega) + f^\prime(r)Y_{lm}(\Omega) \hat{\mathbf{e}}_r
\end{equation}
with $f^\prime=\frac{\partial f}{\partial r}$ and the unit vector $\hat{\mathbf{e}}_r$ pointing in the direction of the point $\mathbf{r}$. We point out that the index $r$ does not denote a dependency and the unit vector is, in fact, a function $\hat{\mathbf{e}}_r(\Omega)$ only of $\Omega$. The Hessian is given by
\begin{equation}\label{eq:Hessian_Y}
\begin{split}
    \nabla\otimes\nabla \big( f(r) Y_{lm}(\Omega)\big) = &\frac{f(r)}{r^2} \, r^2 \nabla\otimes\nabla Y_{lm}(\Omega) \\
    &+ \frac{f^\prime(r)}{r}\,\Big(\hat{\mathbf{e}}_r\otimes r \nabla Y_{lm}(\Omega) + r \nabla Y_{lm}(\Omega) \otimes \hat{\mathbf{e}}_r+ Y_{lm}(\Omega) \big(\mathbf{1}- \hat{\mathbf{e}}_r \otimes\hat{\mathbf{e}}_r \big) \Big) \\
    &+f^{\prime\prime}(r)\,Y_{lm}(\Omega)\hat{\mathbf{e}}_r \otimes\hat{\mathbf{e}}_r
\end{split}
\end{equation}
with the $3\times 3$ identity matrix $\mathbf{1}$ and $f^{\prime\prime}=\frac{\partial^2 f}{\partial r^2}$. We want $\delta \widetilde{\rho}^\text{smooth}_{at,lm}(\mathbf{r})$ to be continuous and twice continuously differentiable. Because $f(r)$ fulfills these continuities $\forall \mathbf{r}$ and $Y_{lm}(\Omega)$ fulfills them $\forall\mathbf{r}\neq \mathbf{0}$, their product also fulfills them $\forall \mathbf{r}$ as long as it fulfills them for $\mathbf{r}=\mathbf{0}$. In other words, we want the limit of $\delta \widetilde{\rho}^\text{smooth}_{at,lm}(\mathbf{r})$ and its gradient and Hessian for $r\to 0^+$ to exist and be independent of $\Omega$:
\begin{equation}\label{eq:limits}
    \lim_{r\to 0^+} F(\mathbf{r}) \bigg|_{\Omega=\Omega_1} = \lim_{r\to 0^+} F(\mathbf{r}) \bigg|_{\Omega=\Omega_2} \quad \forall \, \Omega_1, \Omega_2
\end{equation}
where for $F(\mathbf{r})$ we insert \cref{eq:f_times_Y,eq:gradient_Y,eq:Hessian_Y}. We note that while $Y_{lm}(\Omega)$ and $\hat{\mathbf{e}}_r$ are discontinuous at $\mathbf{r}=\mathbf{0}$, they do not depend on $r$ and can be separated out of the limit. The gradient of the spherical harmonics is given by
\begin{equation}
    \nabla Y_{lm}(\Omega) = \left( \hat{\mathbf{e}}_r\frac{\partial}{\partial r} +\frac{1}{r}\left( \hat{\mathbf{e}}_\theta\frac{\partial}{\partial \theta} + \hat{\mathbf{e}}_\phi\frac{1}{\sin \theta}\frac{\partial}{\partial \phi} \right)\right) Y_{lm}(\Omega)
\end{equation}
where $\theta$ and $\phi$ are the polar and azimuthal components of $\Omega$ and $\hat{\mathbf{e}}_{\theta/\phi}$ are the corresponding unit vectors. We point out that, like $\hat{\mathbf{e}}_r$, these unit vectors are functions only of $\Omega$ but not of $r$. Realizing that the derivative with respect to $r$ vanishes because $Y_{lm}(\Omega)$ does not depend on it, we find that the gradient can be rewritten as
\begin{equation}
    \nabla Y_{lm}(\Omega) = \frac{1}{r} \mathbf{G}_{lm}(\Omega)
\end{equation}
with some vector valued function $\mathbf{G}_{lm}(\Omega)$ which does not depend on $r$. For the same reasons of vanishing dependencies, we find the Hessian of the spherical harmonics to be
\begin{equation}
    \nabla\otimes\nabla Y_{lm}(\Omega) = -\frac{1}{r^2} \hat{\mathbf{e}}_r \otimes \mathbf{G}_{lm}(\Omega) + \frac{1}{r^2} \mathbf{H}_{lm}(\Omega)
\end{equation}
with some matrix valued function $\mathbf{H}_{lm}(\Omega)$ which does not depend on $r$. We notice that the limits
\begin{subequations}
\begin{align}
    \lim\limits_{r\to 0^+} r \nabla Y_{lm}(\Omega) &= \mathbf{G}_{lm}(\Omega)\\
    \lim\limits_{r\to 0^+} r^2 \nabla\otimes\nabla Y_{lm}(\Omega) &= -\hat{\mathbf{e}}_r \otimes \mathbf{G}_{lm}(\Omega) + \mathbf{H}_{lm}(\Omega)
\end{align}
\end{subequations}
exist because all dependencies on $r$ cancel. This means that the limits in \cref{eq:limits} exist as long as the $r$-dependent terms in \cref{eq:f_times_Y,eq:gradient_Y,eq:Hessian_Y}
\begin{subequations}
\begin{align}
    f(r) &= a_0 + a_1 r + a_2 r^2 + \hdots \\
    \frac{f(r)}{r} &= \frac{a_0}{r} + a_1 + a_2 r + \hdots \\
    f^\prime(r) &= a_1 + 2 a_2 r + \hdots \\
    \frac{f(r)}{r^2} &= \frac{a_0}{r^2} + \frac{a_1}{r} + a_2 + \hdots \\
    \frac{f^\prime(r)}{r} &= \frac{a_1}{r} + 2a_2 + \hdots \\
    f^{\prime\prime}(r) &= 2a_2 + \hdots
\end{align}
\end{subequations}
do not diverge at $r=0$. This is the case if $a_0=a_1=0$. This is also sufficient for the limits of the value and gradient of $\delta \widetilde{\rho}^\text{smooth}_{at,lm}(\mathbf{r})$ to be zero at the origin, fulfilling \cref{eq:limits}. If, additionally, $a_2=0$, the Hessian is also zero and therefore continuous at the origin.

While $a_0=a_1=a_2=0$ is a sufficient condition for $\delta \widetilde{\rho}^\text{smooth}_{at,lm}(\mathbf{r})$ to be continuous and twice continuously differentiable, it may not always be a necessary one. This can be shown using a substitution
\begin{equation}\label{eq:sub_f_g}
    g(r) = N_l^{-1}\frac{f(r)}{r^l} = N_l^{-1}\left(a_0 r^{-l} + a_1 r^{1-l} + a_2 r^{2-l} + \hdots \right)
\end{equation}
We can rewrite
\begin{equation}\label{eq:g_times_R}
    \delta \widetilde{\rho}^\text{smooth}_{at,lm}(\mathbf{r}) = g(r) \,\mathcal{R}_{lm}(\mathbf{r})
\end{equation}
with the regular solid harmonic function
\begin{equation}
    \mathcal{R}_{lm}(\mathbf{r}) = N_l\, r^l \,Y_{lm}(\Omega)
\end{equation}
where $N_l$ is a normalization constant. $\mathcal{R}_{lm}(\mathbf{r})$ are continuous and continuously differentiable to infinite order $\forall \mathbf{r}$, including $\mathbf{r}=\mathbf{0}$. The gradient of \cref{eq:g_times_R} is given by
\begin{equation}\label{eq:gradient_R}
    \nabla \big( g(r) \mathcal{R}_{lm}(\mathbf{r}) \big) = g(r)\nabla\mathcal{R}_{lm}(\mathbf{r}) + \frac{g^\prime(r)}{r} \mathcal{R}_{lm}(\mathbf{r}) \mathbf{r}
\end{equation}
The Hessian is given by
\begin{equation}\label{eq:Hessian_R}
\begin{split}
    \nabla \otimes \nabla \big( g(r) \mathcal{R}_{lm}(\mathbf{r}) \big) =& g(r) \nabla\otimes\nabla \mathcal{R}_{lm}(\mathbf{r}) \\
    & +  \frac{g^\prime(r)}{r} \Big(\mathbf{r}\otimes \nabla \mathcal{R}_{lm}(\mathbf{r}) + \nabla \mathcal{R}_{lm}(\mathbf{r})\otimes\mathbf{r} + \mathcal{R}_{lm}(\mathbf{r})\mathbf{1}\Big) \\
    & + \left(\frac{g^{\prime\prime}(r)}{r^2}-\frac{g^\prime(r)}{r^3}\right)\mathcal{R}_{lm}(\mathbf{r})\mathbf{r}\otimes\mathbf{r}
\end{split}
\end{equation}
There is a difference in \cref{eq:f_times_Y,eq:gradient_Y,eq:Hessian_Y} compared to \cref{eq:g_times_R,eq:gradient_R,eq:Hessian_R}. While the limits $r\to 0^+$ of the terms $\hat{\mathbf{e}}_r$, $Y_{lm}(\Omega)$, $r\nabla Y_{lm}(\Omega)$ and $r^2 \nabla\otimes\nabla Y_{lm}(\Omega)$ exist, they do depend on $\Omega$. In other words, these terms are discontinuous at $\mathbf{r}=\mathbf{0}$. By contrast, the terms $\mathbf{r}$, $\mathcal{R}_{lm}(\mathbf{r})$, $\nabla \mathcal{R}_{lm}(\mathbf{r})$ and $\nabla\otimes\nabla \mathcal{R}_{lm}(\mathbf{r})$ are continuous at the origin, making the check for a condition equivalent to \cref{eq:limits} trivial. We can directly see that \cref{eq:g_times_R,eq:gradient_R,eq:Hessian_R} are continuous if the terms
\begin{subequations}
\begin{align}
    g(r) =& N_l^{-1}\left(\frac{a_0}{r^l} + \frac{a_1} {r^{l-1}} + \hdots \right.  \nonumber \\
    &+ \left.\frac{a_{l-1}}{r} + a_l + a_{l+1}r + a_{l+2}r^2 + a_{l+3}r^3 + a_{l+4}r^4 + \hdots \right) \\
    \frac{g^\prime(r)}{r} =& N_l^{-1}\left(-l\frac{a_0}{r^{l+2}} -(l-1) \frac{a_1} {r^{l+1}} + \hdots \right.\nonumber \\
    &-\left.\frac{a_{l-1}}{r^3} + \frac{a_{l+1}}{r} + 2a_{l+2} + 3a_{l+3}r + 4a_{l+4}r^2 + \hdots \right) \\
    \frac{g^{\prime\prime}(r)}{r^2}-\frac{g^\prime(r)}{r^3} =& N_l^{-1}\left( l(l+2)\frac{a_0}{r^{l+4}} + (l+1)(l-1)\frac{a_1}{r^{l+3}} + \hdots\right. \nonumber \\
    &+\left. 3\frac{a_{l-1}}{r^5} - \frac{a_{l+1}}{r^3} +3 \frac{a_{l+3}}{r} + 8 a_{l+4} + \hdots \right)
\end{align}
\end{subequations}
are continuous. There are three kinds of coefficients $a_n$ which never occur as coefficients of a diverging term in the above formulas. First, $a_l$ occurs only as a constant in $g(r)$ and vanishes in the derivatives. Second, $a_{l+2}$ occurs as the coefficient of $r^2$ in $g(r)$, as a constant in $\frac{g^\prime(r)}{r}$, and cancels in $\frac{g^{\prime\prime}(r)}{r^2}-\frac{g^\prime(r)}{r^3}$. Finally, $a_n$ with $n\geq l+4$ occur as coefficients of $r^{n^\prime}$ with $n^\prime \geq 0$ in all three terms.

Therefore, $\delta \widetilde{\rho}^\text{smooth}_{at,lm}(\mathbf{r})$ is continuous and twice continuously differentiable at $\mathbf{r}=\mathbf{0}$ if $a_n=0\quad\forall n \in \{n \, : \, n < l+4 \} \setminus \{l,l+2\}$. This is, again, a sufficient but not necessary condition. For most $l$, this is more restrictive than $a_0=a_1=a_2=0$, as we determined using \cref{eq:f_times_Y,eq:gradient_Y,eq:Hessian_Y,eq:limits}. For $l\leq 2$, however, we find additional terms which may be non-zero. By separating the low and high order terms
\begin{equation}
    \delta \widetilde{\rho}^\text{smooth}_{at,lm}(\mathbf{r}) = \widetilde{g}(r) \,\mathcal{R}_{lm}(\mathbf{r}) + \widetilde{f}(r)Y_{lm}(\Omega)
\end{equation}
with
\begin{subequations}
\begin{align}
    \widetilde{g}(r) &= N_l^{-1}\left(a_0 r^{-l} + a_1 r^{1-l} + a_2 r^{2-l} \right) \\
    \widetilde{f}(r) &= a_3 r^3 + a_4 r^4 + \hdots + a_{n_\text{max}}r^{n_\text{max}}
\end{align}
\end{subequations}
we can see that $\widetilde{f}(r)Y_{lm}(\Omega)$ is continuous and twice continuously differentiable at the origin. $\widetilde{g}(r) \,\mathcal{R}_{lm}(\mathbf{r})$ and its derivatives are trivially continuous if $a_0=a_1=a_2=0$. In the special cases of $l\leq 2$, however, $\widetilde{g}(r) \,\mathcal{R}_{lm}(\mathbf{r})$ is still continuous if $a_l$ (and $a_{l+2}$ for $l=0$) is non-zero.

We have four additional conditions to enforce, namely the continuity of the value, gradient, and Hessian of $\delta \widetilde{\rho}^\text{smooth}_{at,lm}(\mathbf{r})$ at $r=r_{at}^\text{cut}$ and the conservation of the long range multipole moment. Note that the continuities in angular direction are always fulfilled due to the continuity of $Y_{lm}(\Omega)$, so only the continuity in $r$ needs to be enforced. In order to have 4 free parameters, we choose $n_\text{max}=4$ for $l=0$, $n_\text{max}=5$ for $l=1$ and $l=2$, and $n_\text{max}=6$ for $l\geq 3$.

\section{Energy in vacuum}\label{sec:E_vac}

The total energy of a KS DFT calculation in vacuum, i.e., without continuum embedding, is given by
\begin{equation}\label{eq:E_vac_def}
    E^\text{vac} = \frac{1}{2} \int d\mathbf{r} \rho(\mathbf{r})\Phi^\text{vac}(\mathbf{r}) + T[\rho^\text{el}] + E_\text{xc}[\rho^\text{el}]
\end{equation}
where $\rho$ is the full explicit charge density $\rho^\text{el}+\rho^\text{nuc}$ with the nuclear charge density $\rho^\text{nuc}$, $T$ is the Kohn-Sham kinetic energy of the electrons, $E_\text{xc}$ is the xc energy, and $\Phi$ is the electrostatic potential generated by $\rho$. It is implied here and in the following that the first integral and similar terms do not include the interaction of any nucleus with its own singularity. In practice, $E^\text{vac}$ is obtained from the sum of KS eigenvalues $\epsilon_i$ weighted by occupation numbers $f_i$
\begin{equation}\label{eq:sum_eigval}
    \sum_i f_i \epsilon_i = \int d\mathbf{r} \rho^\text{el}(\mathbf{r})V_\text{eff}(\mathbf{r}) + T[\rho^\text{el}]
\end{equation}
which follows from the KS equations and contains $T$ as well as the interaction of the electrons with the effective potential $V_\text{eff}$ which enters the KS operator. \Cref{eq:sum_eigval} holds for any $V_\text{eff}$. To arrive at \cref{eq:E_vac_def} in a vacuum calculation, the interaction with $v_\text{xc}$ is replaced with $E_\text{xc}$, the double counting of the electron-electron interaction is removed, and the electrostatic interaction between the nuclei is added.
\begin{equation}\label{eq:E_vac}
\begin{split}
    E^\text{vac} &= \sum_i f_i \epsilon_i - \int d\mathbf{r}\, \rho^\text{el}(\mathbf{r})\left(V_\text{eff}(\mathbf{r})-\Phi^\text{vac}(\mathbf{r}) \right) + E_\text{xc}[\rho^\text{el}] \\
    &\quad -\frac{1}{2} \int d\mathbf{r}\, \rho^\text{el}(\mathbf{r})\Phi^\text{vac}(\mathbf{r}) + \frac{1}{2} \int d\mathbf{r}\, \rho^\text{nuc}(\mathbf{r})\Phi^\text{vac}(\mathbf{r})
\end{split}
\end{equation}
The first integral is more general than the more conventional formulation $\int d\mathbf{r}\, \rho^\text{el}(\mathbf{r})v_\text{xc}(\mathbf{r})$, and ensures that \cref{eq:E_vac} yields \cref{eq:E_vac_def} for any $V_\text{eff}$, not just \cref{eq:V_eff_vac}. In practice, ${V_\text{eff}-\Phi^\text{vac}}$ is not computed as a difference but rather by considering individually any terms that enter $V_\text{eff}$, except $\Phi^\text{vac}$. In the case of a vacuum calculation using \cref{eq:V_eff_vac}, the conventional formulation is recovered. In the case of an embedded calculation using \cref{eq:V_eff_solv}, however, an additional term $-\int d\mathbf{r}\, \rho^\text{el}(\mathbf{r})\Delta \Bar{V}_\text{eff}^\text{solv}(\mathbf{r})$ emerges.

FHI-aims uses the above formulation for the second and third integral which conveniently uses only the full electrostatic potential $\Phi^\text{vac}$ rather than considering the electronic and nuclear potentials separately. To arrive at this formulation, half of the electron-nuclear interaction is subtracted from the electron-electron double counting correction and added to the nuclear-nuclear term, effectively adding zero to the overall expression.

In practice, replacing the electron density $\rho^\text{el}$ with the multipole density $\rho^\text{MP}$ in the double counting correction for the electron-electron interaction as well as the solvent potential correction leads to a faster convergence with the multipole expansion order, as discussed in ref.~\citenum{blum2009} as well as \cref*{main-sec:conv_lmax} in the main article.

\section{Separation of smoothing region from solvent}\label{sec:separation_smoothing_solvent}

We tested whether our choice of $r^\text{cut}_{at}$ safely remains \textit{inside} for the standard parameterizations of SCCS\cite{andreussi2012} and SSCS,\cite{fisicaro2017} a reparameterization of SCCS for electrochemical applications\cite{hoermann2019}, and the field-aware modification of SSCS.\cite{truscott2019} The test systems were a water molecule and a NaF dimer at 2~{\AA} separation. The field-aware SSCS was only tested for the water molecule, as no parameterization for systems containing both anions and cations was available. Additionally, we tested the anion/cation specific parameterizations of SCCS,\cite{dupont2013} SSCS, and field-aware SSCS on a single F$^-$/Na$^+$ ion, respectively. We did not test the solvent-aware modification\cite{andreussi2019_2} of either method as it generally only excludes points from the solvent and assigns them to the cavity but not vice versa. For these initial tests, we used values of $e_\text{cut}=350\,\text{Ry}$ for the $\mathbf{g}$ space cutoff, $p_1=8\cdot\pi/3$, $p_2=5\cdot\pi/3$ for the low-pass filter, and the conjugate-gradient electrostatic solver throughout, with one exception described in \cref{fig:s_f_cation}. As shown in \cref*{main-sec:conv_ecut} in the main article, we later found that these settings are not sufficient to yield converged energies and forces for some of the parameterizations and systems. However, $r^\text{cut}_{at}$ does not depend on these parameters, and $f^\text{switch}_{at,l\neq 0}$ will only get narrower with increased $e_\text{cut}$, cf.~\cref{main-eq:d_diag} in the main article. Therefore, these tests are sufficient to demonstrate that $r^\text{cut}_{at}$ remains \textit{inside}.

The results for the water H atom in SCCS and SSCS are shown in \cref*{main-fig:regions} in the main article. The results for the O atom in the same systems, as well as for all other systems and solvent parameterizations are shown below. The only cases where the solvent region was found to penetrate into $r^\text{cut}_{at}$ were the ion specific parameterizations of field-aware SSCS. However, these were originally parameterized using only a single OH$^-$/H$_3$O$^+$ ion as a reference. Once a general parameterization of field-aware SSCS becomes available, it may be worth revisiting our choice of $r^\text{cut}_{at}$. At the present time, we refer the user to the option of manually choosing $r^\text{cut}_{at}$ if any issues arise. Even with our default choice, the overlap of the smoothing region with the cavity function is rather small. For all other cases, we found $r^\text{cut}_{at}$ to be at a safe distance from the cavity boundary.

We do, however, observe an overlap of $f^\text{switch}_{at,l\neq0}$ with the solvent in some cases. This is generally not a desirable feature because, as elaborated in \cref*{main-sec:switch} in the main article, one of the purposes of the switching function is to exclude integration errors \textit{outside} which cannot be exactly compensated by corrections inside $r^\text{solv}_{at}$. Nonetheless, we point out that $f^\text{switch}_{at,l\neq0}$ enters \cref*{main-eq:multipole_condition_numeric} in the main article merely as a weighting function for integration errors. The errors themselves may still be systematically reduced by choosing a denser grid in Environ. In addition, this will also lead to a narrower transition region of $f^\text{switch}_{at,l\neq0}$ beyond $r^\text{solv}_{at}$, reducing its overlap with the solvent.

\FloatBarrier

\newpage

\begin{figure}[!htb]
    \centering
    \includegraphics[width=0.9\linewidth]{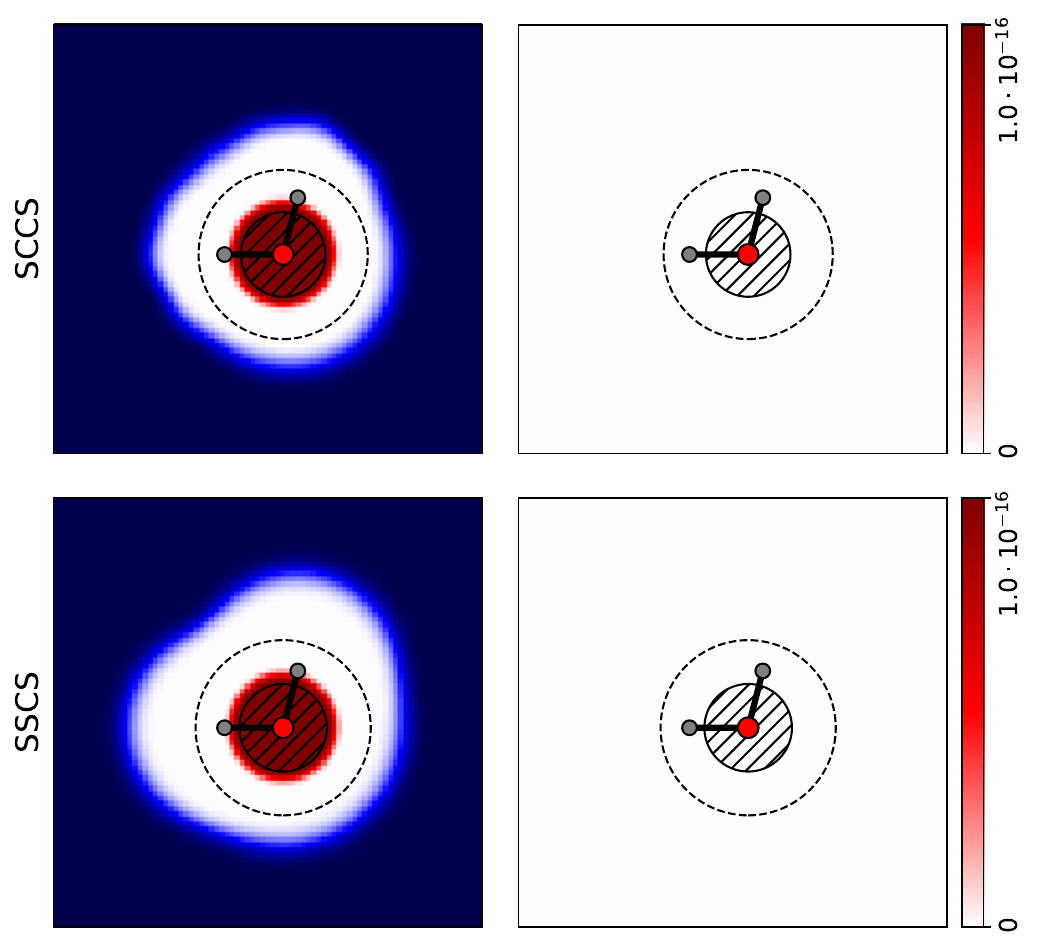}
    \caption{Like \cref*{main-fig:regions} in the main article but with the switching function and radii of the O atom. Contrast in left subfigures identical to \cref*{main-fig:regions}. Contrast in right subfigures set to maximum value of displayed data and indicated by colorbar (or $10^{-16}$ if numerically zero as is the case here, but non-zero in some of the cases below).}
    \label{fig:s_f_water_O}
\end{figure}

\begin{figure}[p]
    \centering
    \includegraphics[width=0.9\linewidth]{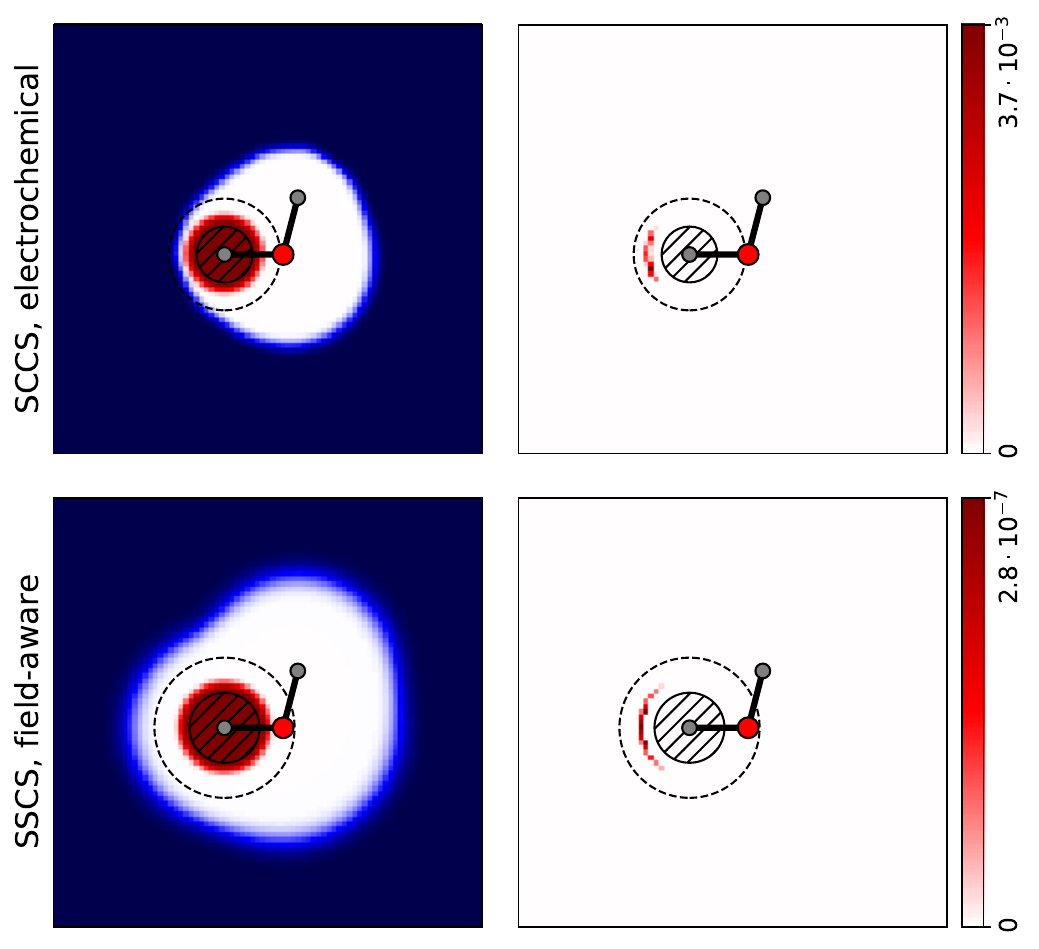}
    \caption{Like \cref{fig:s_f_water_O} with the switching function and radii of one H atom. Electrochemical parameterization of SCCS from ref.~\citenum{hoermann2019} (top) and field-aware modification of SSCS, cf.~ref.~\citenum{truscott2019}, `water' parameters in Table 1 therein (bottom).}
    \label{fig:s_f_water_H_extra}
\end{figure}

\begin{figure}[p]
    \centering
    \includegraphics[width=0.9\linewidth]{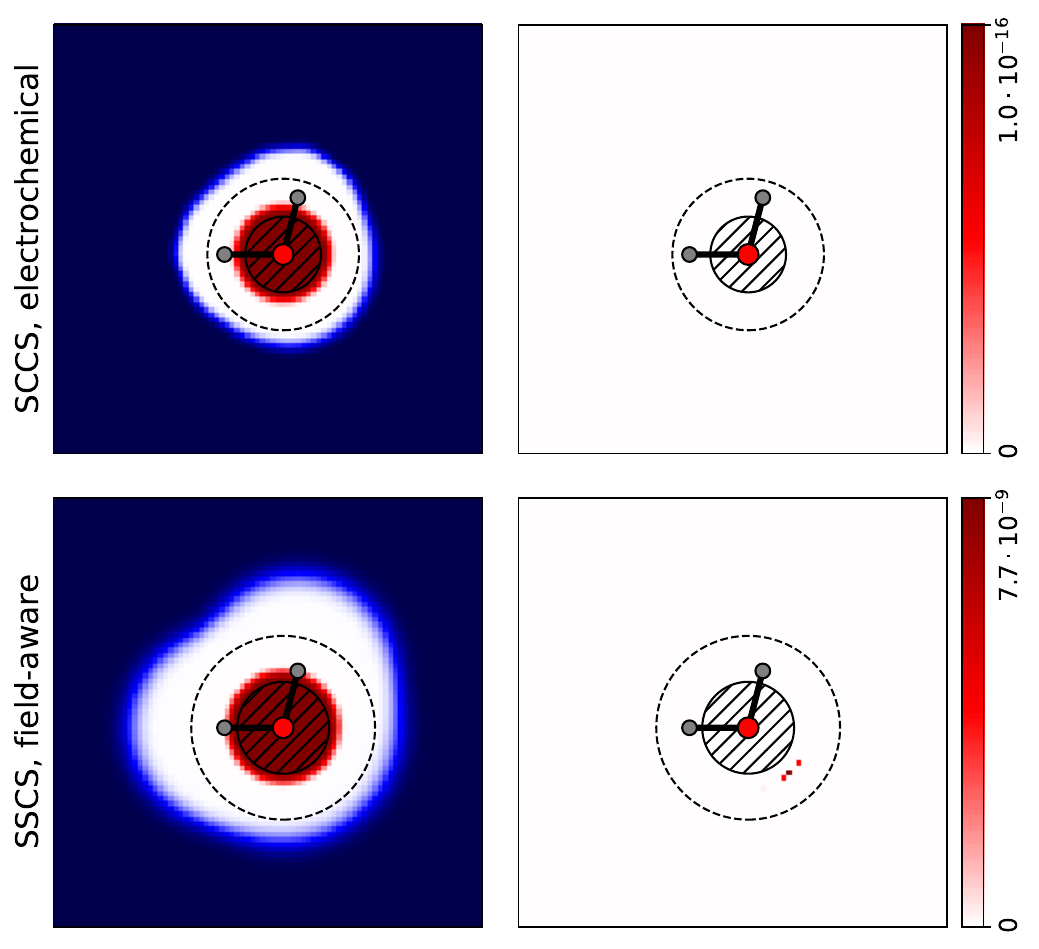}
    \caption{Like \cref{fig:s_f_water_H_extra} with the switching function and radii of the O atom.}
    \label{fig:s_f_water_O_extra}
\end{figure}

\begin{figure}[p]
    \centering
    \includegraphics[width=0.9\linewidth]{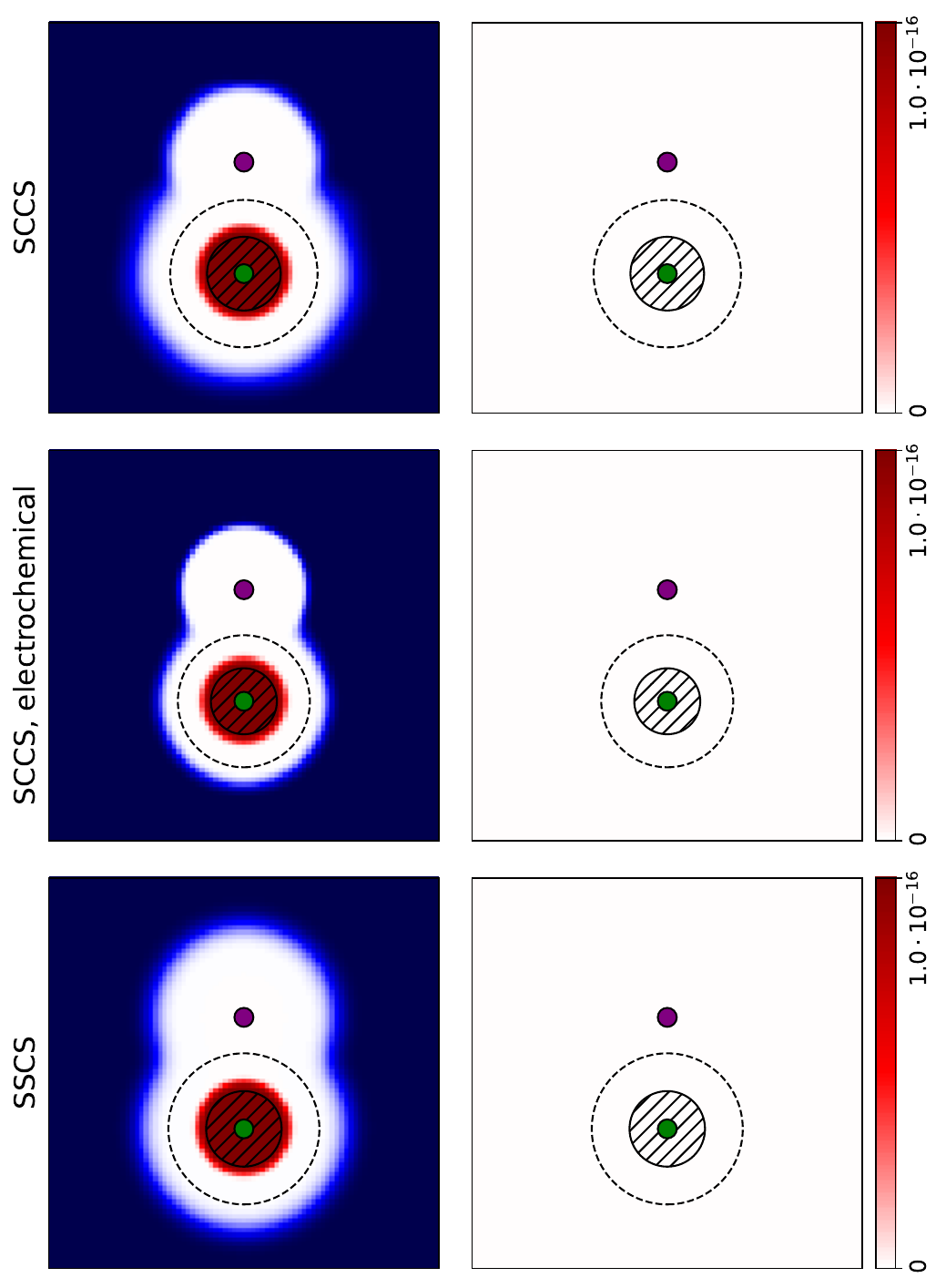}
    \caption{Like \cref{fig:s_f_water_H_extra} for a NaF dimer at 2~{\AA} separation (Na: violet, F: green). SCCS (top), electrochemical reparameterization of SCCS\cite{hoermann2019} (middle), and SSCS (bottom). Switching function and radii of F atom.}
    \label{fig:s_f_dimer_F}
\end{figure}

\begin{figure}[p]
    \centering
    \includegraphics[width=0.9\linewidth]{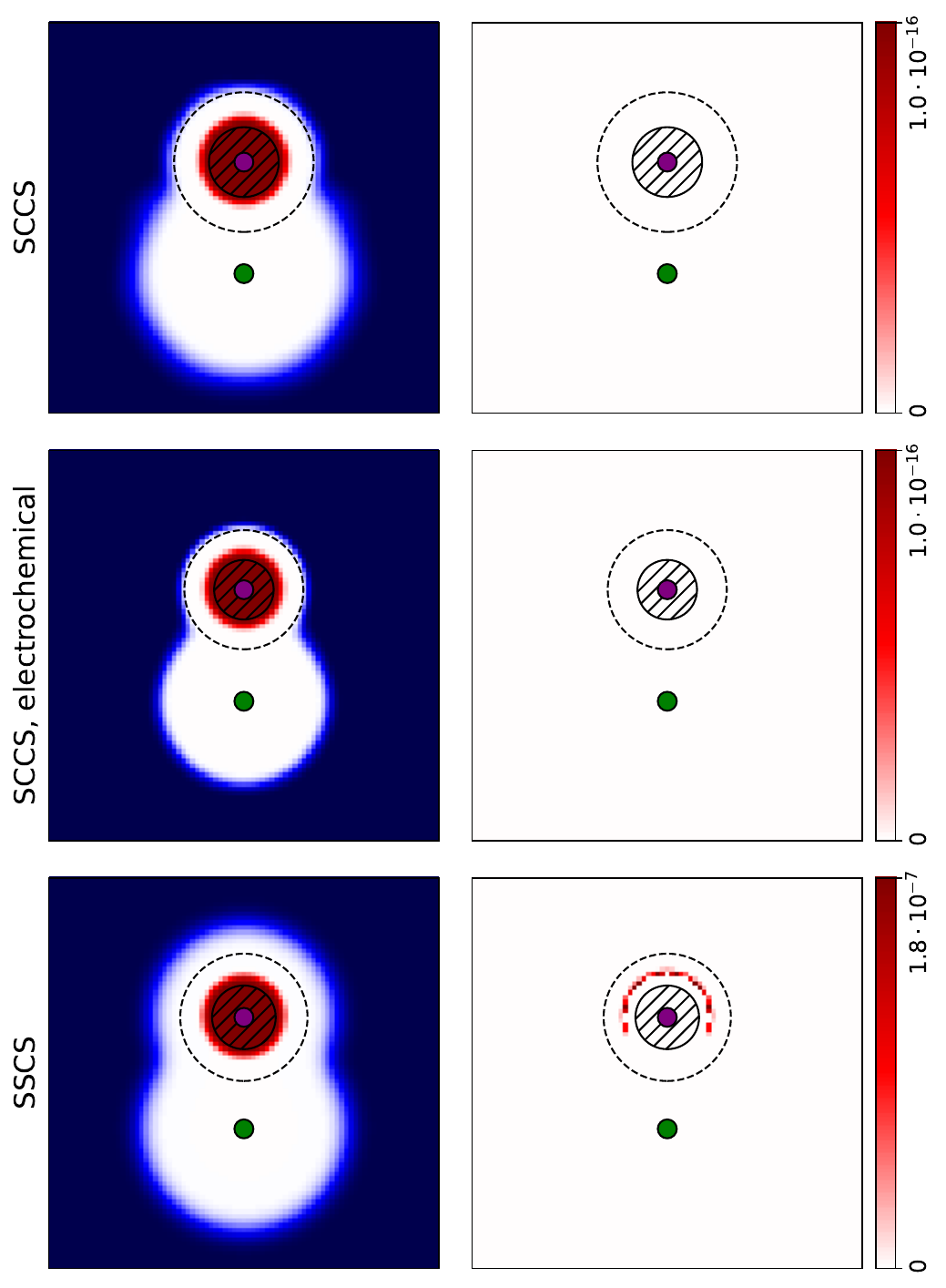}
    \caption{Like \cref{fig:s_f_dimer_F} with the switching function and radii of the Na atom.}
    \label{fig:s_f_dimer_Na}
\end{figure}

\begin{figure}[p]
    \centering
    \includegraphics[width=0.9\linewidth]{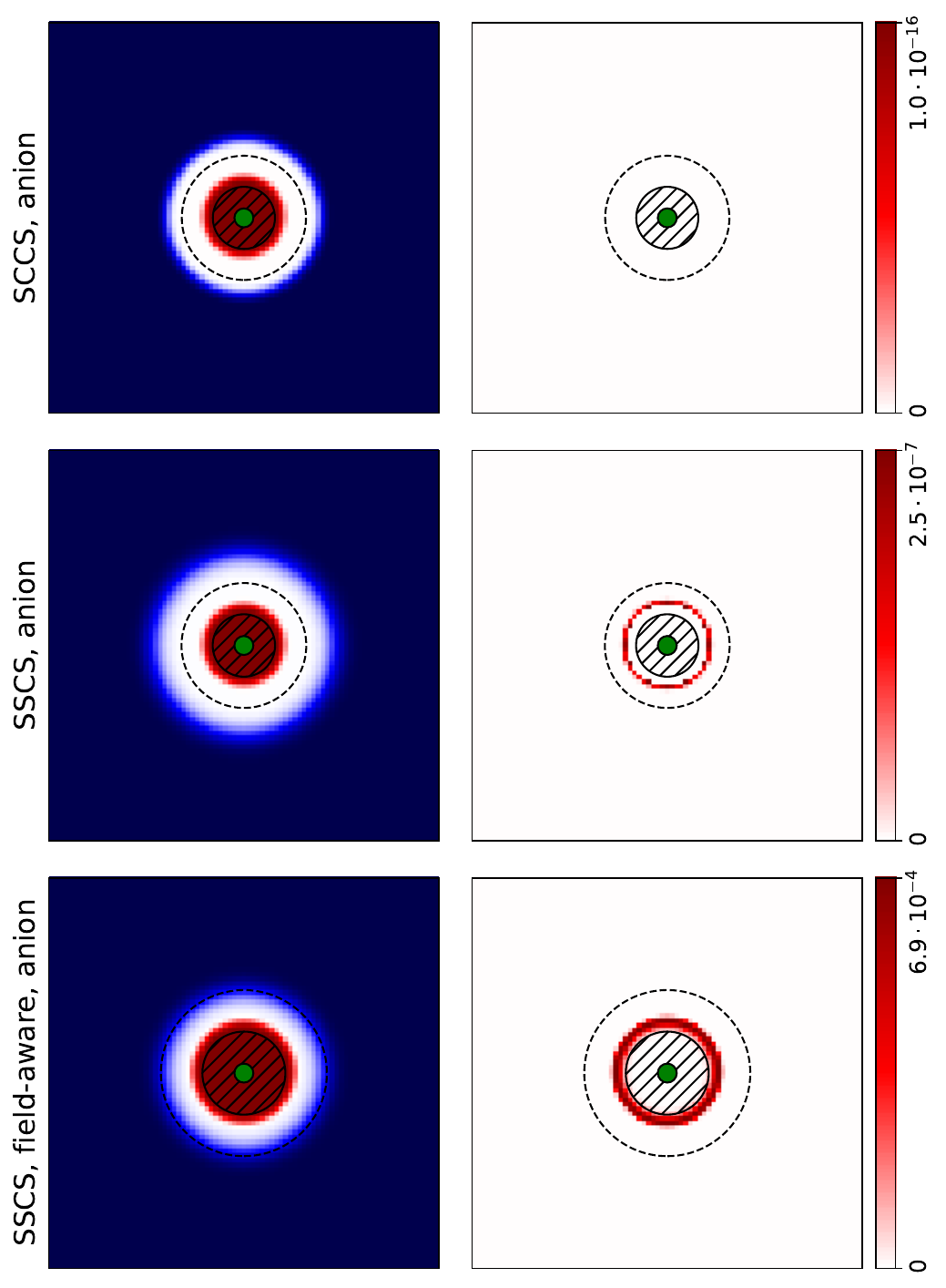}
    \caption{Like \cref{fig:s_f_water_O} for an F$^-$ anion. SCCS with anion specific parameters\cite{dupont2013} (top), SSCS with anion specific parameters\cite{fisicaro2017} (middle), field-aware SSCS with `water (-)' parameters\cite{truscott2019} (bottom).}
    \label{fig:s_f_anion}
\end{figure}

\begin{figure}[p]
    \centering
    \includegraphics[width=0.9\linewidth]{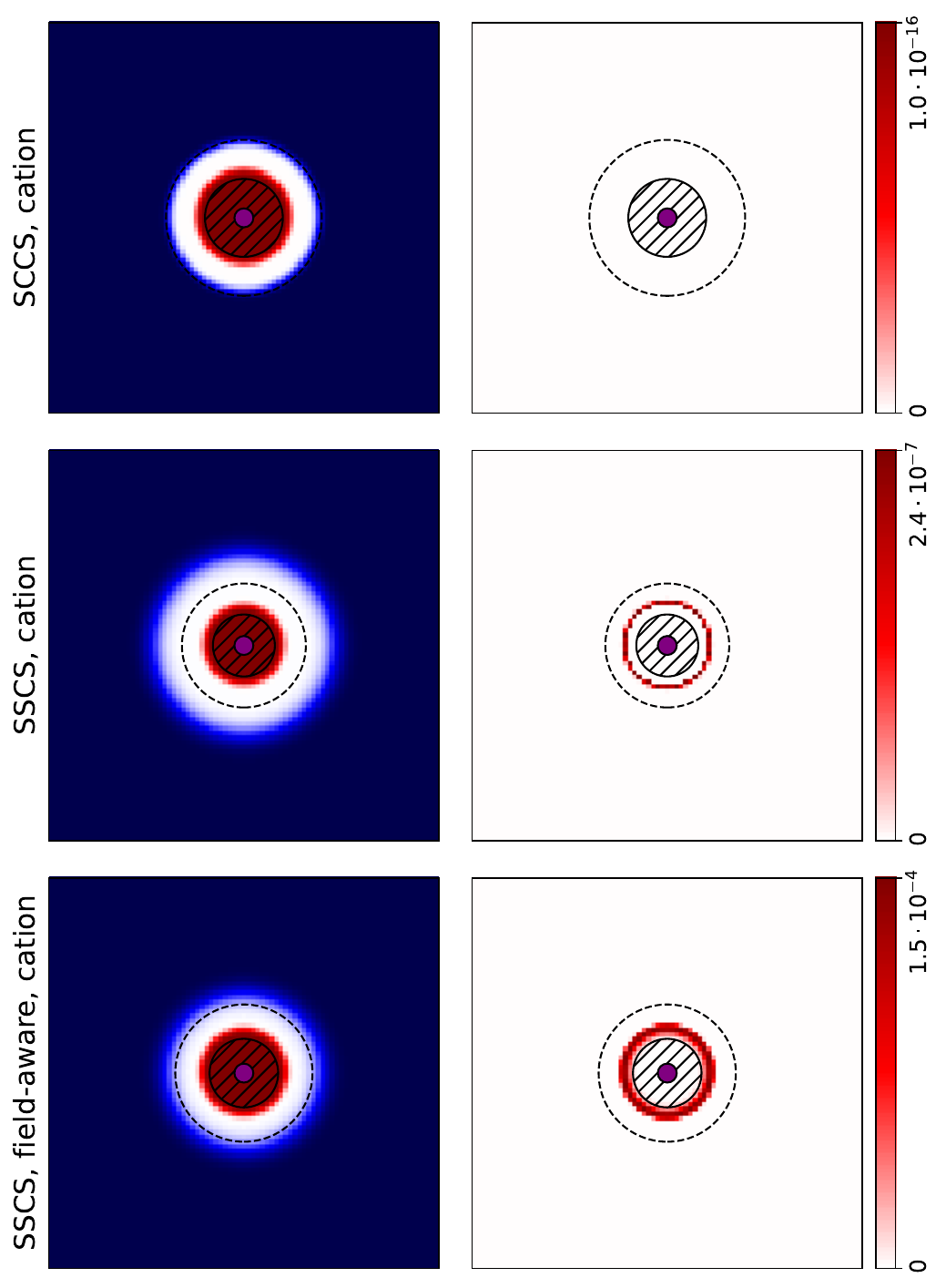}
    \caption{Like \cref{fig:s_f_water_O} for an Na$^+$ cation. SCCS with cation specific parameters\cite{dupont2013} (top), SSCS with cation specific parameters\cite{fisicaro2017} (middle), field-aware SSCS with `water (+)' parameters\cite{truscott2019} (bottom). For SCCS, Environ's `env\_ecut' parameter had to be increased to 450~Ry for the SCF procedure to converge.}
    \label{fig:s_f_cation}
\end{figure}

\FloatBarrier

\newpage

\section{\revision{$\mathbf{k}$ grid convergence for CO on Pt(111)} \label{sec:kgrid}}

\begin{figure}[!htb]
    \centering
    \includegraphics[width=0.70\linewidth]{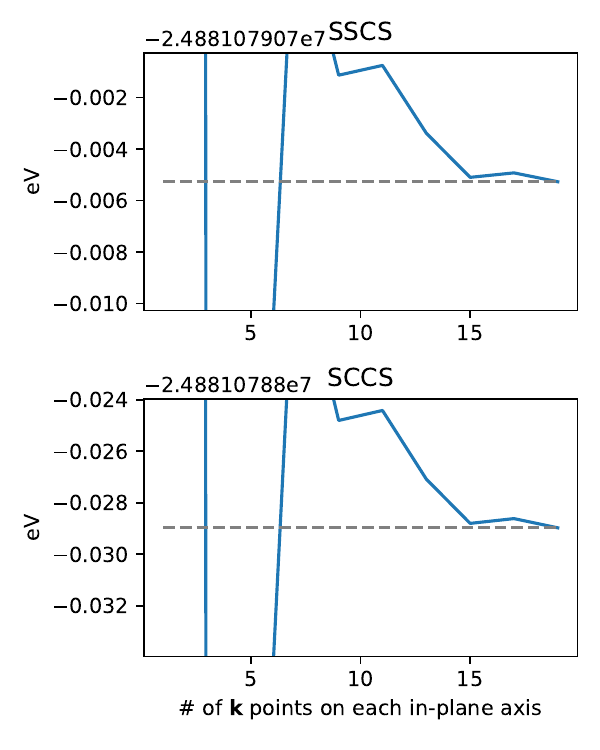}
    \caption{\revision{Convergence of total energy with number of in-plane $\mathbf{k}$ points for CO on Pt(111). Dashed lines at energy for $19\times 19\times 1$ grid as visual guidelines.}}
    \label{fig:kgrid}
\end{figure}

\revision{Starting from the Gamma point, we increased the number of $\mathbf{k}$ grid points on both in-plane axes simultaneously in steps of 2 up to 19. This was tested separately for SSCS and SCCS. As shown in \cref{fig:kgrid}, total energies are converged to 1~meV accuracy with a $15\times 15\times 1$ grid.}

\FloatBarrier

\newpage

\section{Additional figures}\label{sec:add_fig}

\subsection{Energy convergence with grid density, low-pass filter (SSCS)}\label{sec:conv_E_ecut_sscs}

\begin{figure}[!htb]
    \centering
    \includegraphics[width=0.85\linewidth]{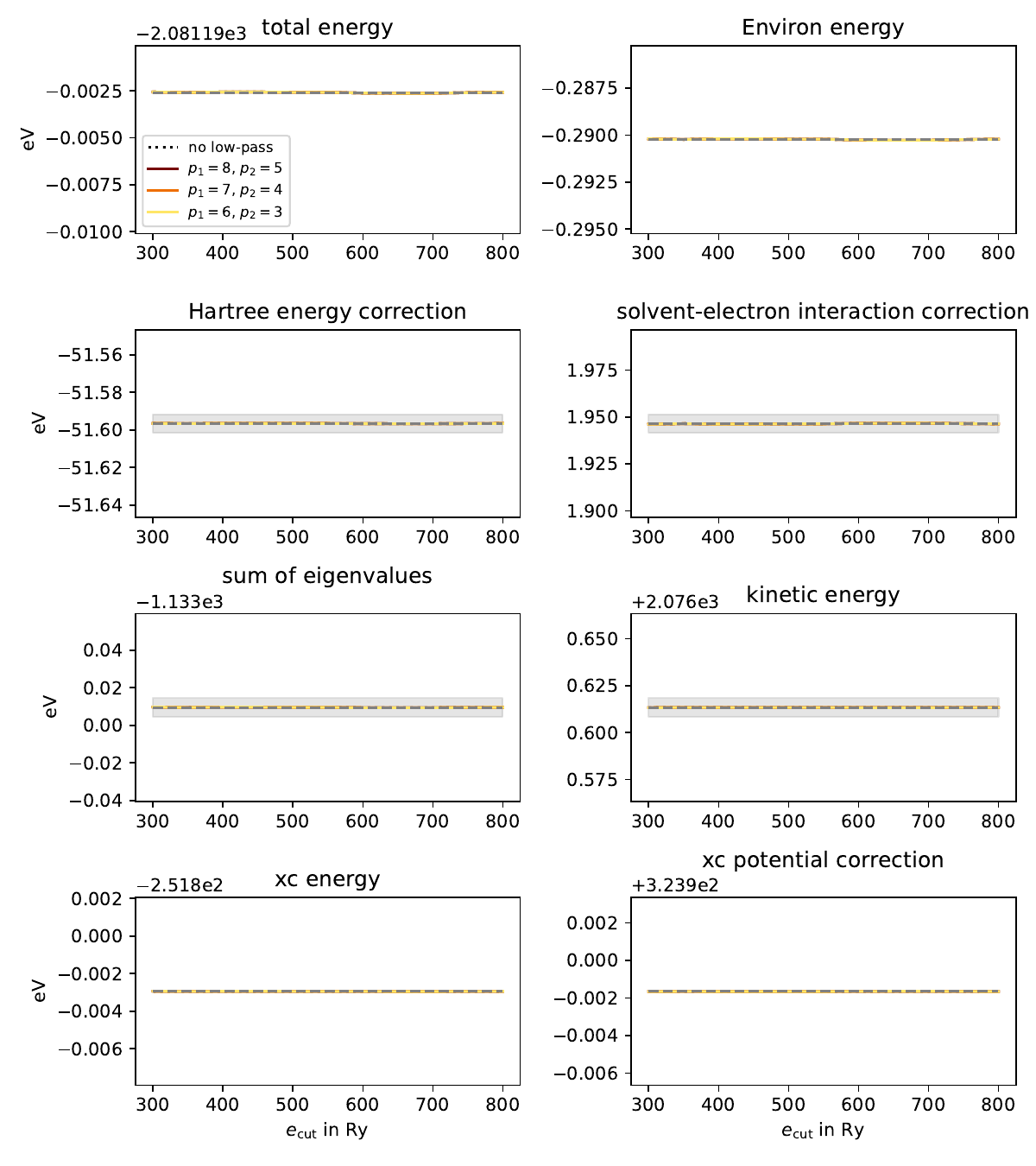}
    \caption{Convergence with $e_\text{cut}$ of total energy and its contributions as reported in the FHI-aims\cite{blum2009} output file using different low-pass filter settings for a water molecule in SSCS. Constant `free atom' contribution not shown. Dashed line shows mean value across $p_1=8$ and $p_1=7$ filters over the three highest $e_\text{cut}$. Note the different energy scales of the second and third row of subfigures. Shaded area indicates the energy scale covered in the other subfigures (10~meV)}
    \label{fig:E_ecut_sscs_water}
\end{figure}

\begin{figure}[p]
    \centering
    \includegraphics[width=0.9\linewidth]{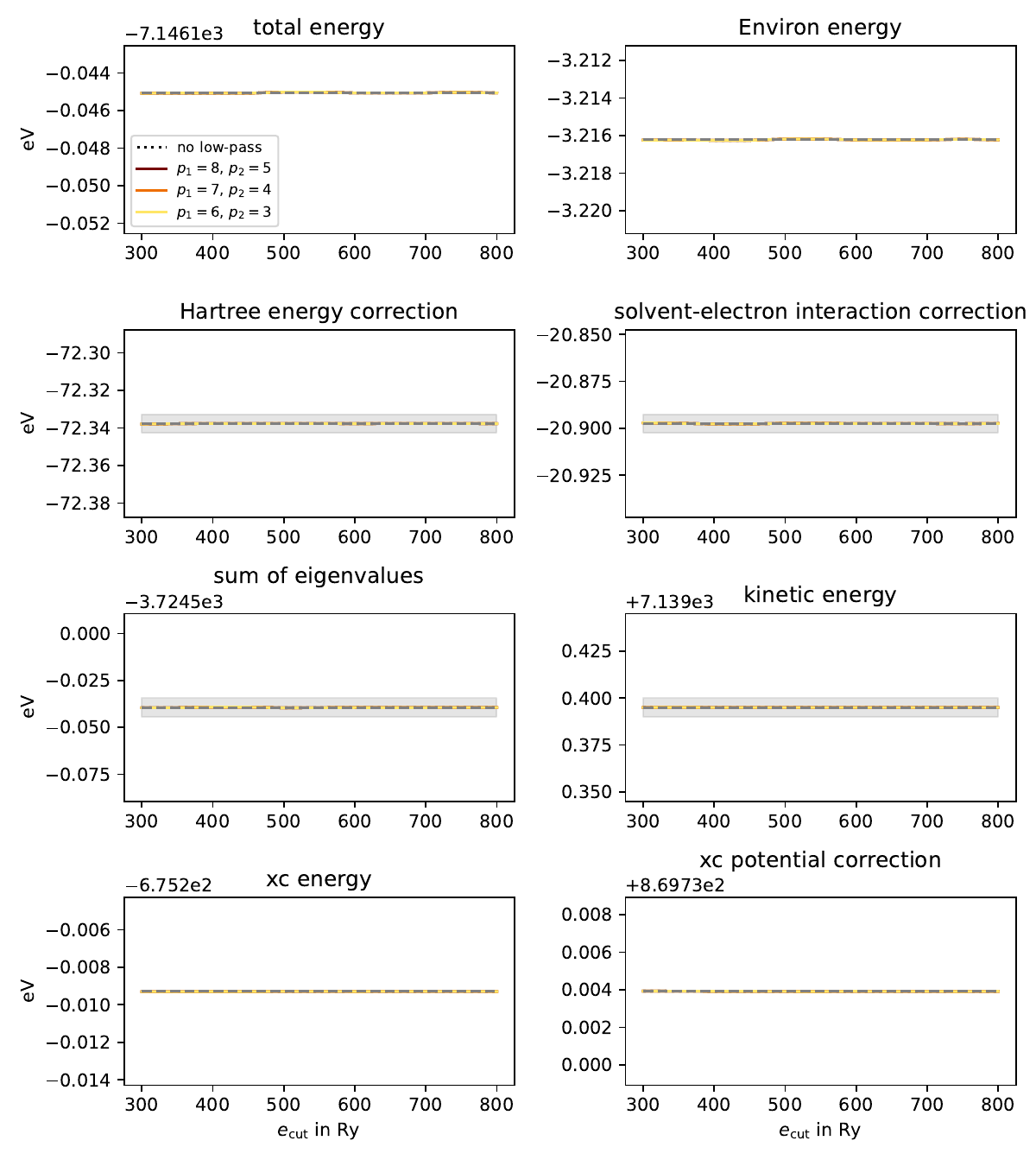}
    \caption{Like \cref{fig:E_ecut_sscs_water}, for a NaF dimer in SSCS.}
    \label{fig:E_ecut_sscs_NaF}
\end{figure}

\begin{figure}[p]
    \centering
    \includegraphics[width=0.9\linewidth]{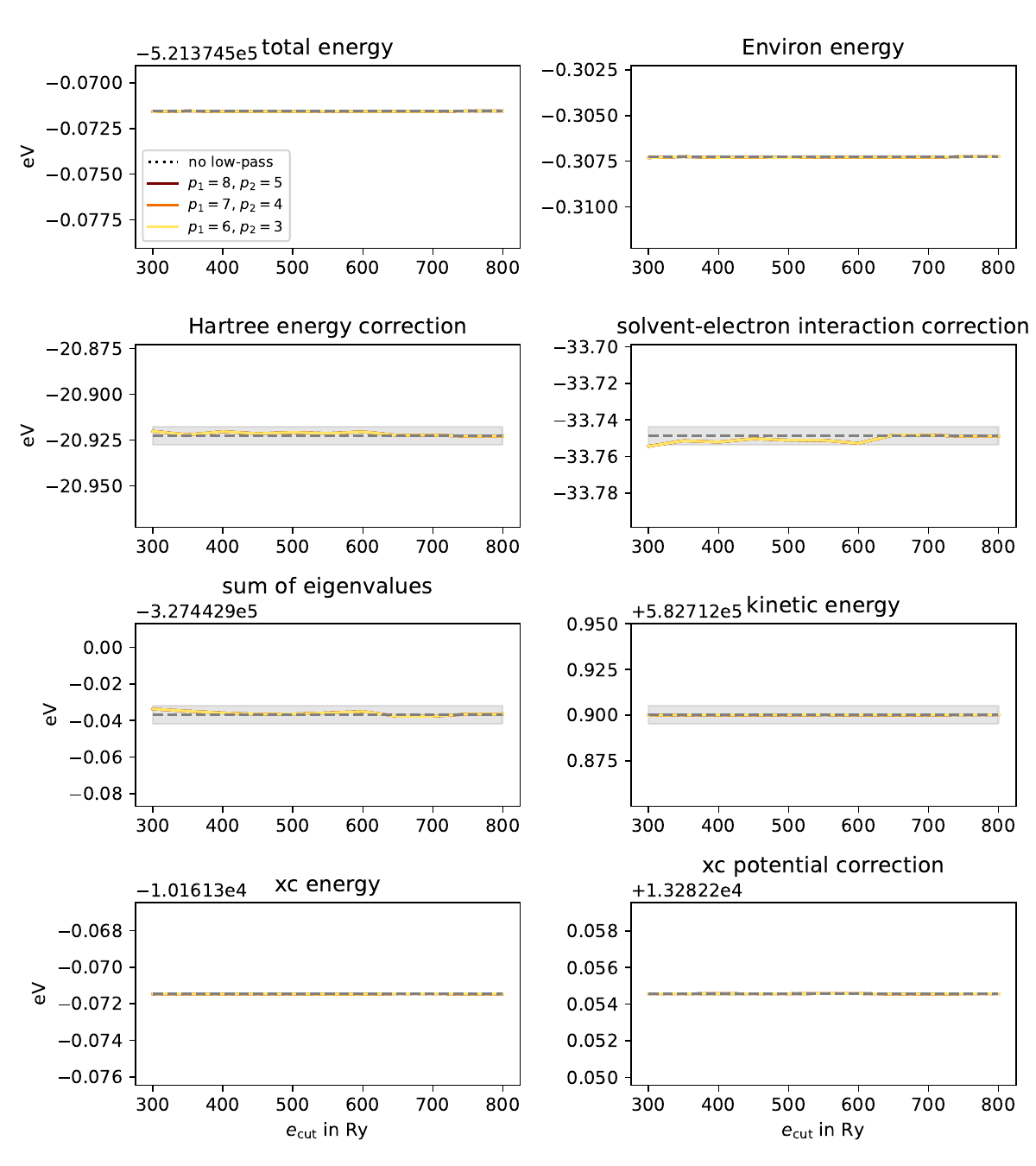}
    \caption{Like \cref{fig:E_ecut_sscs_water}, for a PtCO trimer in SSCS.}
    \label{fig:E_ecut_sscs_PtCO}
\end{figure}

\FloatBarrier

\newpage

\subsection{Force convergence with grid density, low-pass filter (SSCS)}\label{sec:conv_F_ecut_sscs}

\begin{figure}[!htb]
    \centering
    \includegraphics[width=0.9\linewidth]{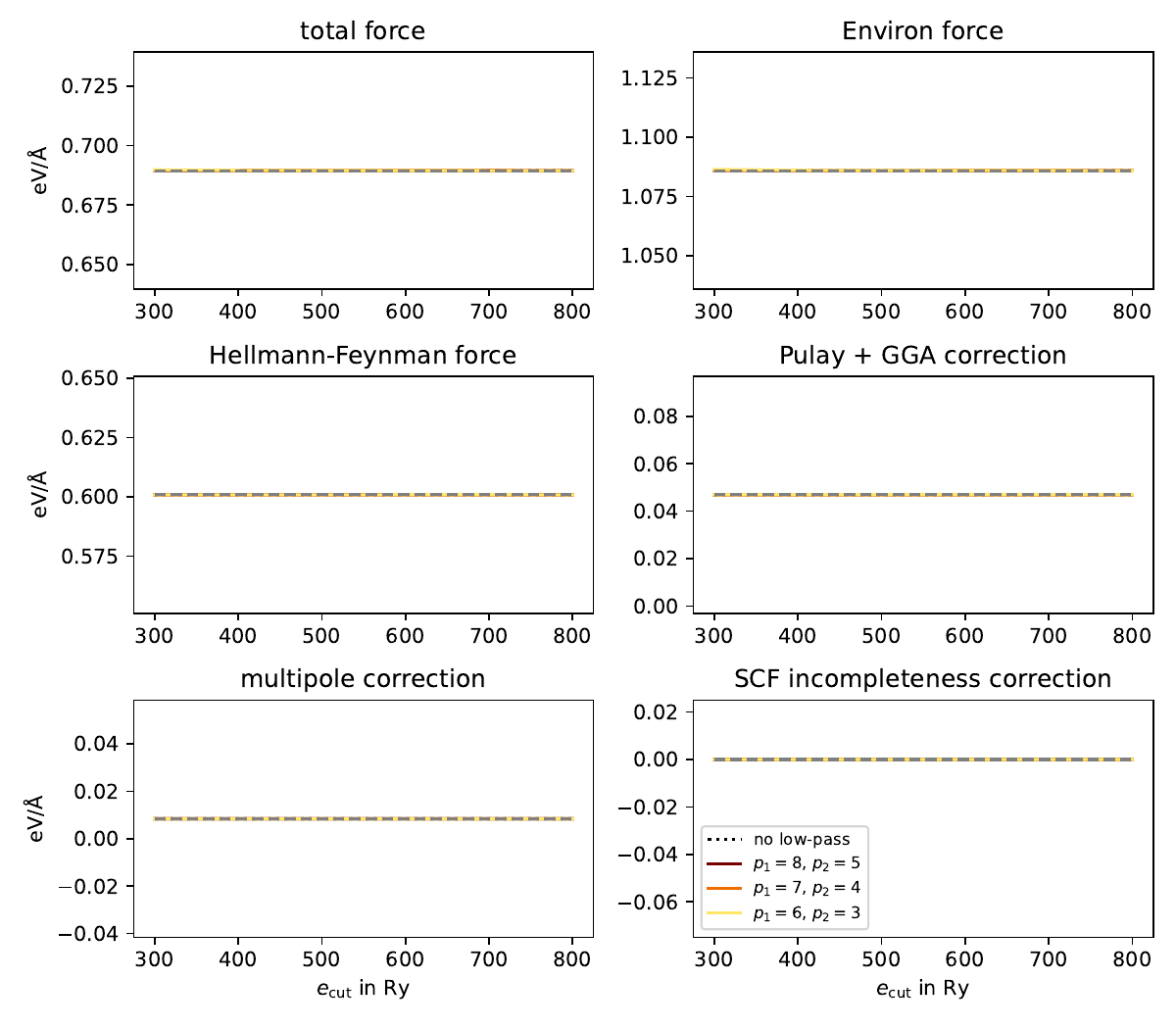}
    \caption{Convergence with $e_\text{cut}$ of total force (Euclidean norm) acting on one H atom (other H atom equivalent) and its contributions as reported in the FHI-aims\cite{blum2009} output file using different low-pass filter settings for a water molecule in SSCS. Dashed line shows mean value across $p_1=8$ and $p_1=7$ filters over the three highest $e_\text{cut}$.}
    \label{fig:F1_ecut_sscs_water}
\end{figure}

\begin{figure}[!htb]
    \centering
    \includegraphics[width=0.9\linewidth]{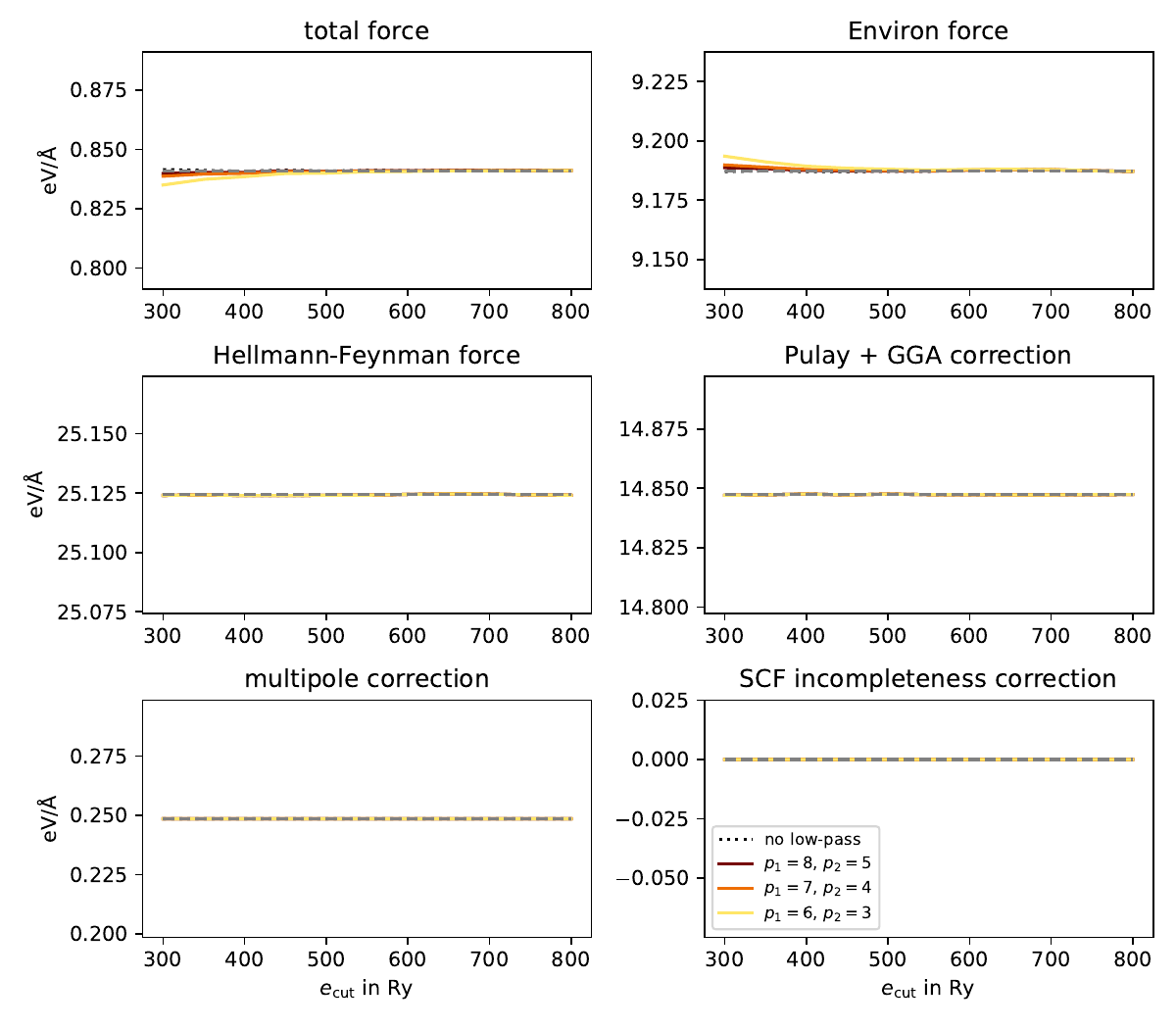}
    \caption{Like \cref{fig:F1_ecut_sscs_water}, for the O atom.}
    \label{fig:F2_ecut_sscs_water}
\end{figure}

\begin{figure}[!htb]
    \centering
    \includegraphics[width=0.9\linewidth]{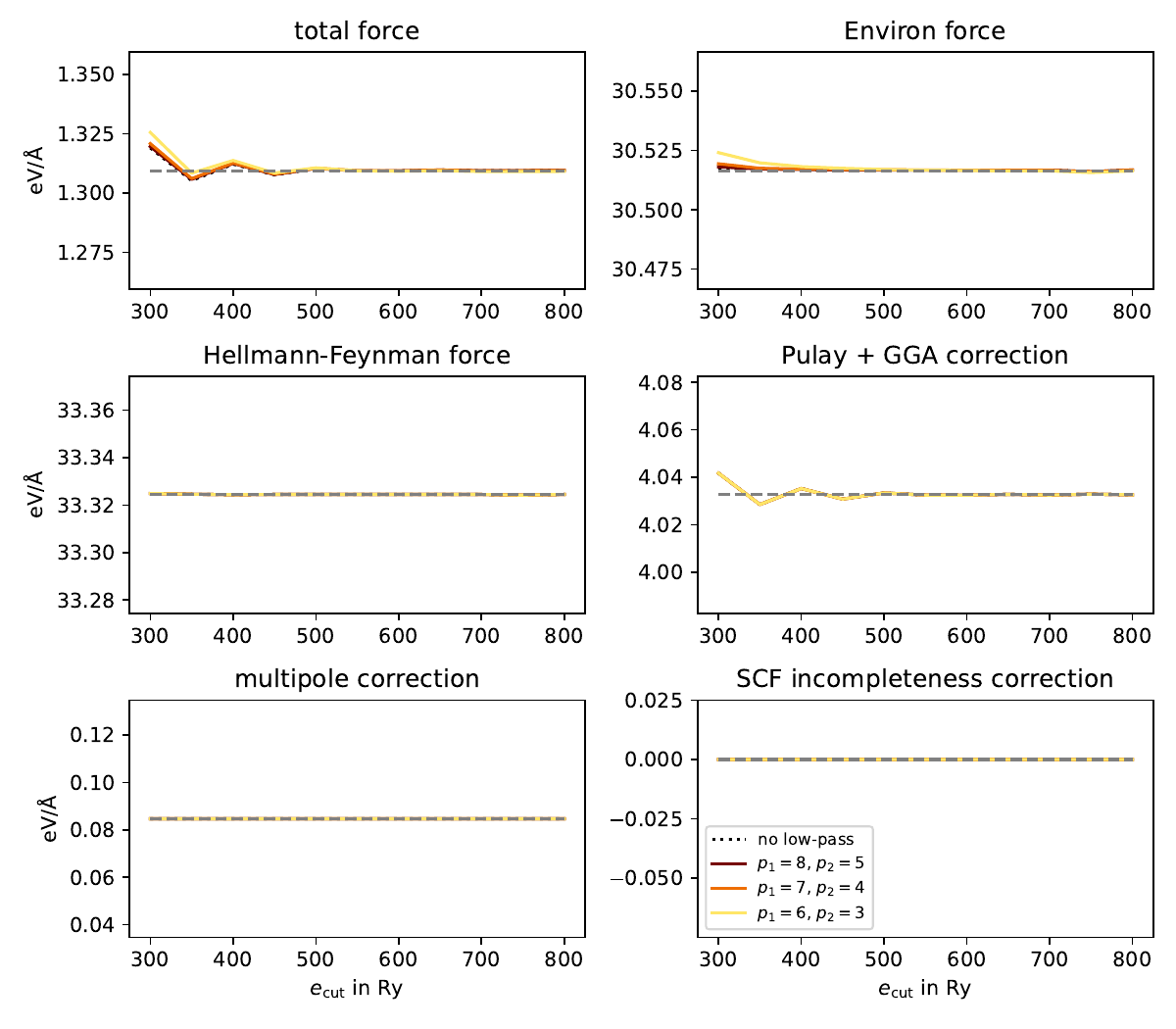}
    \caption{Like \cref{fig:F1_ecut_sscs_water}, for the Na atom in a NaF dimer.}
    \label{fig:F1_ecut_sscs_NaF}
\end{figure}

\begin{figure}[!htb]
    \centering
    \includegraphics[width=0.9\linewidth]{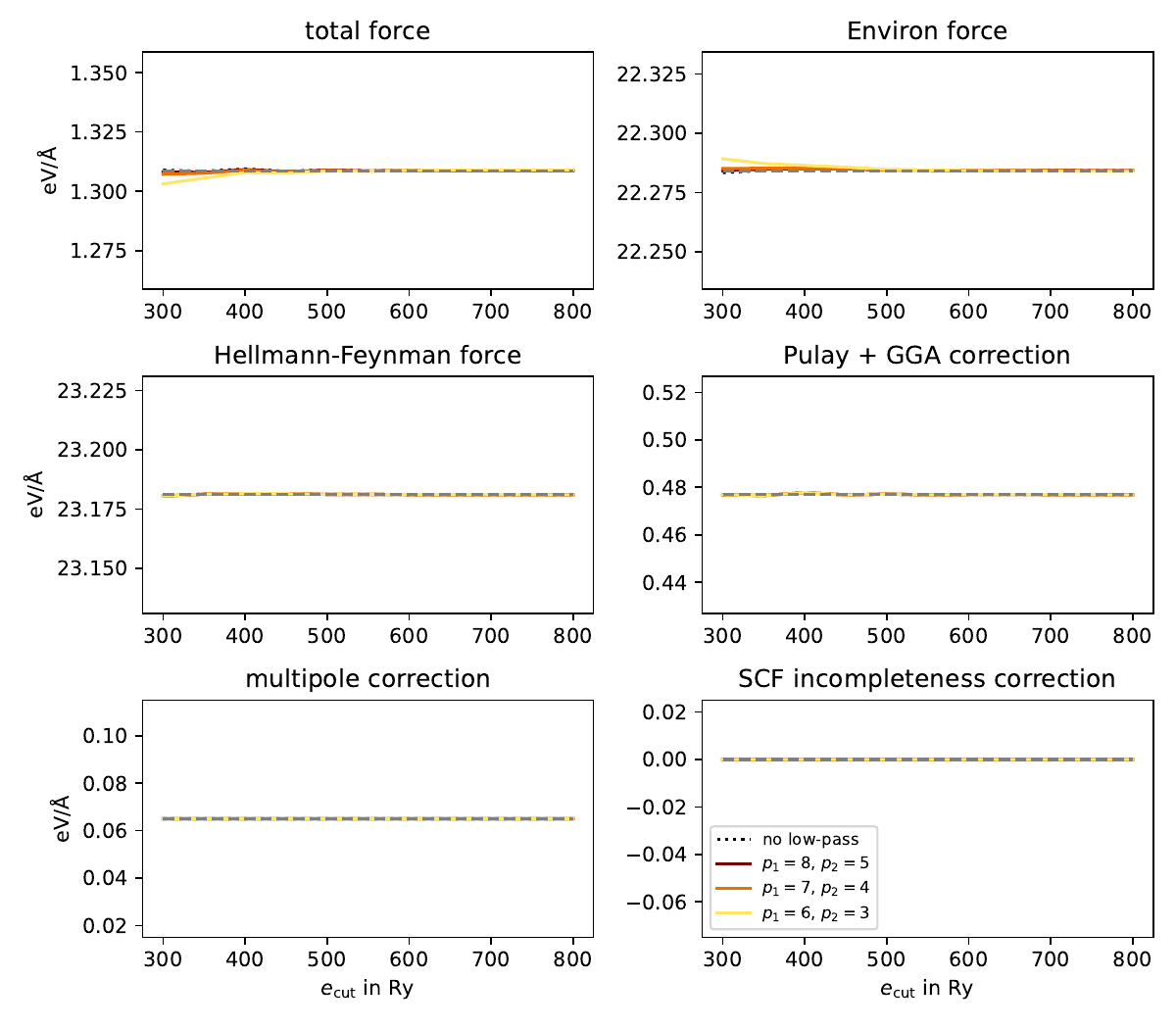}
    \caption{Like \cref{fig:F1_ecut_sscs_water}, for the F atom in a NaF dimer.}
    \label{fig:F2_ecut_sscs_NaF}
\end{figure}

\begin{figure}[!htb]
    \centering
    \includegraphics[width=0.9\linewidth]{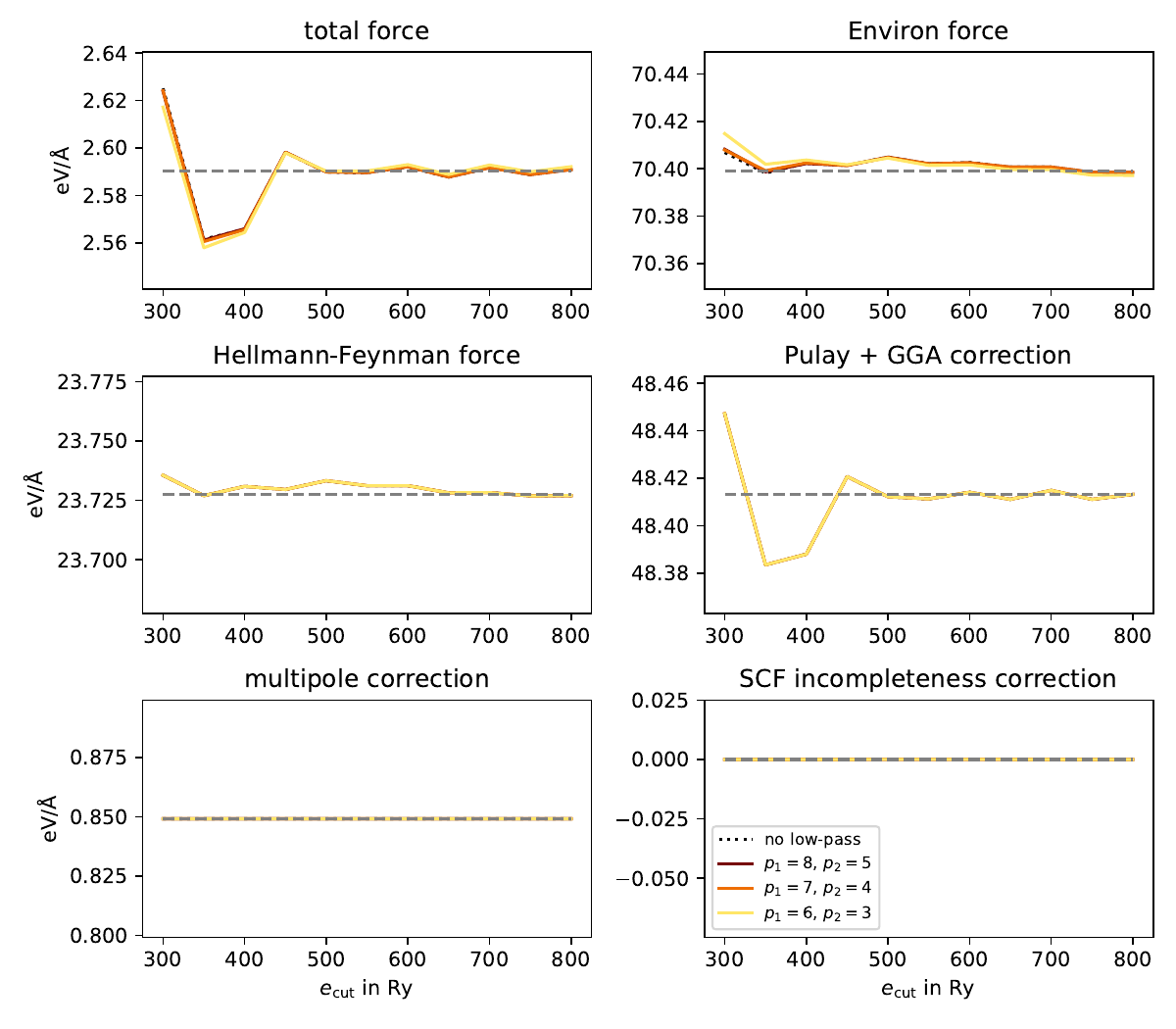}
    \caption{Like \cref{fig:F1_ecut_sscs_water}, for the Pt atom in a PtCO trimer.}
    \label{fig:F1_ecut_sscs_PtCO}
\end{figure}

\begin{figure}[!htb]
    \centering
    \includegraphics[width=0.9\linewidth]{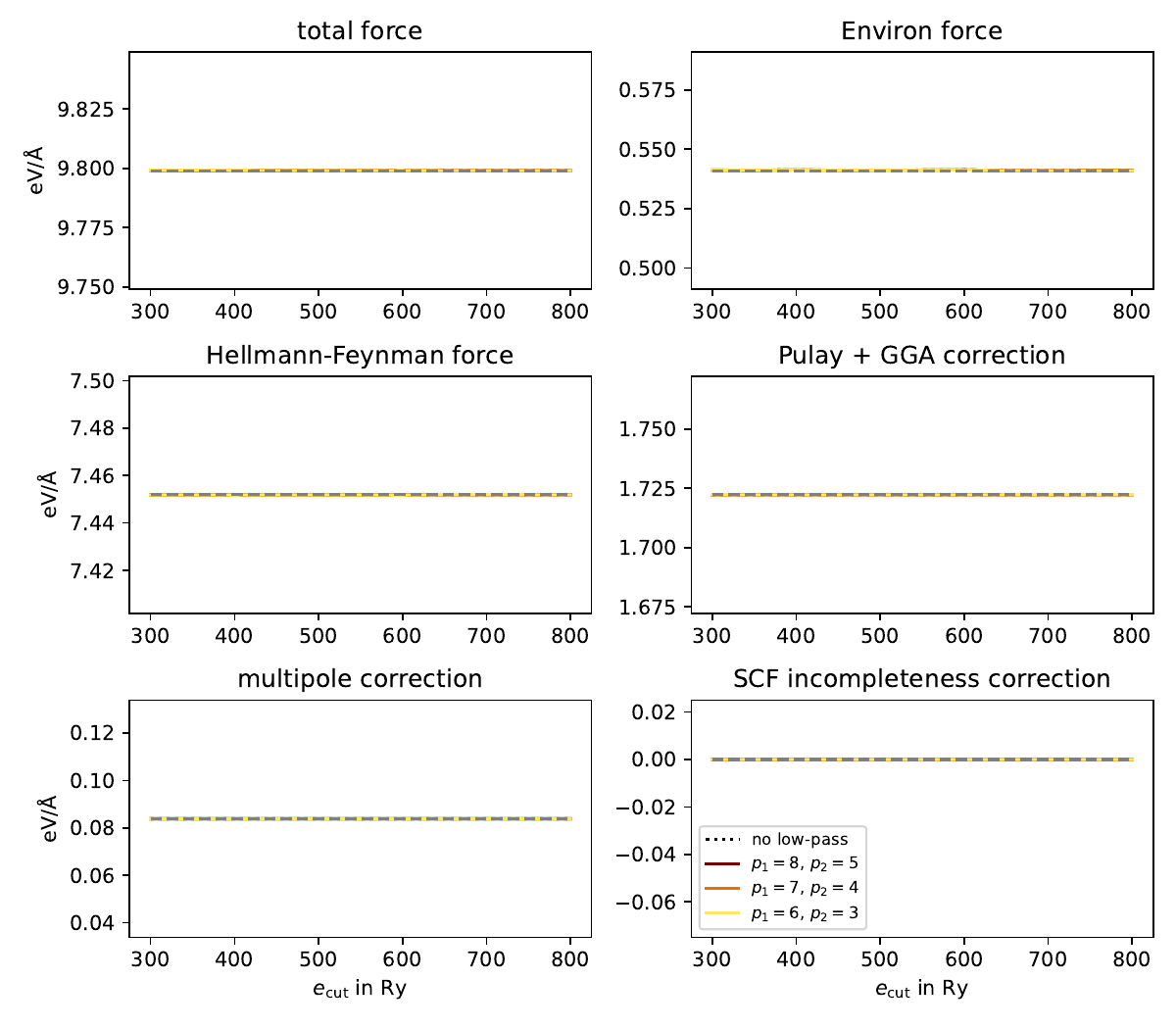}
    \caption{Like \cref{fig:F1_ecut_sscs_water}, for the C atom in a PtCO trimer.}
    \label{fig:F2_ecut_sscs_PtCO}
\end{figure}

\begin{figure}[!htb]
    \centering
    \includegraphics[width=0.9\linewidth]{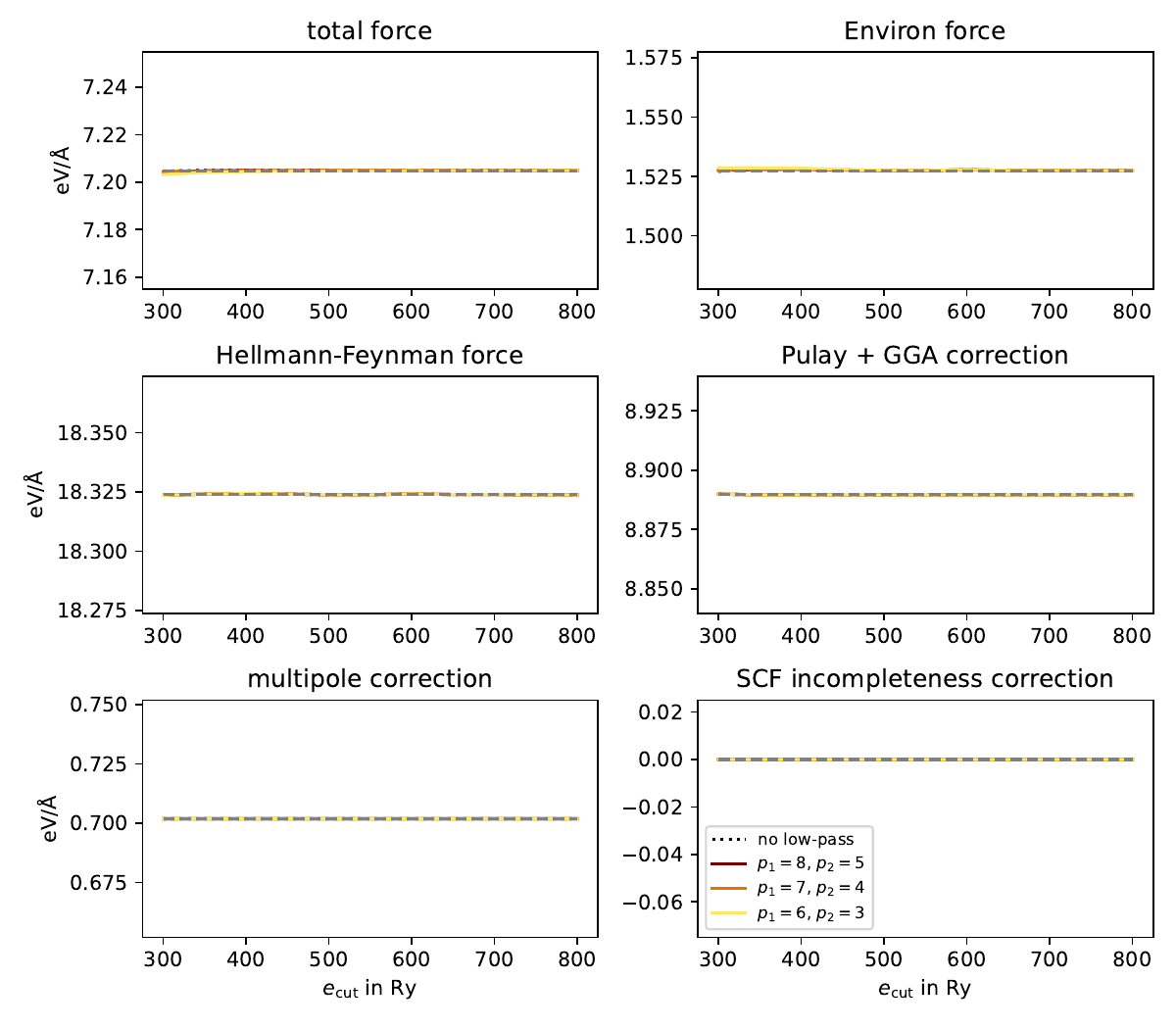}
    \caption{Like \cref{fig:F1_ecut_sscs_water}, for the O atom in a PtCO trimer.}
    \label{fig:F3_ecut_sscs_PtCO}
\end{figure}

\FloatBarrier

\newpage

\subsection{Energy convergence with grid density, low-pass filter (SCCS)}\label{sec:conv_E_ecut_sccs}

\begin{figure}[!htb]
    \centering
    \includegraphics[width=0.9\linewidth]{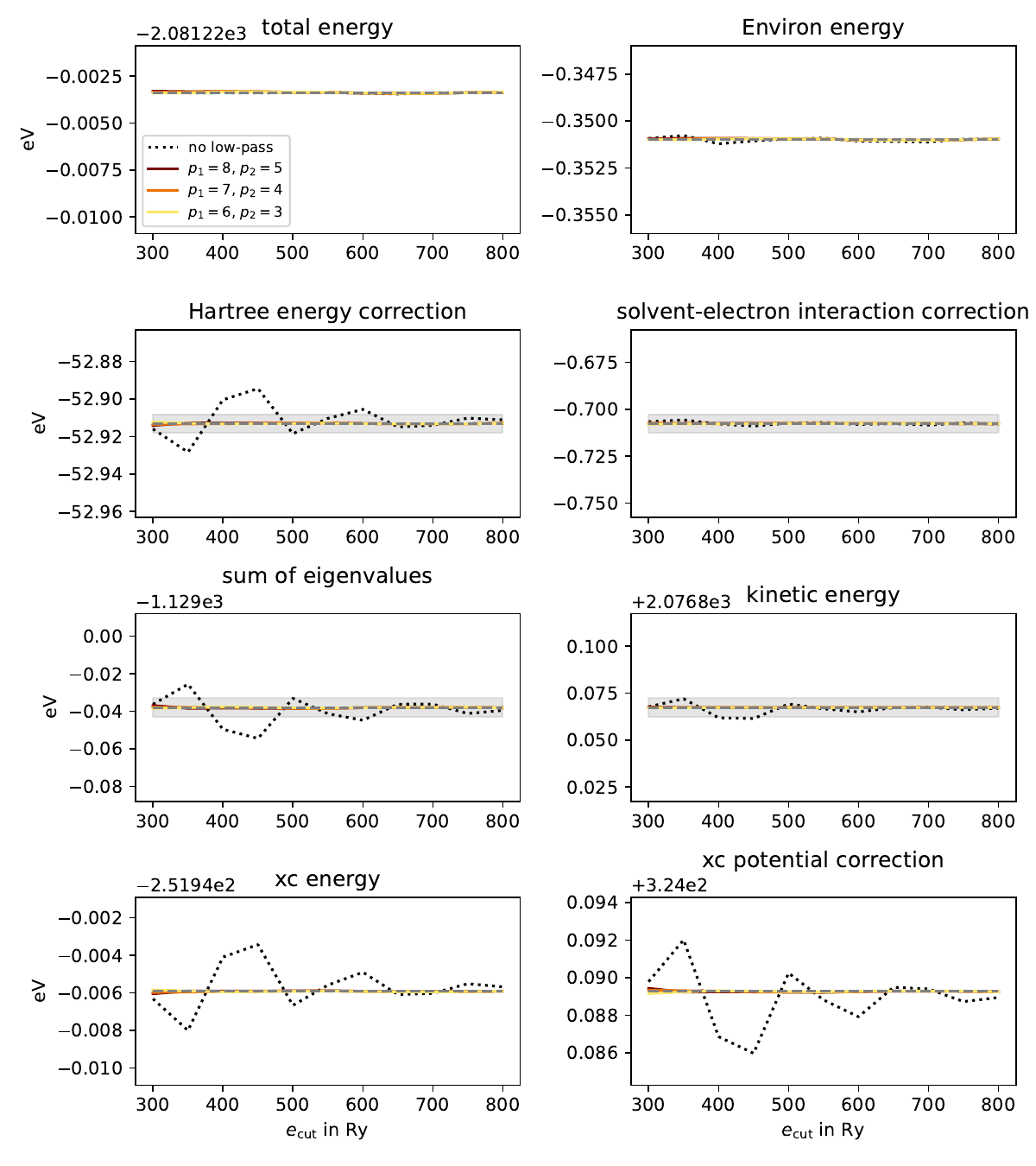}
    \caption{Convergence with $e_\text{cut}$ of total energy and its contributions as reported in the FHI-aims\cite{blum2009} output file using different low-pass filter settings for a water molecule in SCCS. Constant `free atom' contribution not shown. Dashed line shows mean value across $p_1=8$ and $p_1=7$ filters over the three highest $e_\text{cut}$. Note the different energy scales of the second and third row of subfigures. Shaded area indicates the energy scale covered in the other subfigures (10~meV)}
    \label{fig:E_ecut_sccs_water}
\end{figure}

\begin{figure}[p]
    \centering
    \includegraphics[width=0.9\linewidth]{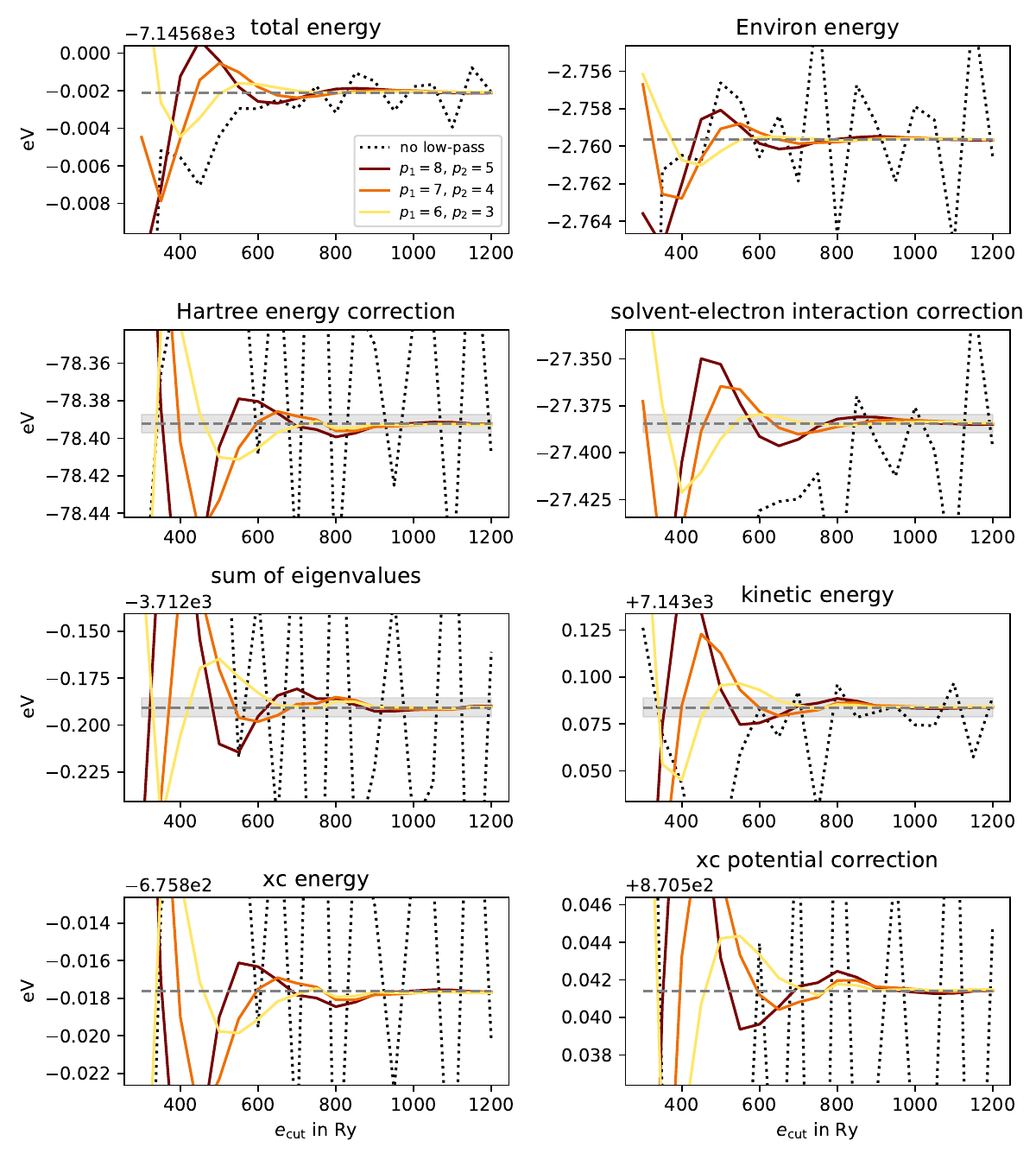}
    \caption{Like \cref{fig:E_ecut_sccs_water}, for a NaF dimer in SCCS.}
    \label{fig:E_ecut_sccs_NaF}
\end{figure}

\begin{figure}[p]
    \centering
    \includegraphics[width=0.9\linewidth]{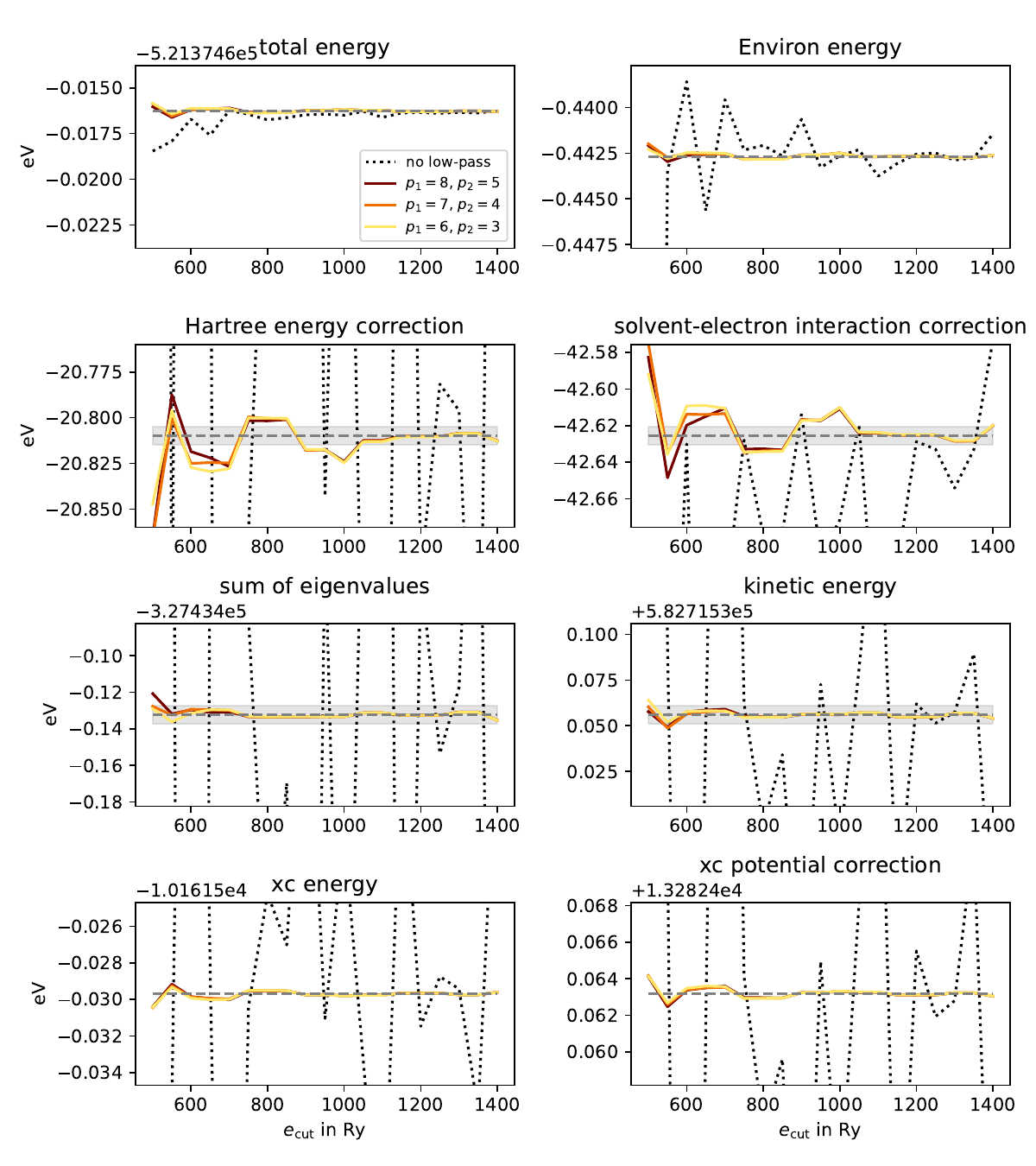}
    \caption{Like \cref{fig:E_ecut_sccs_water}, for a PtCO trimer in SCCS.}
    \label{fig:E_ecut_sccs_PtCO}
\end{figure}

\FloatBarrier

\newpage

\subsection{Force convergence with grid density, low-pass filter (SCCS)}\label{sec:conv_F_ecut_sccs}

\begin{figure}[!htb]
    \centering
    \includegraphics[width=0.9\linewidth]{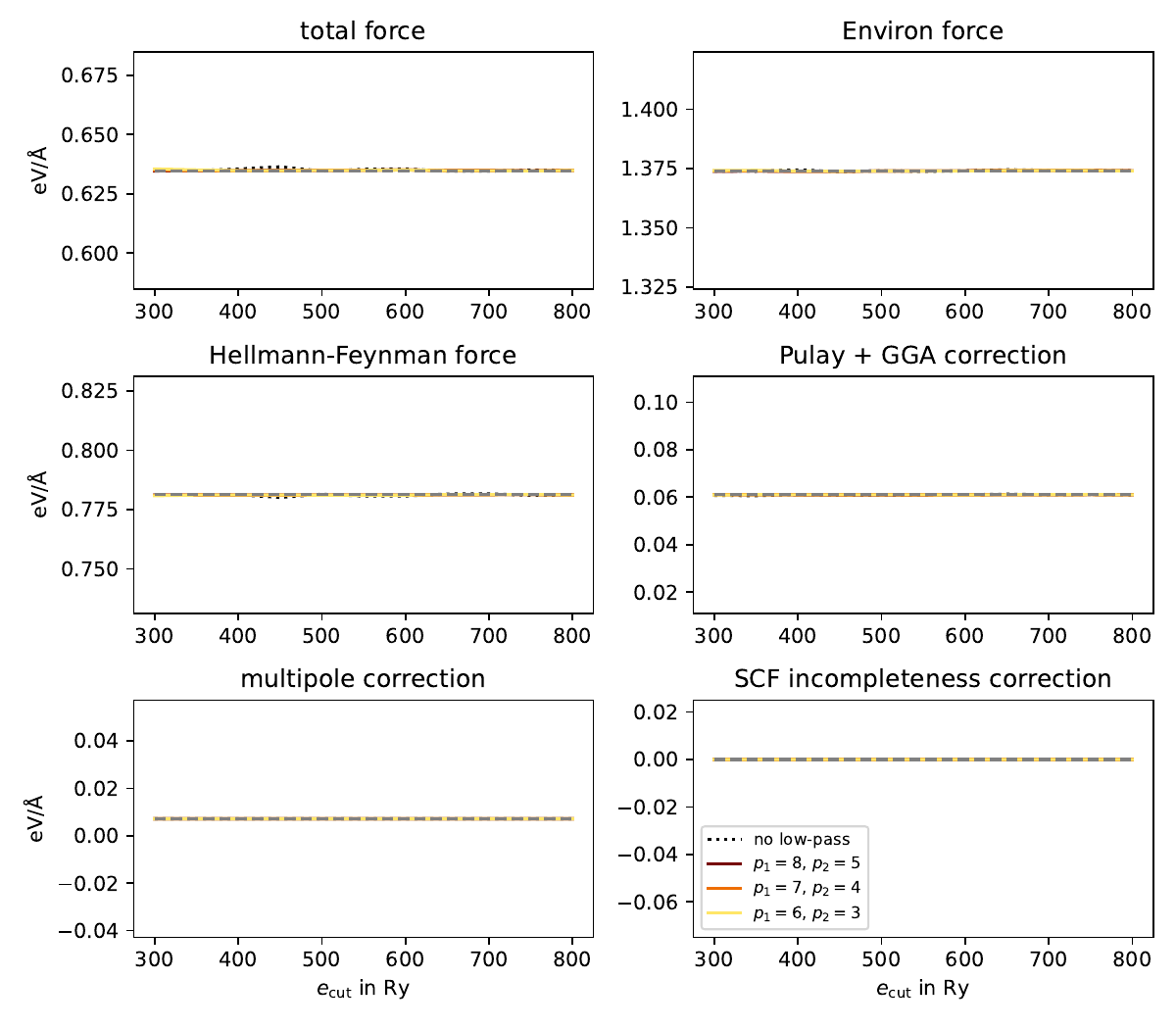}
    \caption{Convergence with $e_\text{cut}$ of total force (Euclidean norm) acting on one H atom (other H atom equivalent) and its contributions as reported in the FHI-aims\cite{blum2009} output file using different low-pass filter settings for a water molecule in SCCS. Dashed line shows mean value across $p_1=8$ and $p_1=7$ filters over the three highest $e_\text{cut}$.}
    \label{fig:F1_ecut_sccs_water}
\end{figure}

\begin{figure}[!htb]
    \centering
    \includegraphics[width=0.9\linewidth]{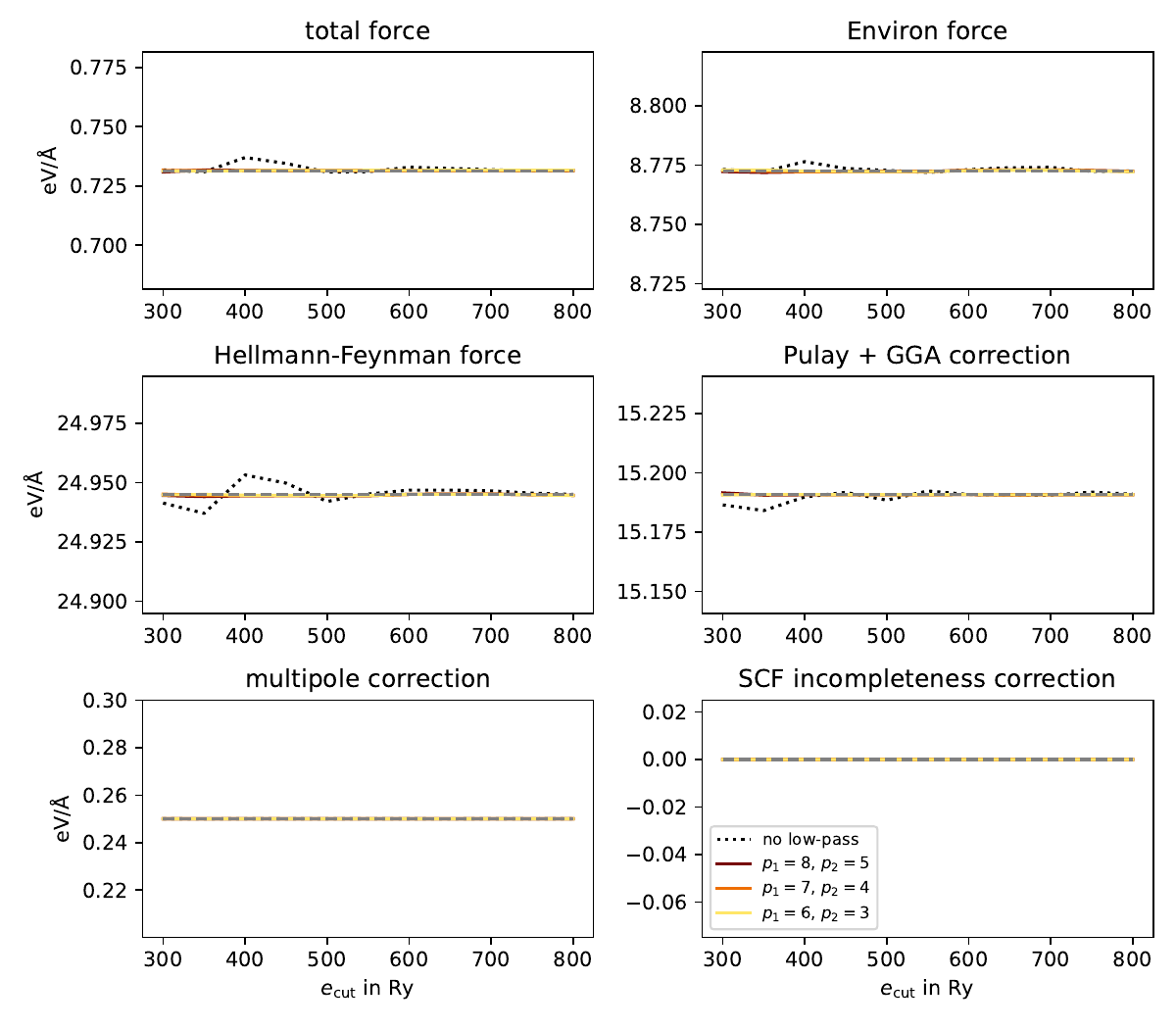}
    \caption{Like \cref{fig:F1_ecut_sccs_water}, for the O atom.}
    \label{fig:F2_ecut_sccs_water}
\end{figure}

\begin{figure}[!htb]
    \centering
    \includegraphics[width=0.9\linewidth]{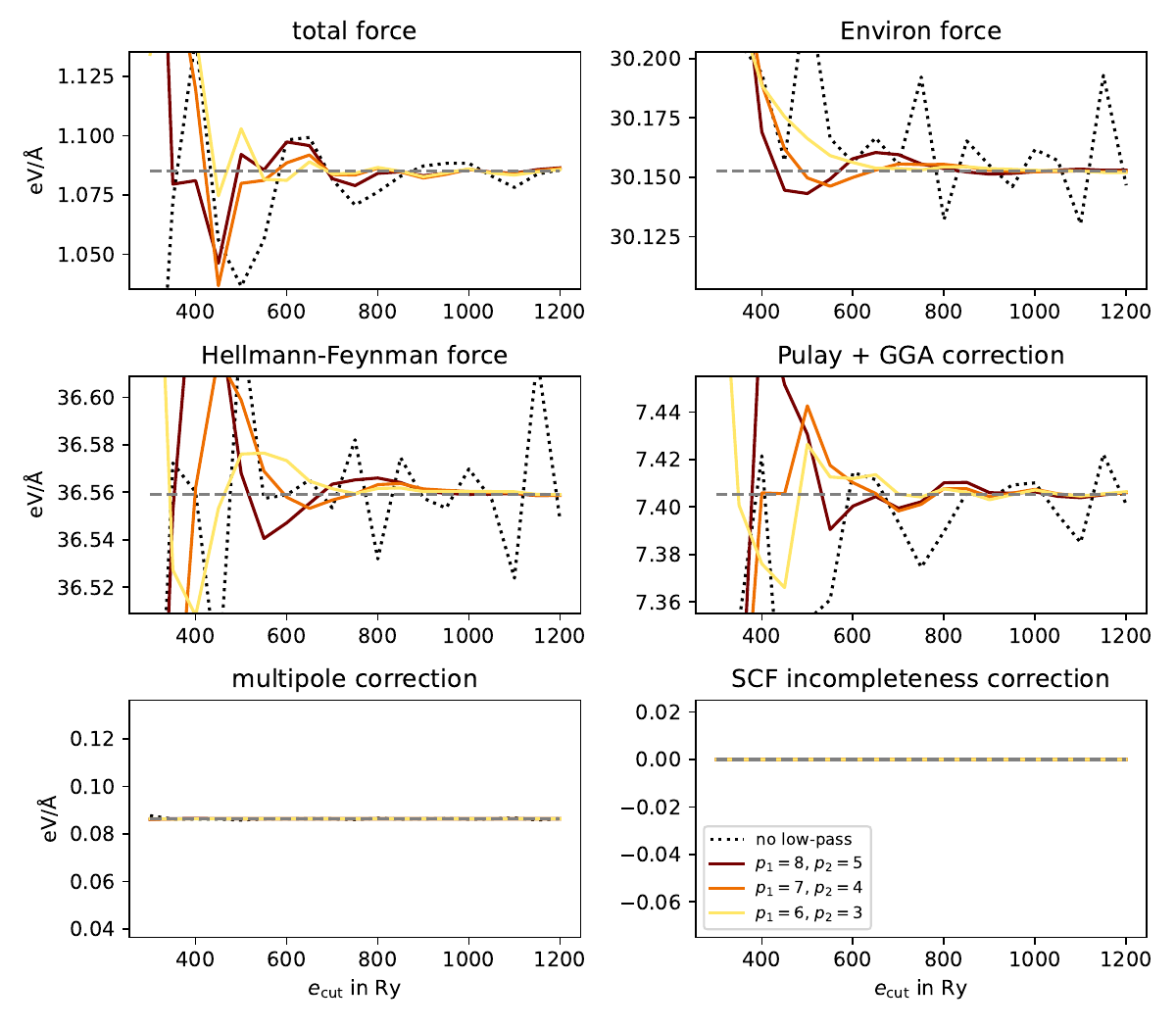}
    \caption{Like \cref{fig:F1_ecut_sccs_water}, for the Na atom in a NaF dimer.}
    \label{fig:F1_ecut_sccs_NaF}
\end{figure}

\begin{figure}[!htb]
    \centering
    \includegraphics[width=0.9\linewidth]{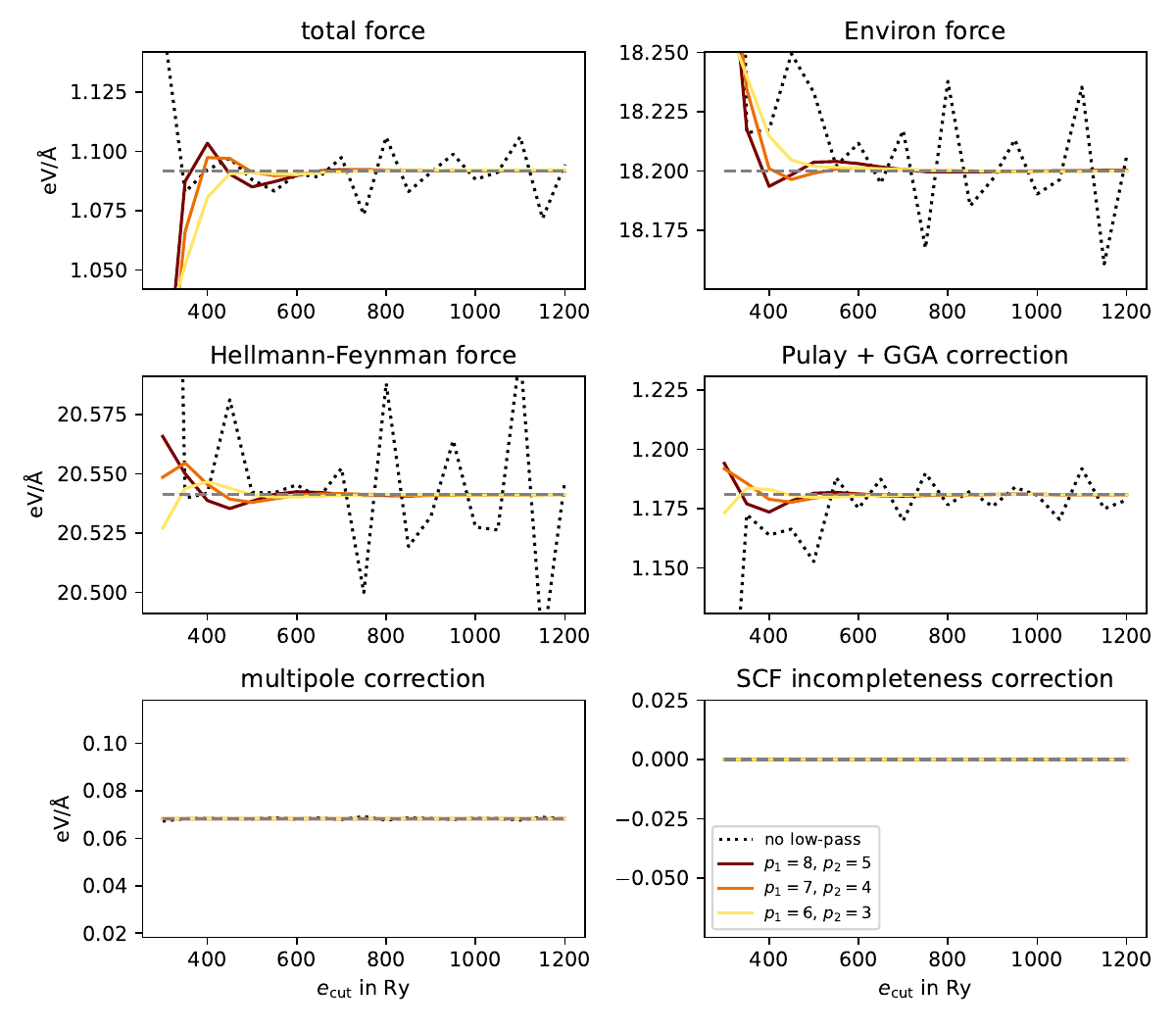}
    \caption{Like \cref{fig:F1_ecut_sccs_water}, for the F atom in a NaF dimer.}
    \label{fig:F2_ecut_sccs_NaF}
\end{figure}

\begin{figure}[!htb]
    \centering
    \includegraphics[width=0.9\linewidth]{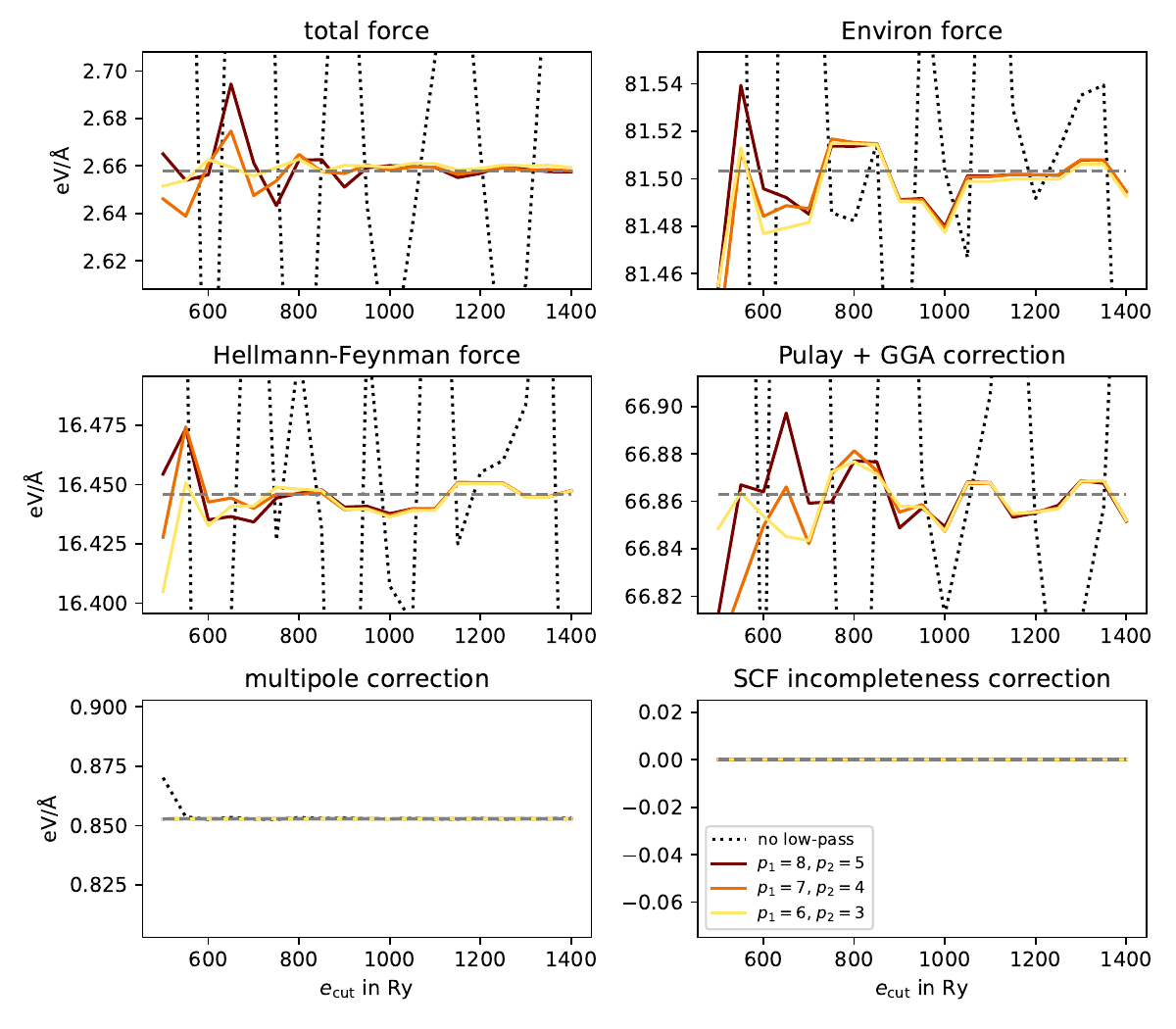}
    \caption{Like \cref{fig:F1_ecut_sccs_water}, for the Pt atom in a PtCO trimer.}
    \label{fig:F1_ecut_sccs_PtCO}
\end{figure}

\begin{figure}[!htb]
    \centering
    \includegraphics[width=0.9\linewidth]{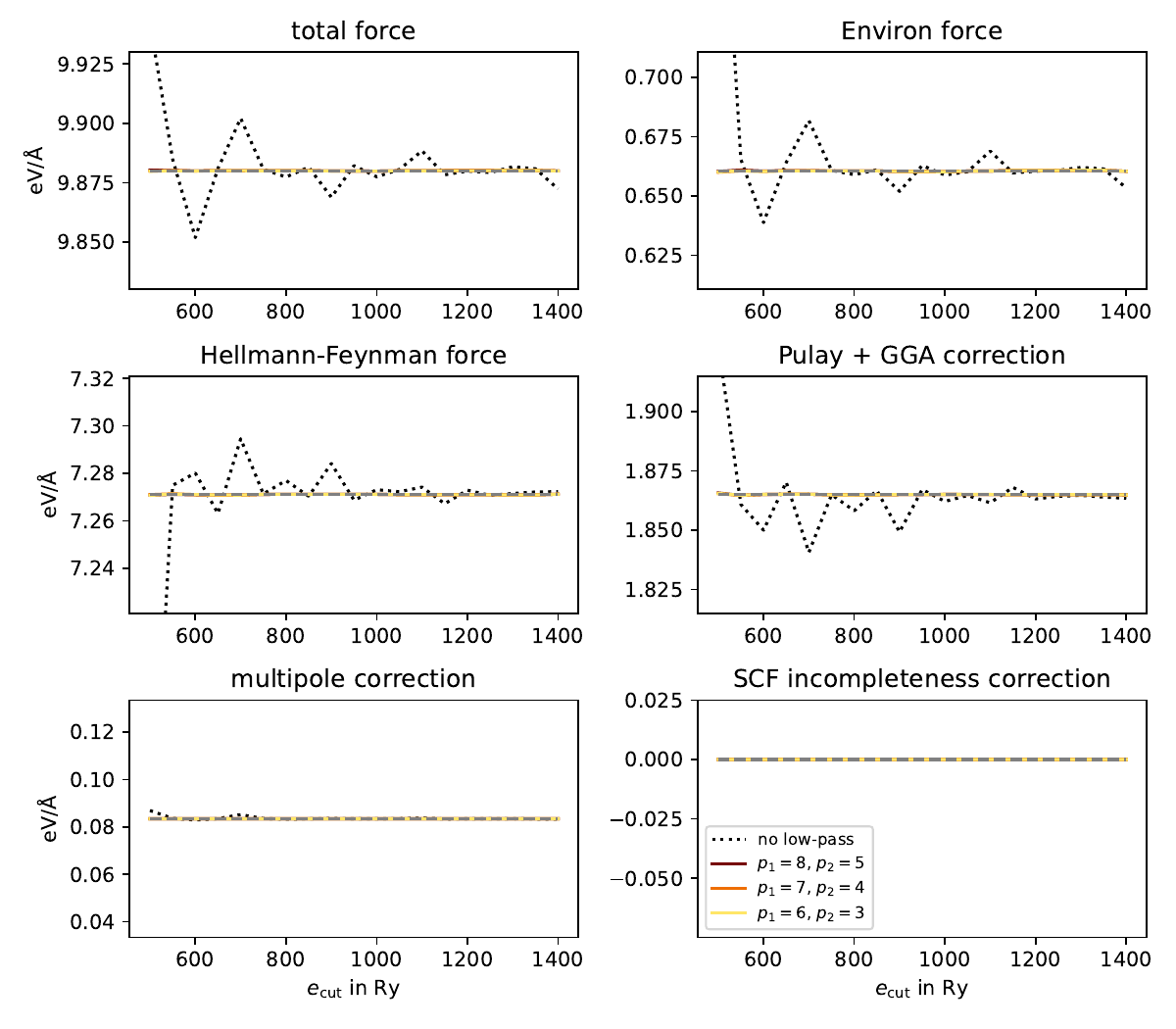}
    \caption{Like \cref{fig:F1_ecut_sccs_water}, for the C atom in a PtCO trimer.}
    \label{fig:F2_ecut_sccs_PtCO}
\end{figure}

\begin{figure}[!htb]
    \centering
    \includegraphics[width=0.9\linewidth]{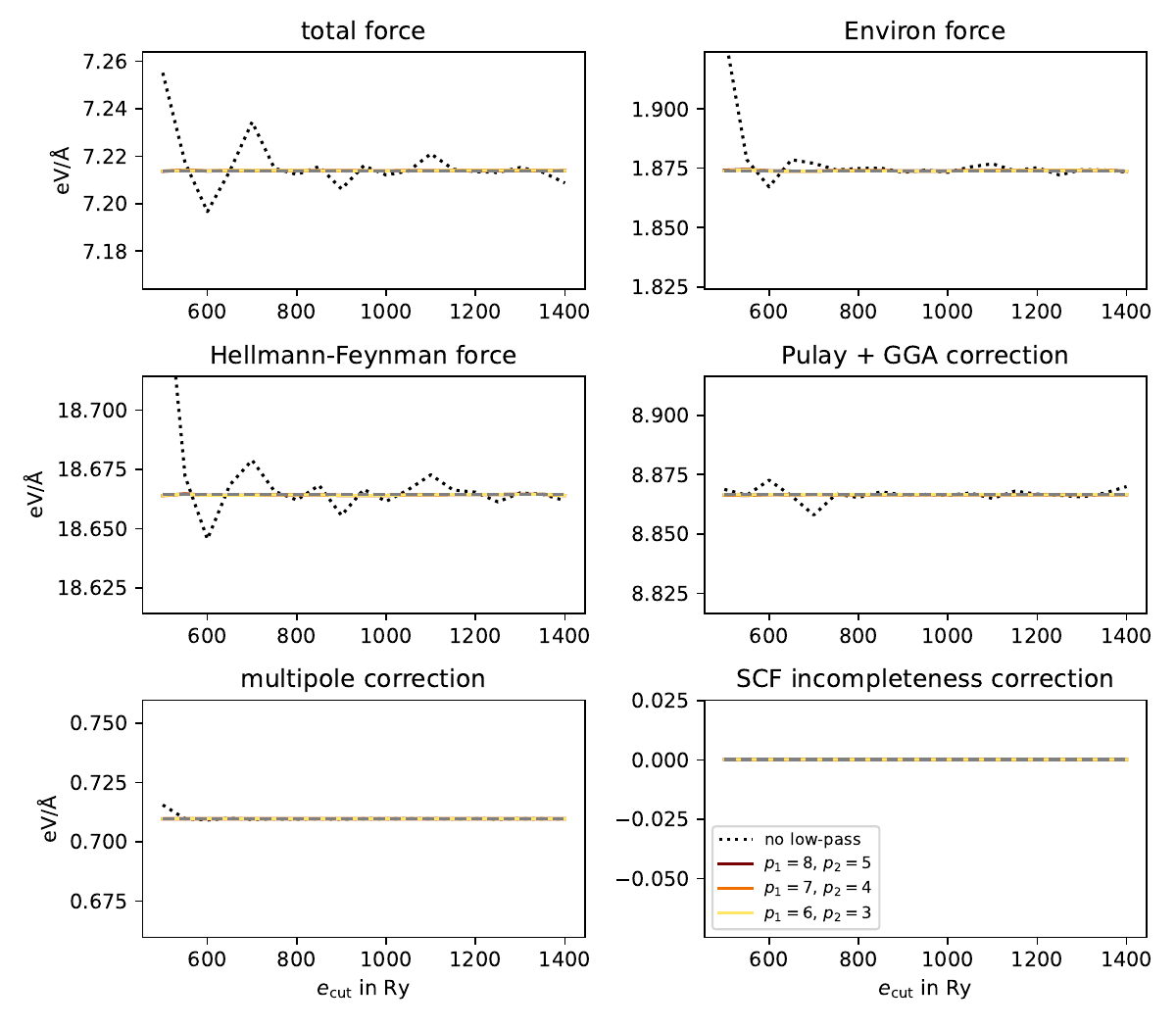}
    \caption{Like \cref{fig:F1_ecut_sccs_water}, for the O atom in a PtCO trimer.}
    \label{fig:F3_ecut_sccs_PtCO}
\end{figure}

\FloatBarrier

\newpage

\subsection{Energy convergence with multipole order (SSCS)}\label{sec:conv_E_lmax_sscs}

\begin{figure}[!htb]
    \centering
    \includegraphics[width=0.9\linewidth]{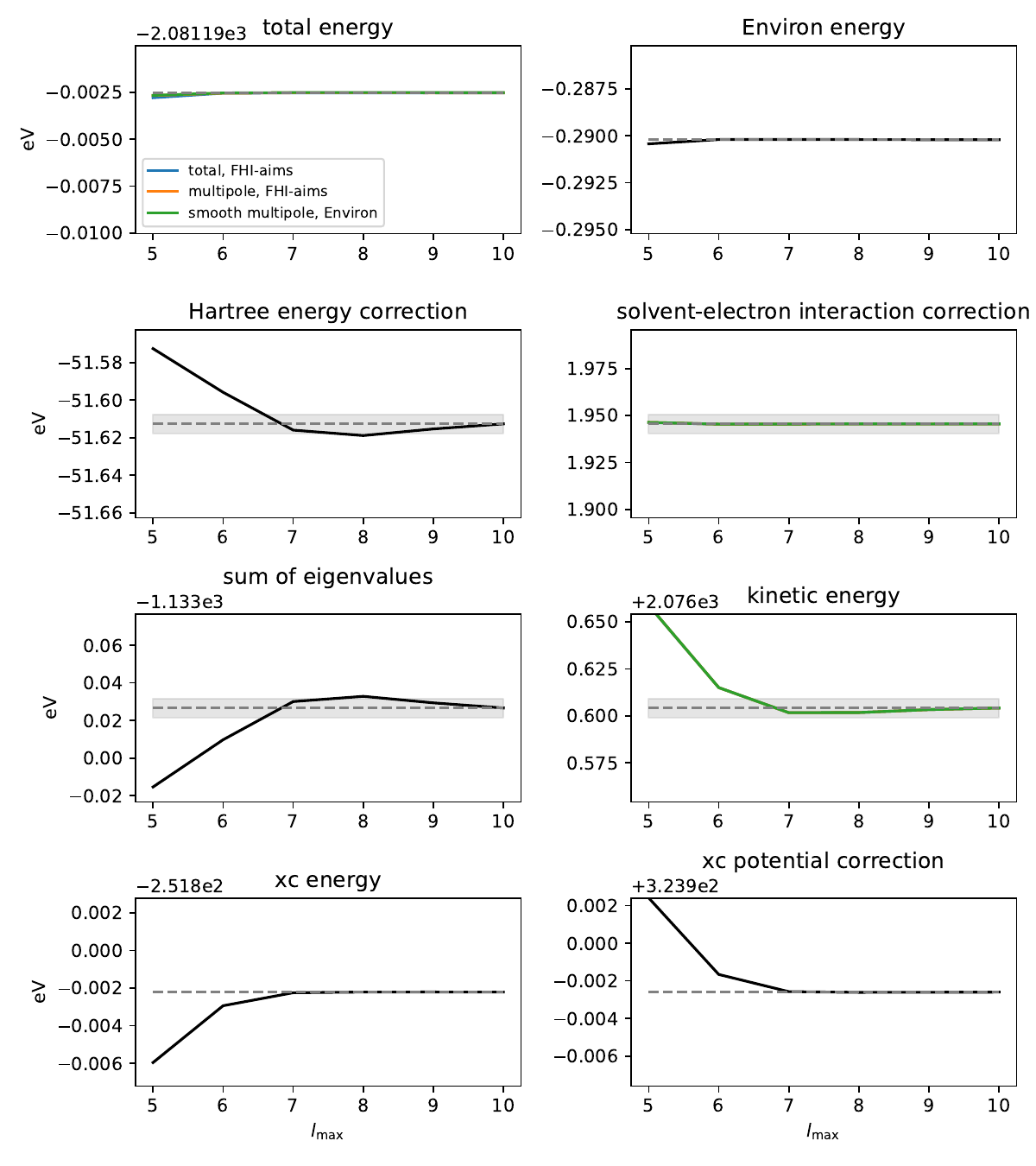}
    \caption{Convergence with $l_\text{max}$ of total energy and its contributions as reported in the FHI-aims\cite{blum2009} output file using different ways of computing the solvent-electron double counting correction for a water molecule in SSCS. Constant `free atom' contribution not shown. Note the different energy scales of the second and third row of subfigures. Shaded area indicates the energy scale covered in the other subfigures (10~meV)}
    \label{fig:E_lmax_sscs_water}
\end{figure}

\begin{figure}[p]
    \centering
    \includegraphics[width=0.9\linewidth]{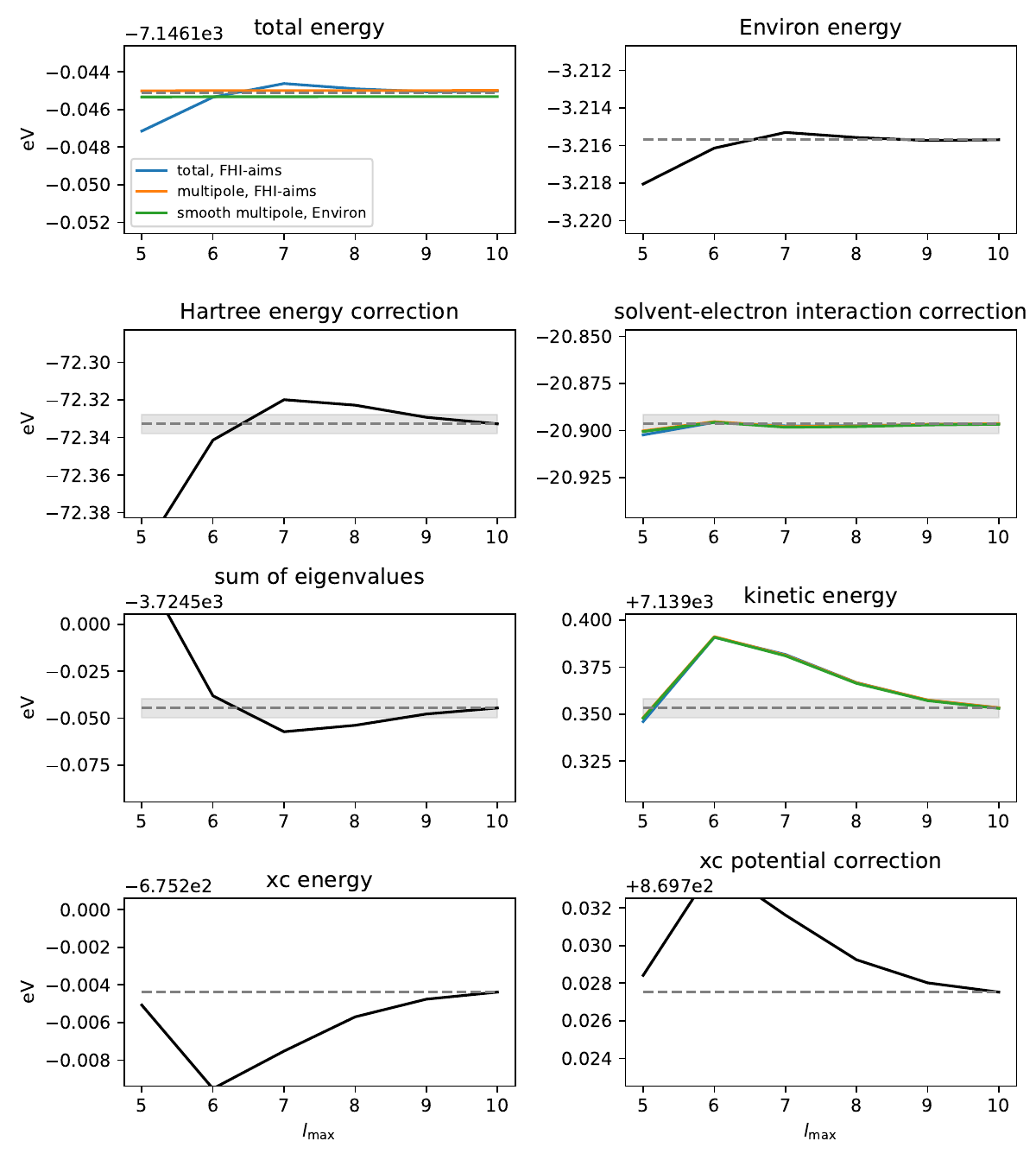}
    \caption{Like \cref{fig:E_lmax_sscs_water}, for a NaF dimer in SSCS.}
    \label{fig:E_lmax_sscs_NaF}
\end{figure}

\begin{figure}[p]
    \centering
    \includegraphics[width=0.9\linewidth]{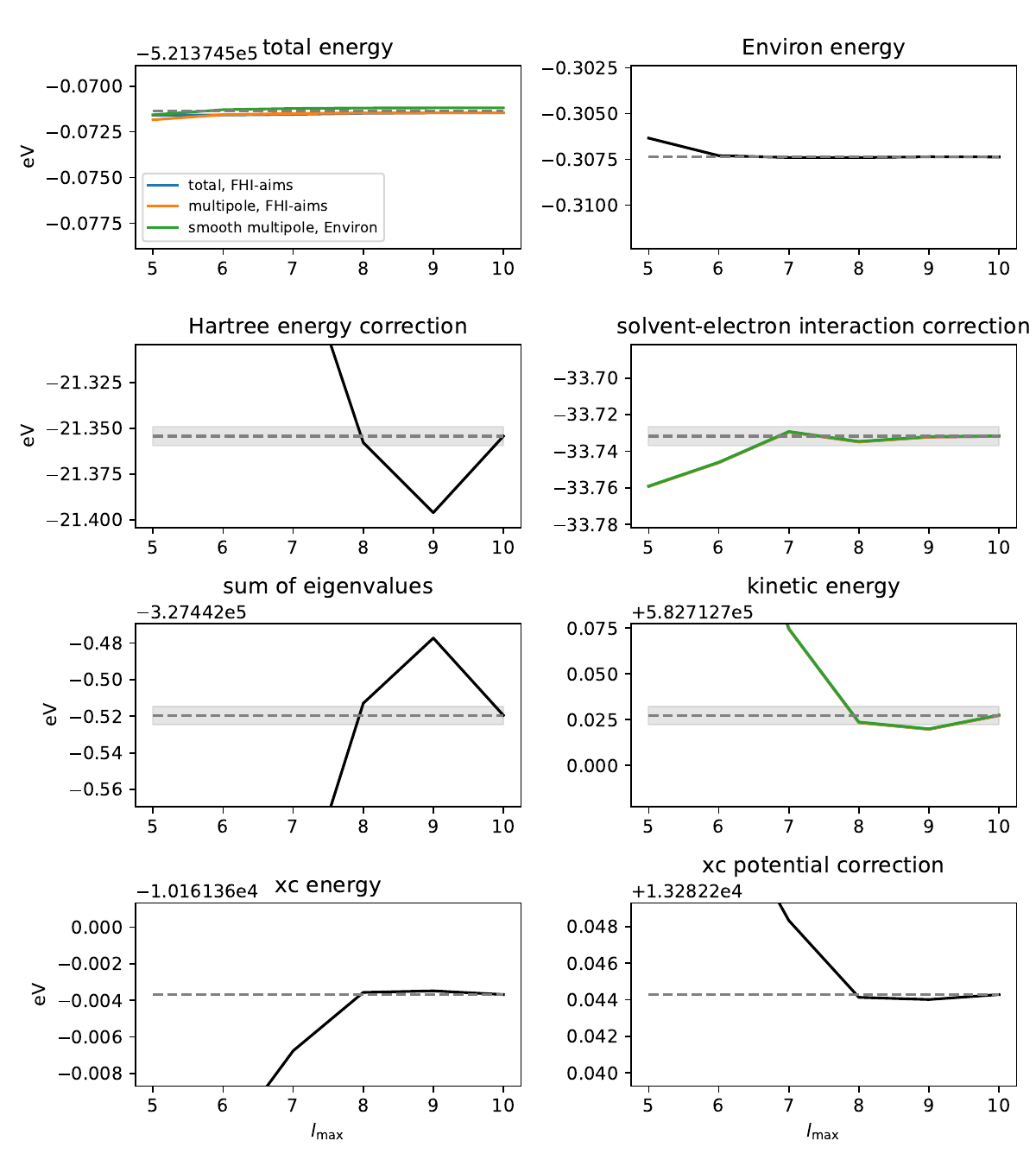}
    \caption{Like \cref{fig:E_lmax_sscs_water}, for a PtCO trimer in SSCS.}
    \label{fig:E_lmax_sscs_PtCO}
\end{figure}

\FloatBarrier

\newpage

\subsection{Force convergence with multipole order (SSCS)}\label{sec:conv_F_lmax_sscs}

\begin{figure}[!htb]
    \centering
    \includegraphics[width=0.9\linewidth]{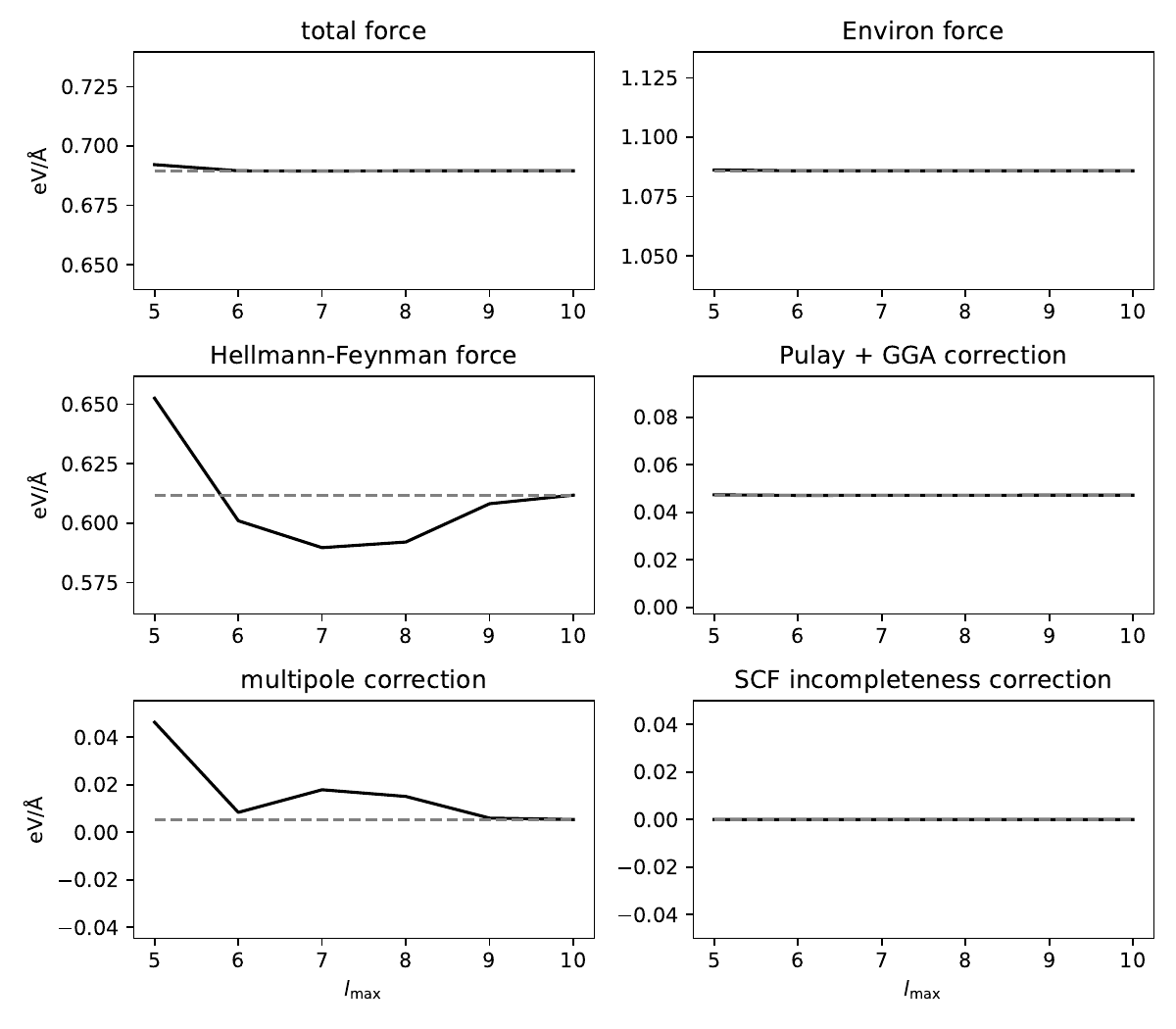}
    \caption{Convergence with $l_\text{max}$ of total force (Euclidean norm) acting on one H atom (other H atom equivalent) and its contributions as reported in the FHI-aims\cite{blum2009} output file using different low-pass filter settings for a water molecule in SSCS.}
    \label{fig:F1_lmax_sscs_water}
\end{figure}

\begin{figure}[!htb]
    \centering
    \includegraphics[width=0.9\linewidth]{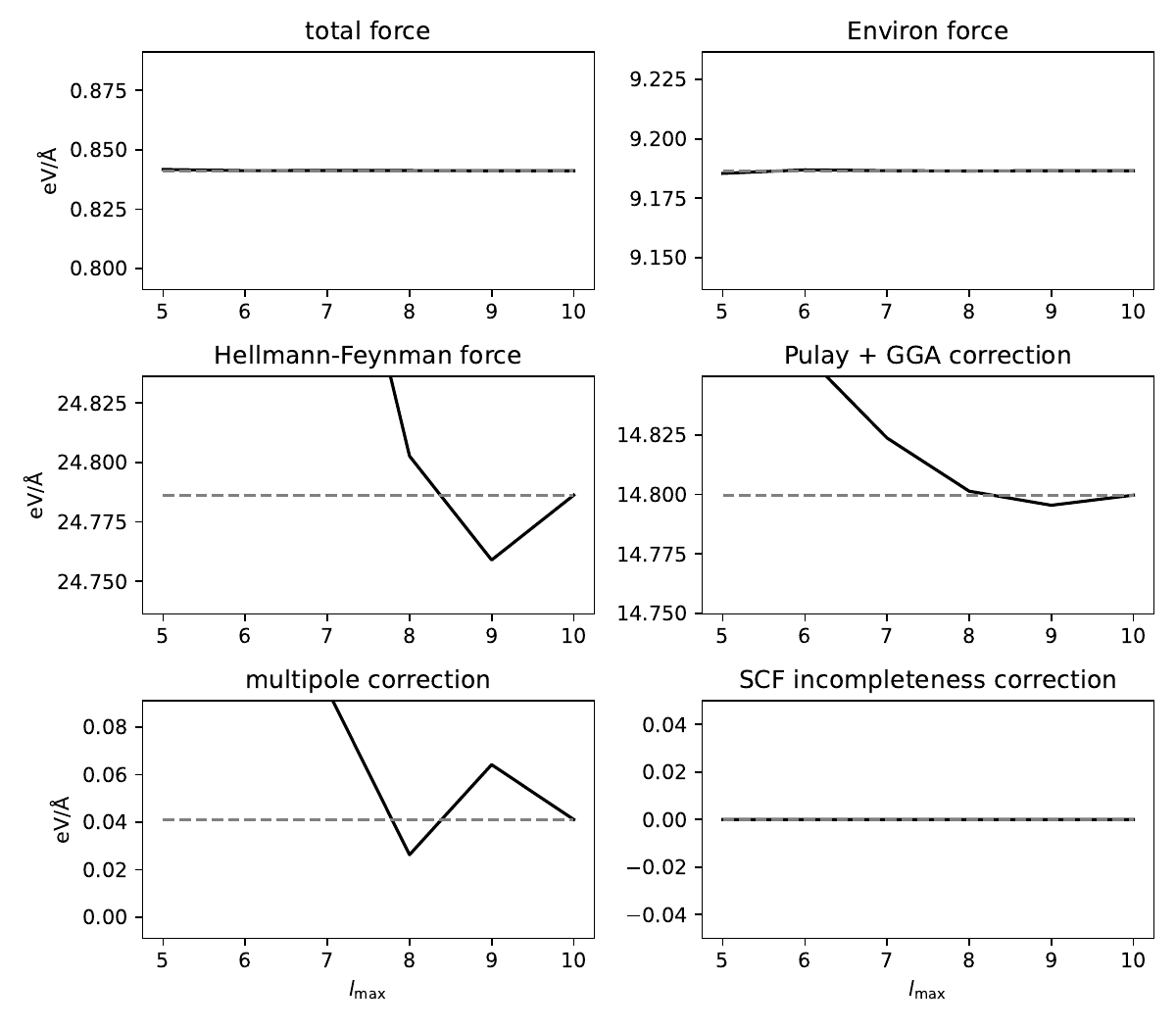}
    \caption{Like \cref{fig:F1_lmax_sscs_water}, for the O atom.}
    \label{fig:F2_lmax_sscs_water}
\end{figure}

\begin{figure}[!htb]
    \centering
    \includegraphics[width=0.9\linewidth]{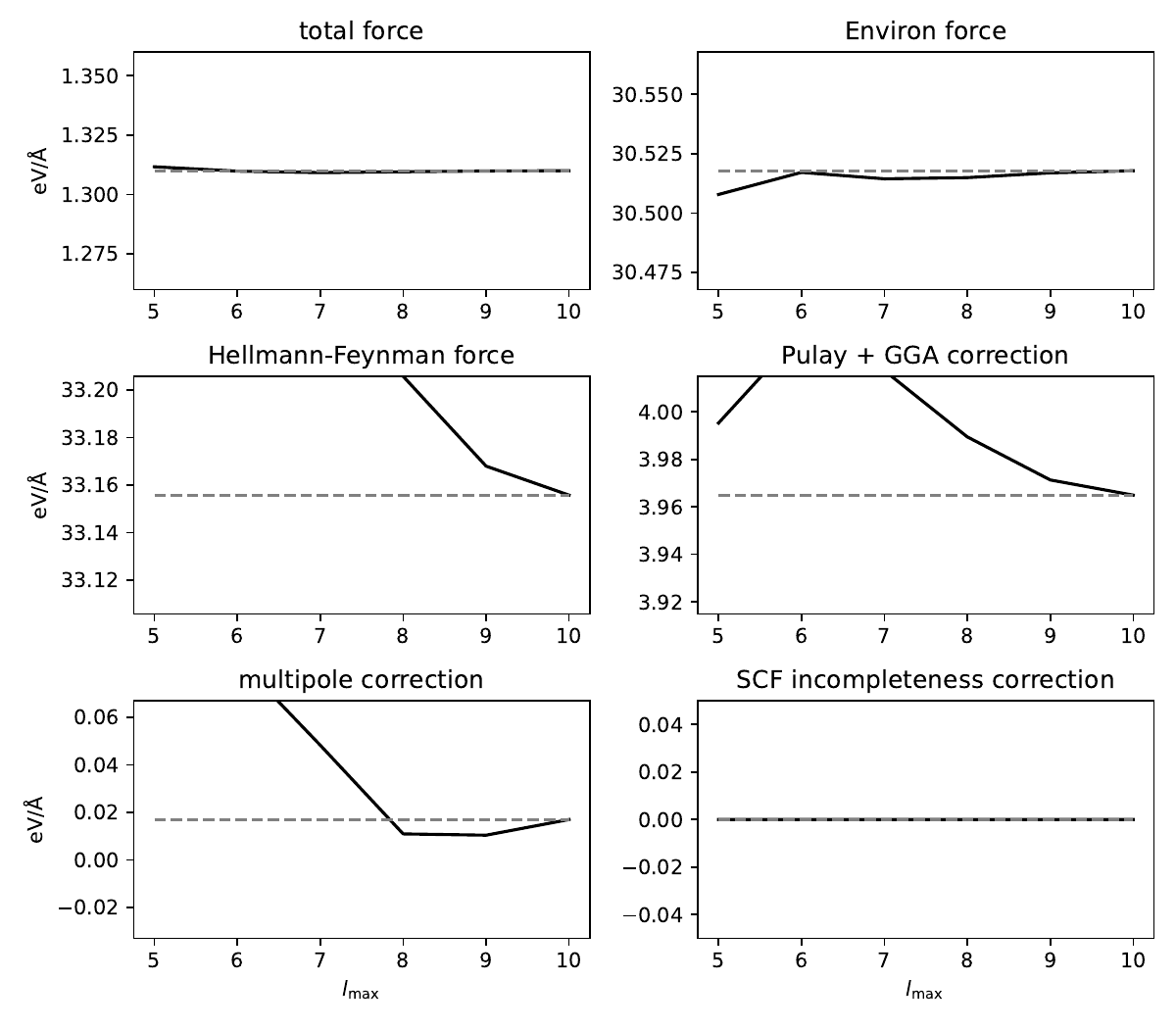}
    \caption{Like \cref{fig:F1_lmax_sscs_water}, for the Na atom in a NaF dimer.}
    \label{fig:F1_lmax_sscs_NaF}
\end{figure}

\begin{figure}[!htb]
    \centering
    \includegraphics[width=0.9\linewidth]{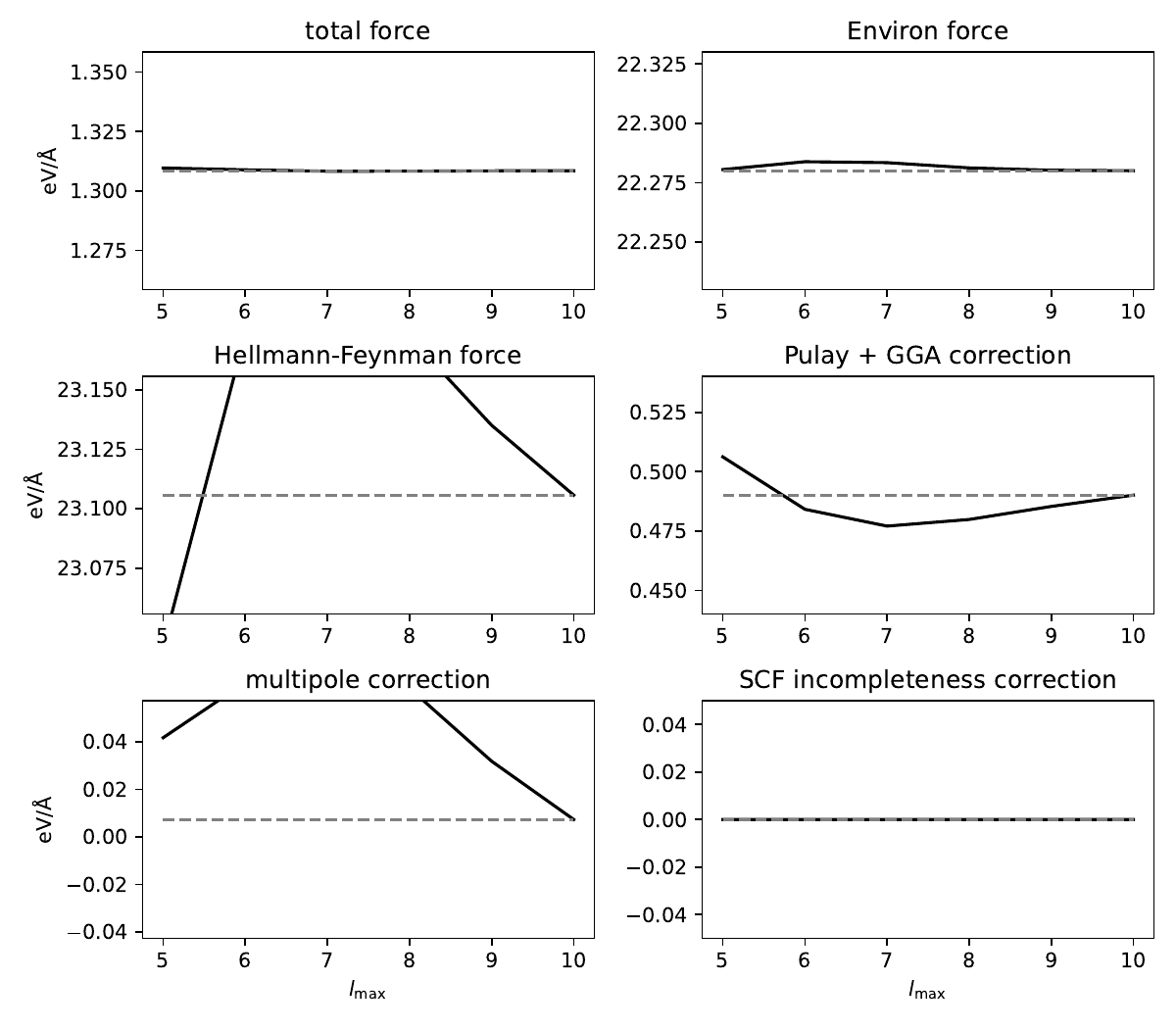}
    \caption{Like \cref{fig:F1_lmax_sscs_water}, for the F atom in a NaF dimer.}
    \label{fig:F2_lmax_sscs_NaF}
\end{figure}

\begin{figure}[!htb]
    \centering
    \includegraphics[width=0.9\linewidth]{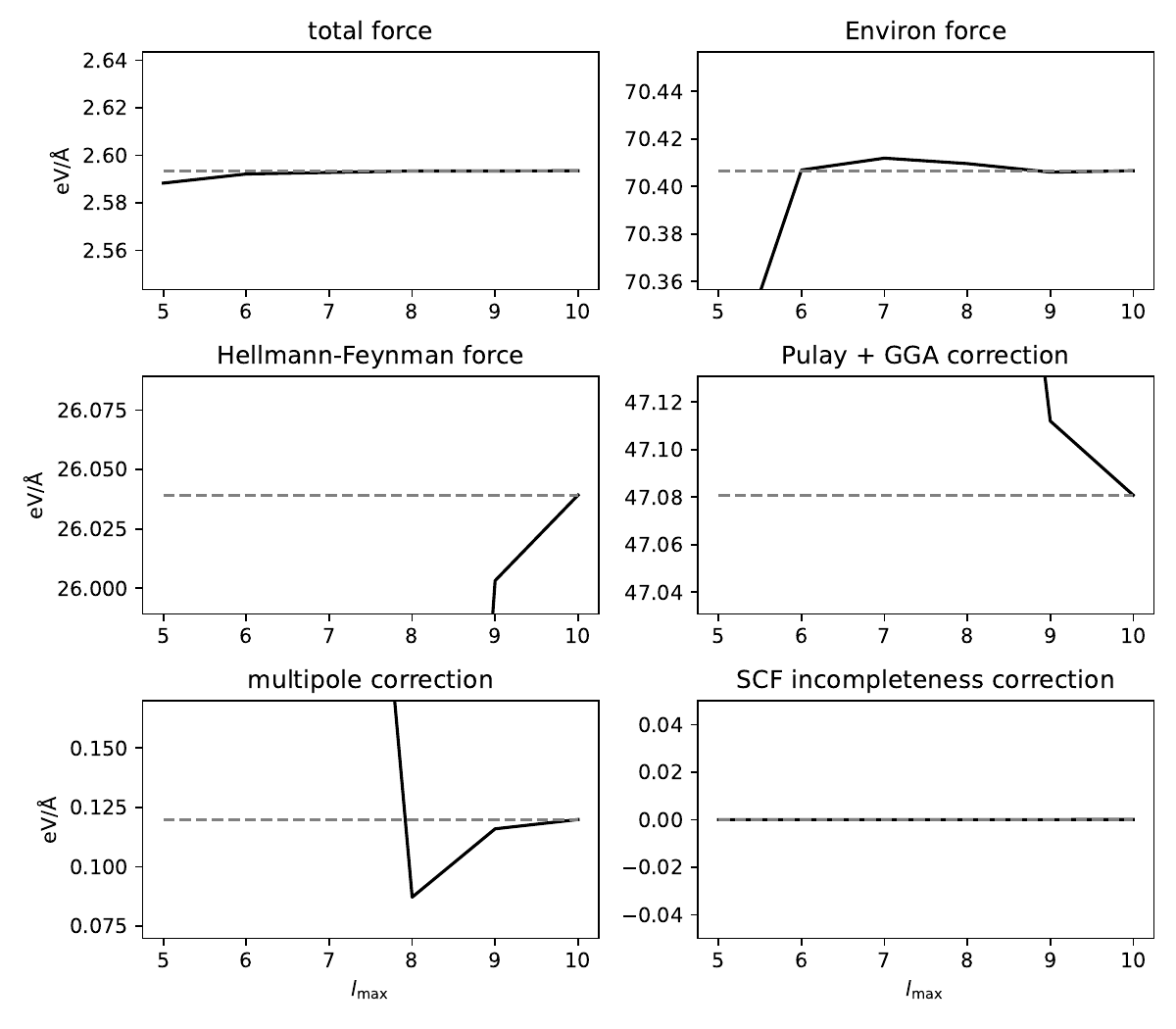}
    \caption{Like \cref{fig:F1_lmax_sscs_water}, for the Pt atom in a PtCO trimer.}
    \label{fig:F1_lmax_sscs_PtCO}
\end{figure}

\begin{figure}[!htb]
    \centering
    \includegraphics[width=0.9\linewidth]{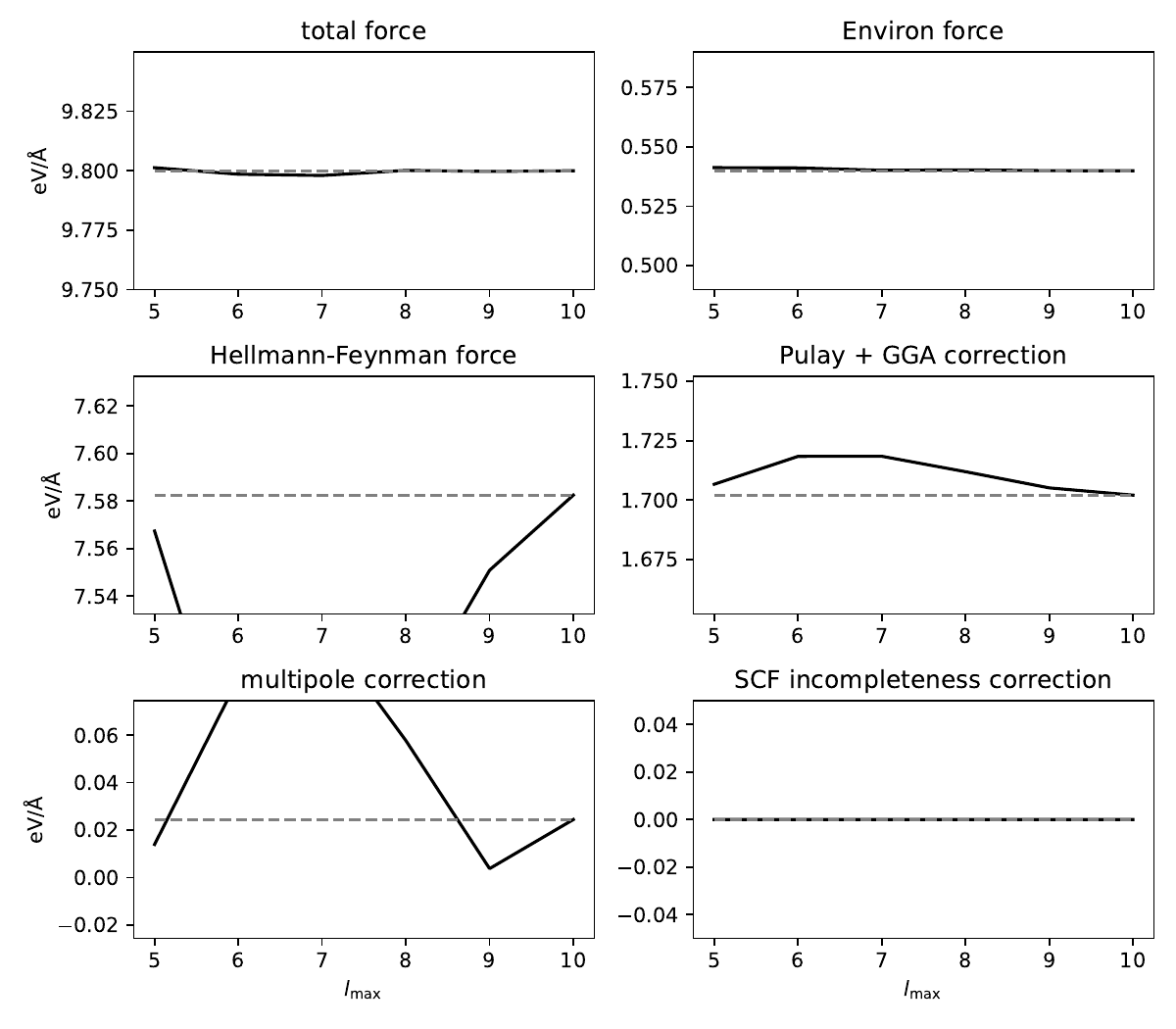}
    \caption{Like \cref{fig:F1_lmax_sscs_water}, for the C atom in a PtCO trimer.}
    \label{fig:F2_lmax_sscs_PtCO}
\end{figure}

\begin{figure}[!htb]
    \centering
    \includegraphics[width=0.9\linewidth]{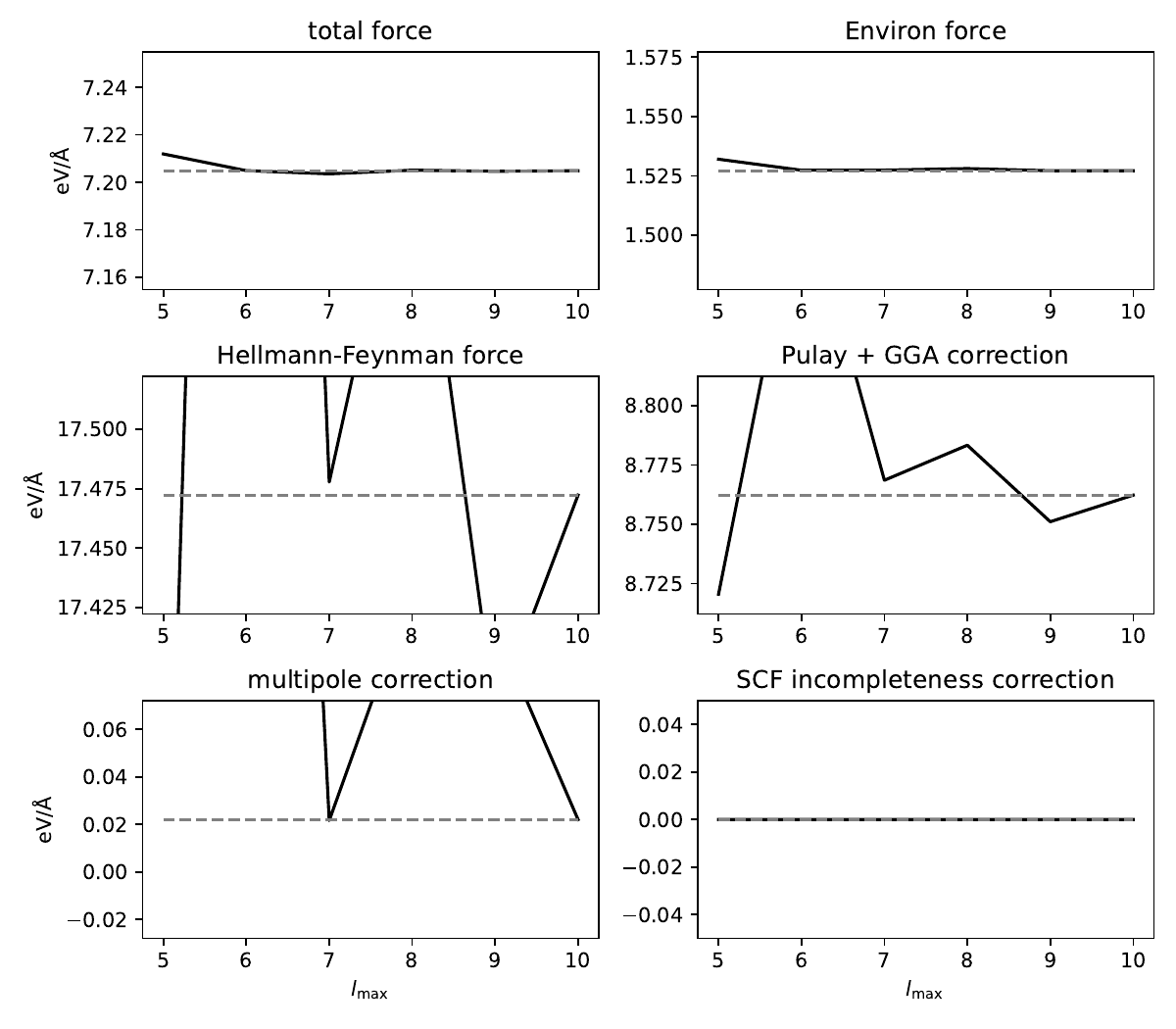}
    \caption{Like \cref{fig:F1_lmax_sscs_water}, for the O atom in a PtCO trimer.}
    \label{fig:F3_lmax_sscs_PtCO}
\end{figure}

\FloatBarrier

\newpage

\subsection{Energy convergence with multipole order (SCCS)}\label{sec:conv_E_lmax_sccs}

\begin{figure}[!htb]
    \centering
    \includegraphics[width=0.9\linewidth]{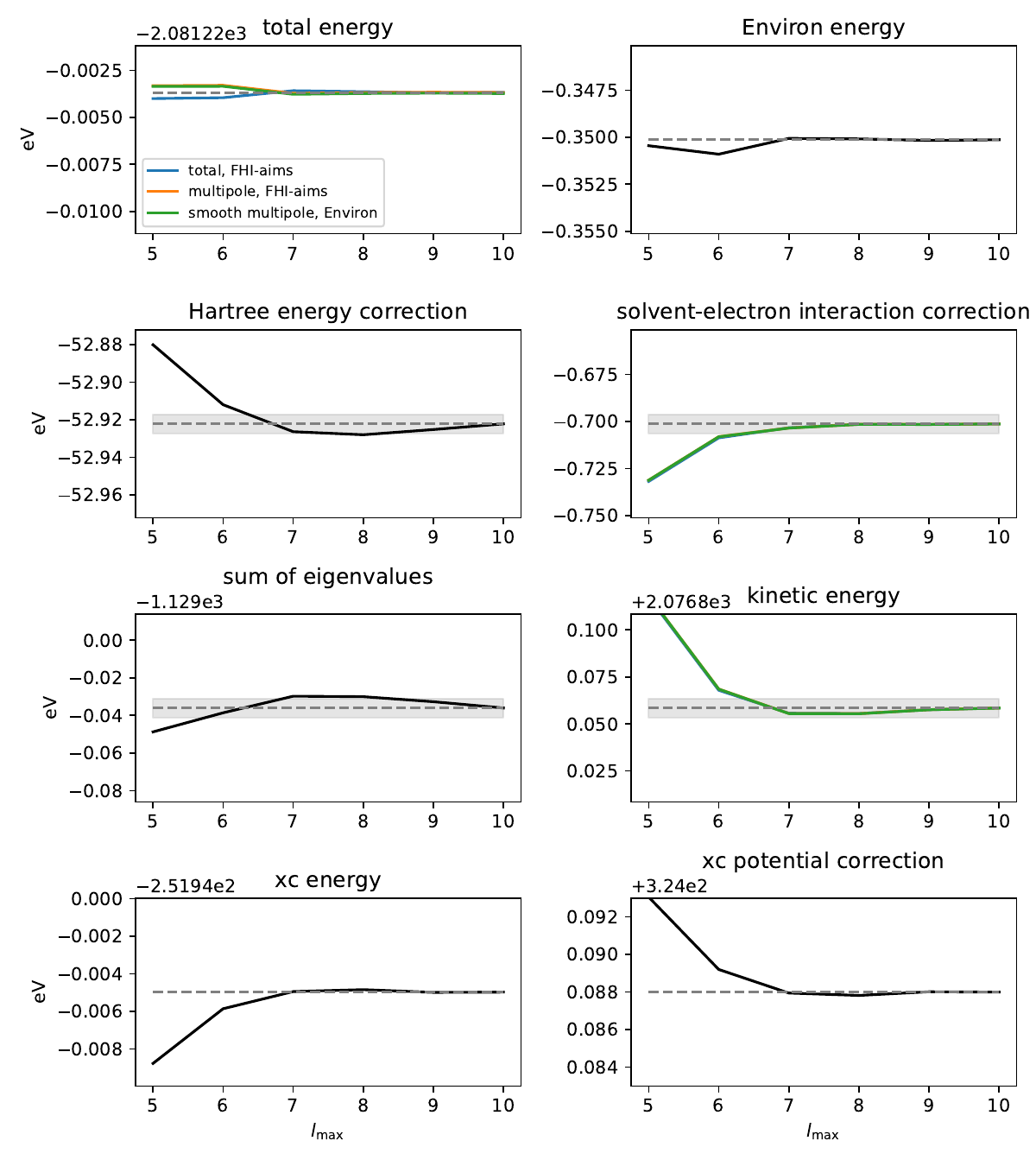}
    \caption{Convergence with $l_\text{max}$ of total energy and its contributions as reported in the FHI-aims\cite{blum2009} output file using different ways of computing the solvent-electron double counting correction for a water molecule in SCCS. Constant `free atom' contribution not shown. Note the different energy scales of the second and third row of subfigures. Shaded area indicates the energy scale covered in the other subfigures (10~meV)}
    \label{fig:E_lmax_sccs_water}
\end{figure}

\begin{figure}[p]
    \centering
    \includegraphics[width=0.9\linewidth]{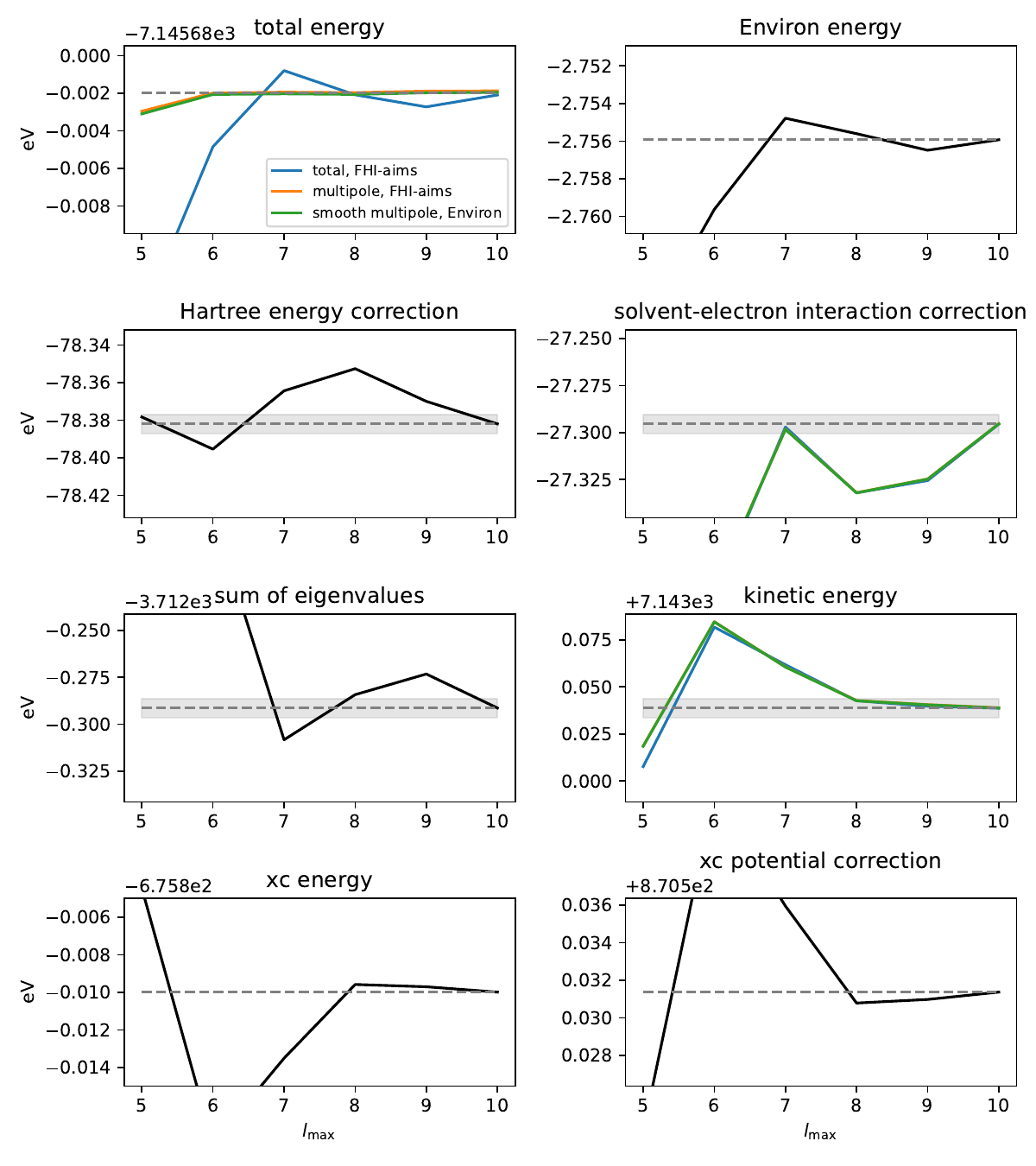}
    \caption{Like \cref{fig:E_lmax_sccs_water}, for a NaF dimer in SCCS.}
    \label{fig:E_lmax_sccs_NaF}
\end{figure}

\begin{figure}[p]
    \centering
    \includegraphics[width=0.9\linewidth]{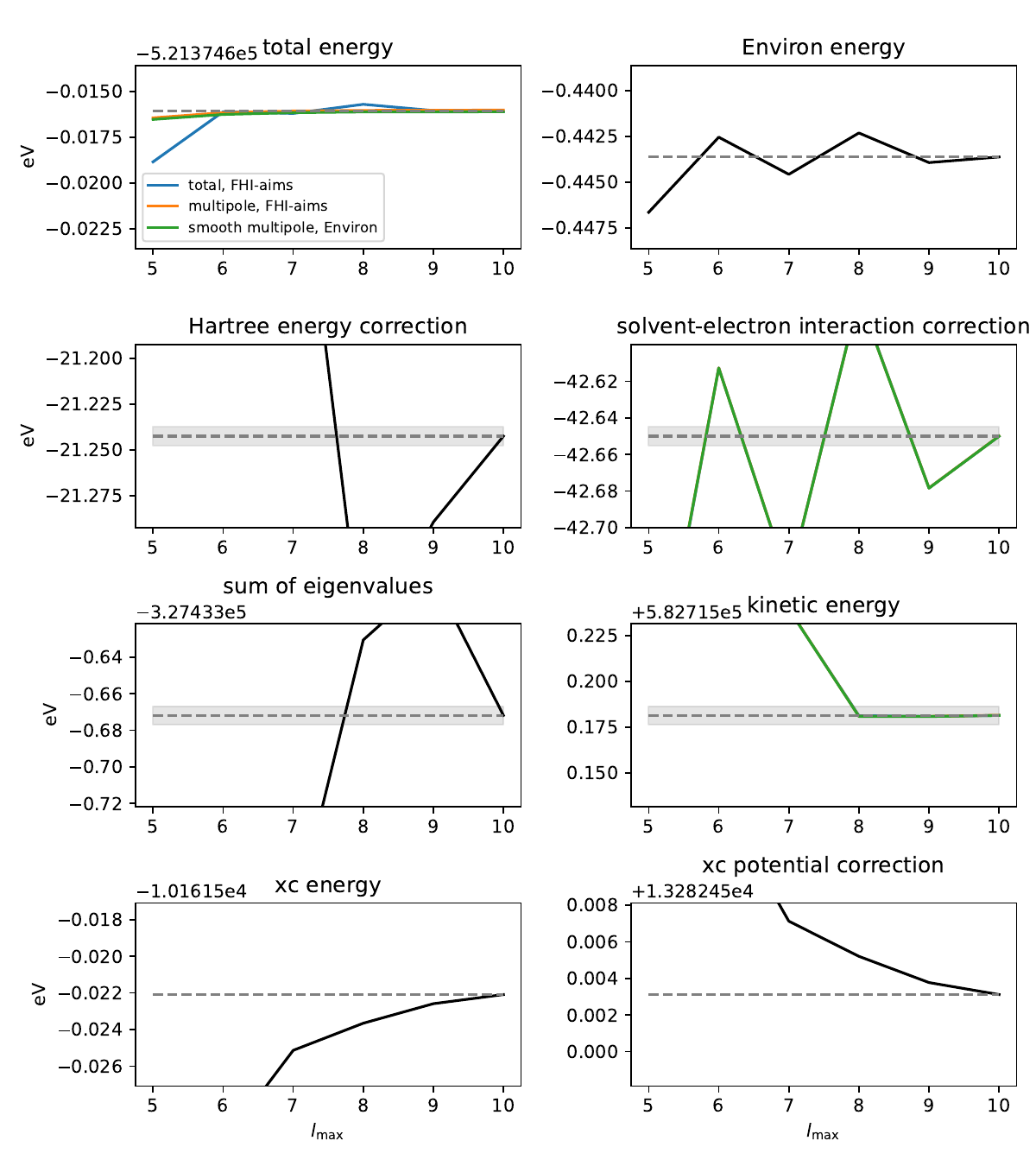}
    \caption{Like \cref{fig:E_lmax_sccs_water}, for a PtCO trimer in SCCS.}
    \label{fig:E_lmax_sccs_PtCO}
\end{figure}

\FloatBarrier

\newpage

\subsection{Force convergence with multipole order (SCCS)}\label{sec:conv_F_lmax_sccs}

\begin{figure}[!htb]
    \centering
    \includegraphics[width=0.9\linewidth]{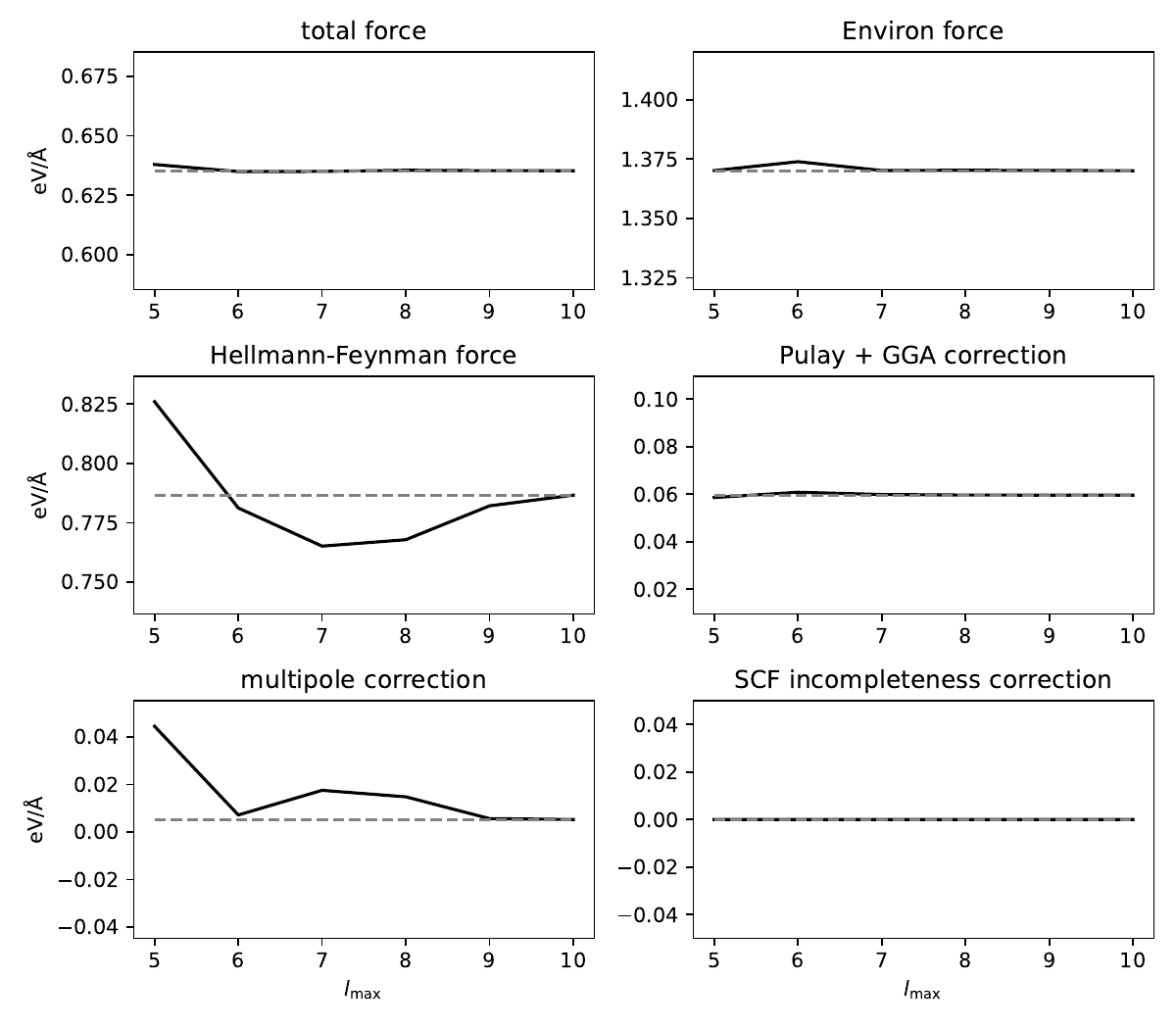}
    \caption{Convergence with $l_\text{max}$ of total force (Euclidean norm) acting on one H atom (other H atom equivalent) and its contributions as reported in the FHI-aims\cite{blum2009} output file using different low-pass filter settings for a water molecule in SCCS.}
    \label{fig:F1_lmax_sccs_water}
\end{figure}

\begin{figure}[!htb]
    \centering
    \includegraphics[width=0.9\linewidth]{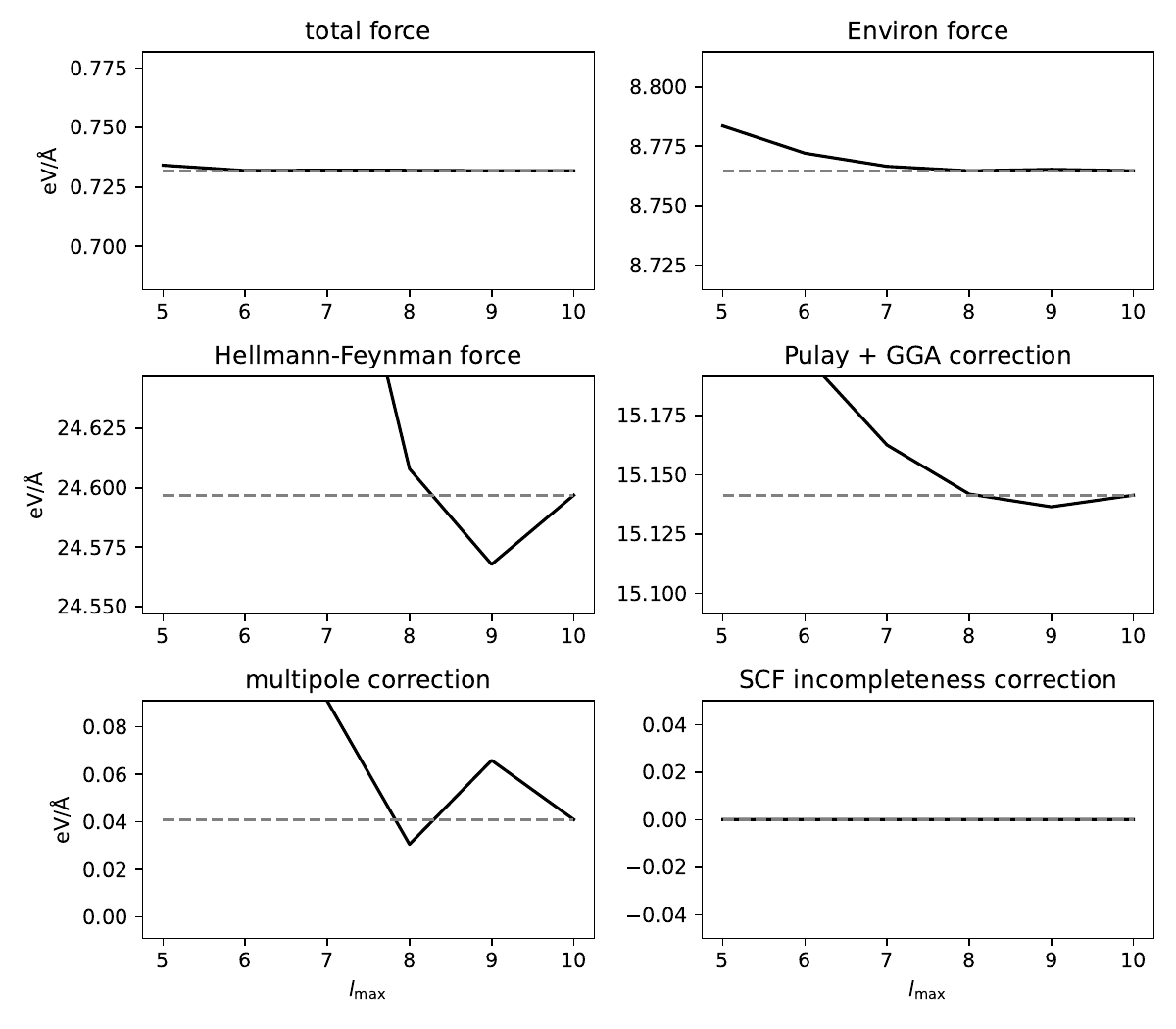}
    \caption{Like \cref{fig:F1_lmax_sccs_water}, for the O atom.}
    \label{fig:F2_lmax_sccs_water}
\end{figure}

\begin{figure}[!htb]
    \centering
    \includegraphics[width=0.9\linewidth]{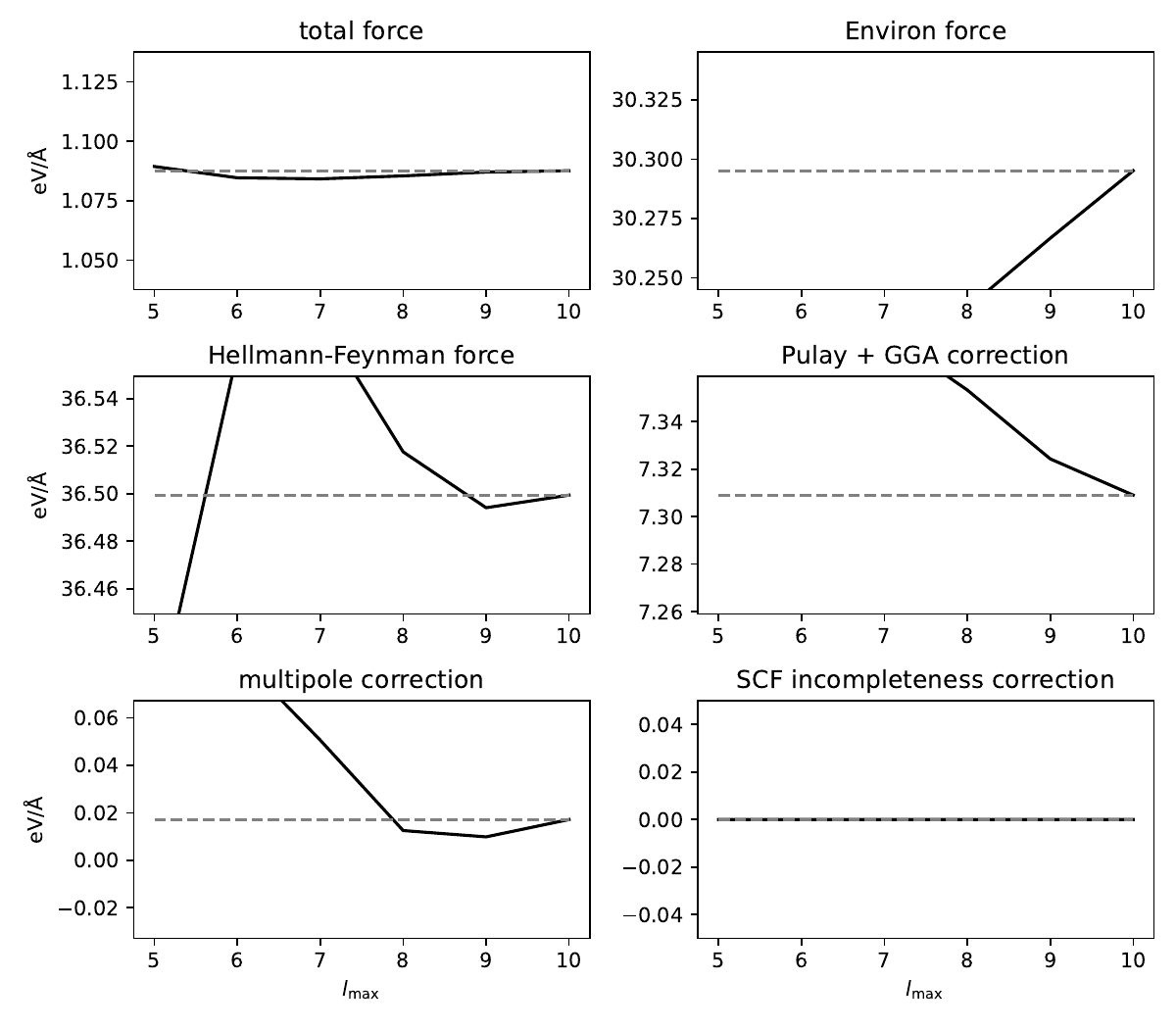}
    \caption{Like \cref{fig:F1_lmax_sccs_water}, for the Na atom in a NaF dimer.}
    \label{fig:F1_lmax_sccs_NaF}
\end{figure}

\begin{figure}[!htb]
    \centering
    \includegraphics[width=0.9\linewidth]{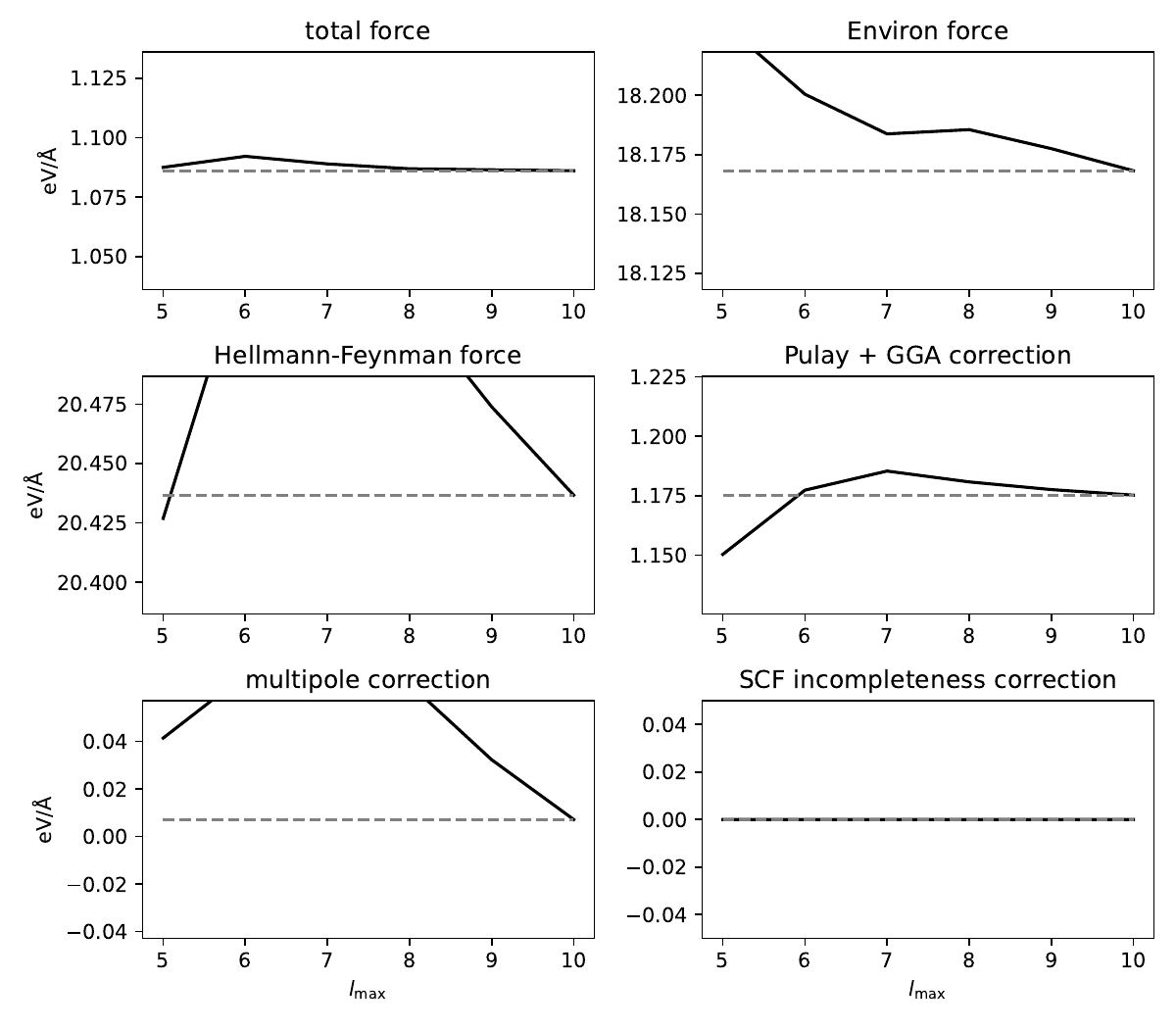}
    \caption{Like \cref{fig:F1_lmax_sccs_water}, for the F atom in a NaF dimer.}
    \label{fig:F2_lmax_sccs_NaF}
\end{figure}

\begin{figure}[!htb]
    \centering
    \includegraphics[width=0.9\linewidth]{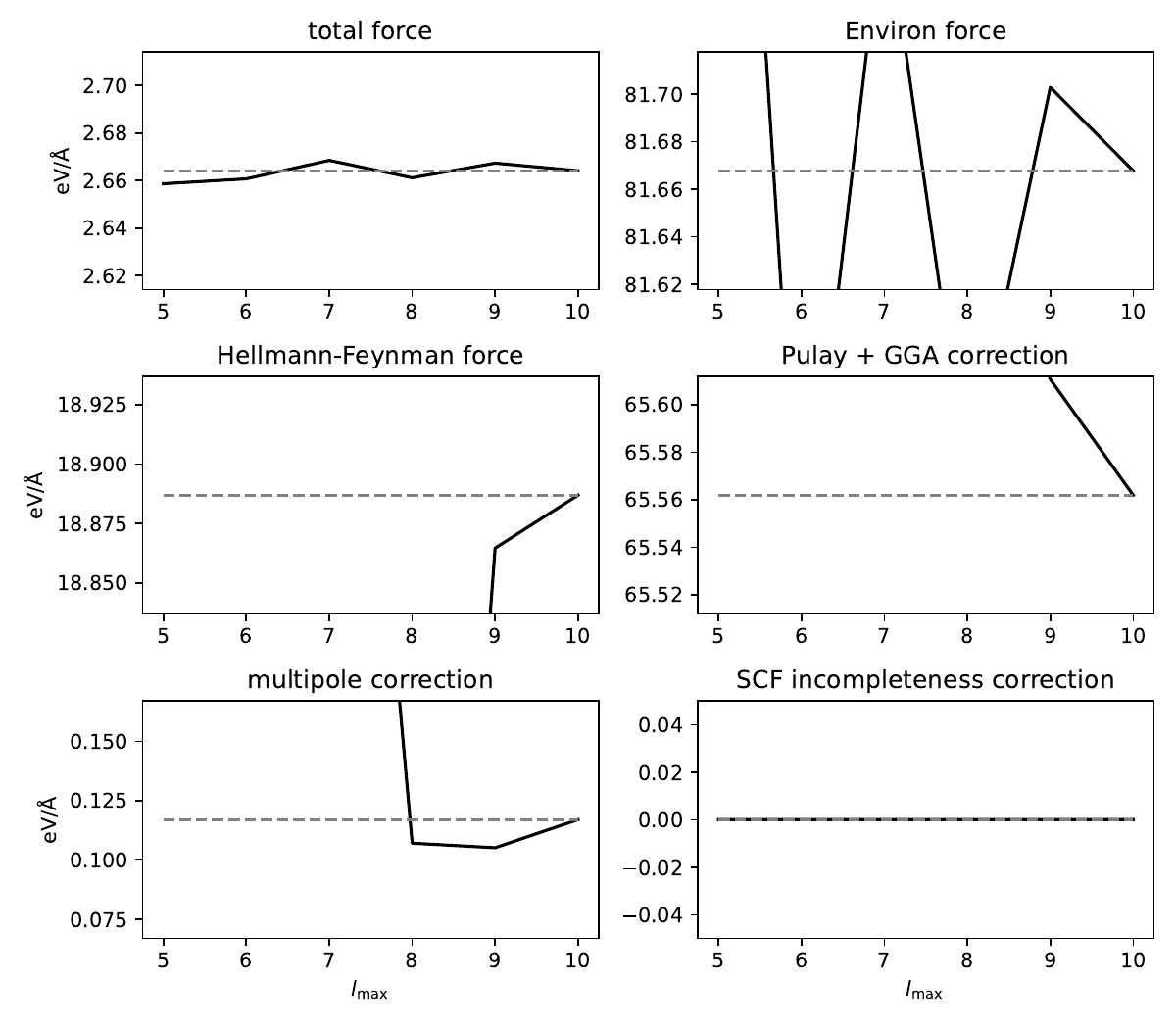}
    \caption{Like \cref{fig:F1_lmax_sccs_water}, for the Pt atom in a PtCO trimer.}
    \label{fig:F1_lmax_sccs_PtCO}
\end{figure}

\begin{figure}[!htb]
    \centering
    \includegraphics[width=0.9\linewidth]{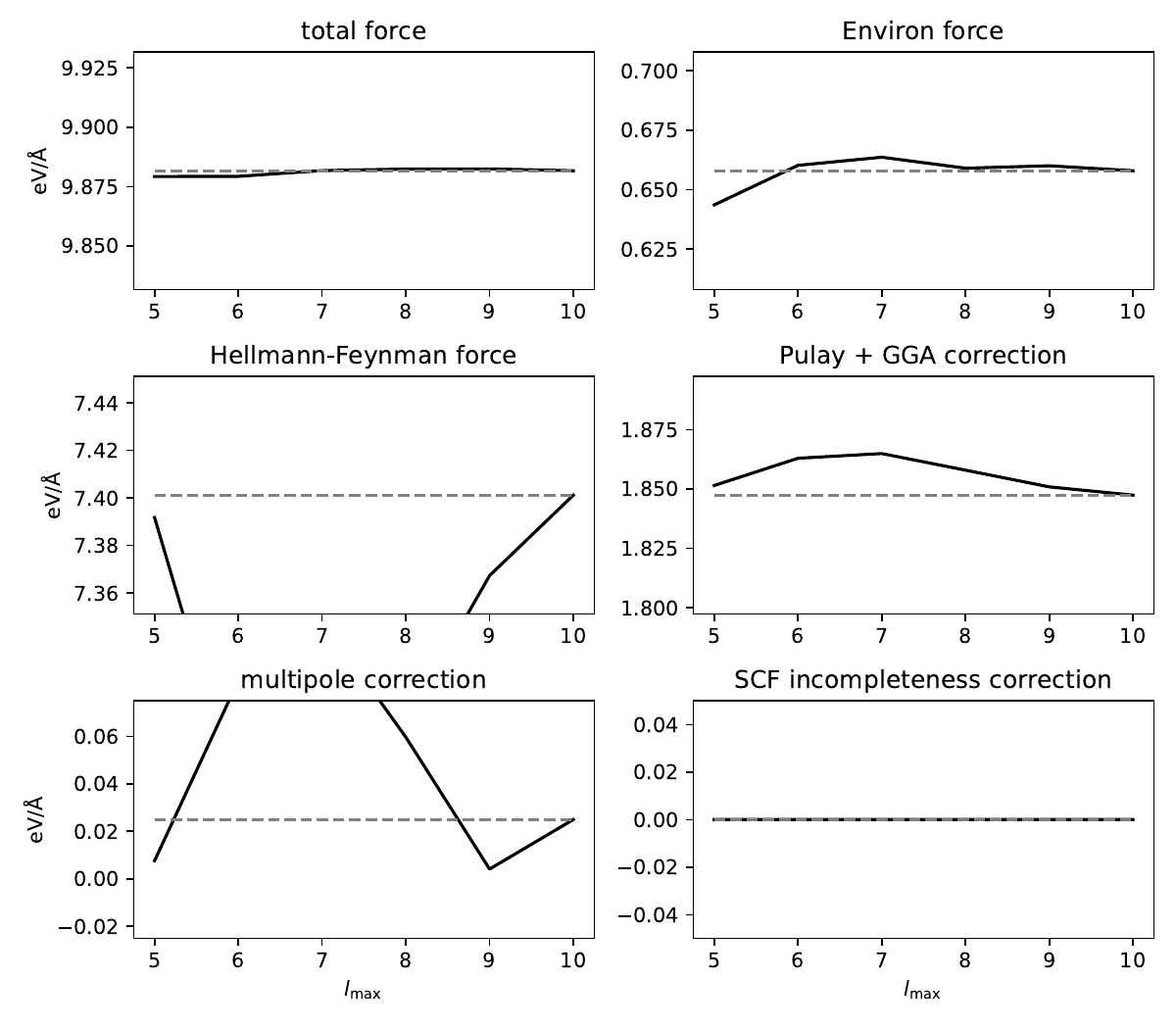}
    \caption{Like \cref{fig:F1_lmax_sccs_water}, for the C atom in a PtCO trimer.}
    \label{fig:F2_lmax_sccs_PtCO}
\end{figure}

\begin{figure}[!htb]
    \centering
    \includegraphics[width=0.9\linewidth]{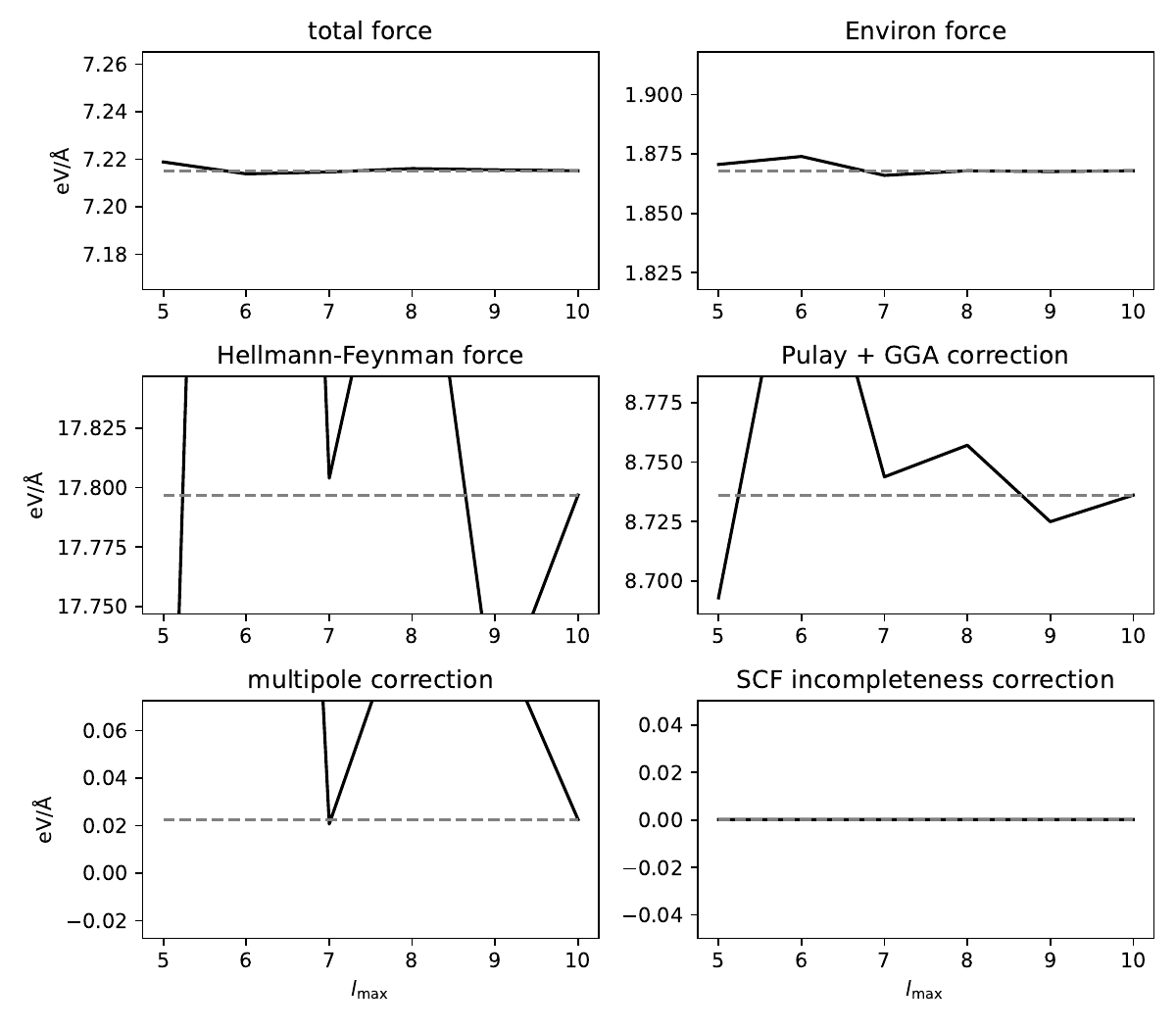}
    \caption{Like \cref{fig:F1_lmax_sccs_water}, for the O atom in a PtCO trimer.}
    \label{fig:F3_lmax_sccs_PtCO}
\end{figure}

\FloatBarrier

\newpage

\subsection{Low-pass filter}

\begin{figure}[!htb]
    \centering
    \includegraphics[width=0.95\linewidth]{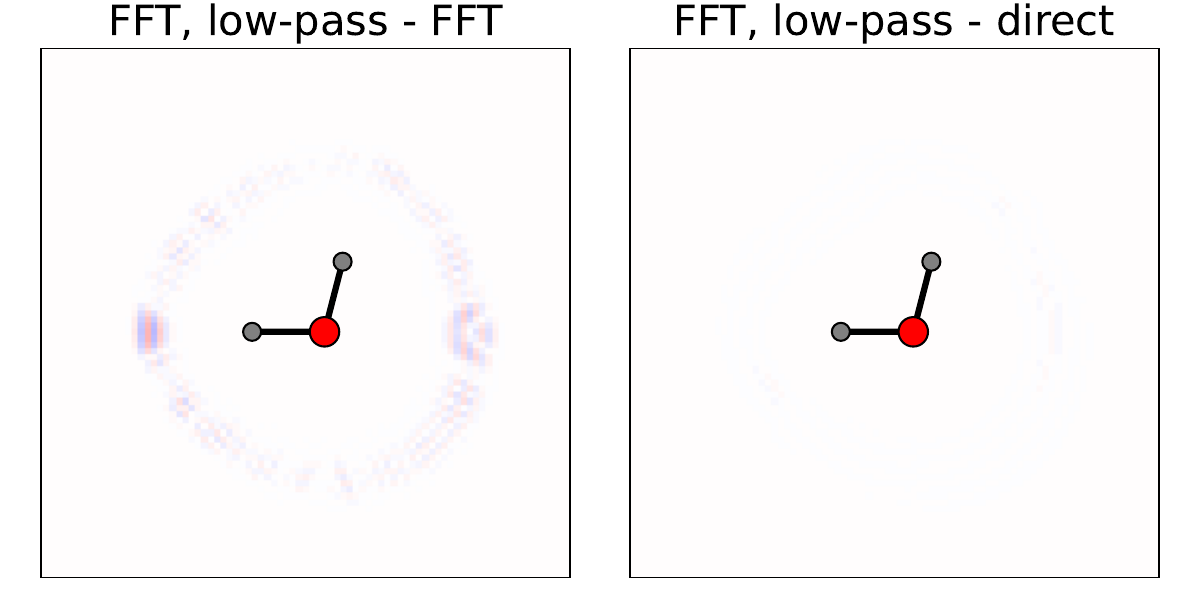}
    \caption{Differences between `FHI-aims, FFT, low-pass' and `FHI-aims, FFT' (left) as well as `FHI-aims, FFT, low-pass' and `FHI-aims, direct' (right) subfigures of \cref*{main-fig:gradients} in the main article in the main article. Contrast of colormaps arbitrary but identical to \cref*{main-fig:gradients} in the main article.}
    \label{fig:gradients_diff}
\end{figure}


\begin{figure}[p]
    \centering
    \includegraphics[width=0.95\linewidth]{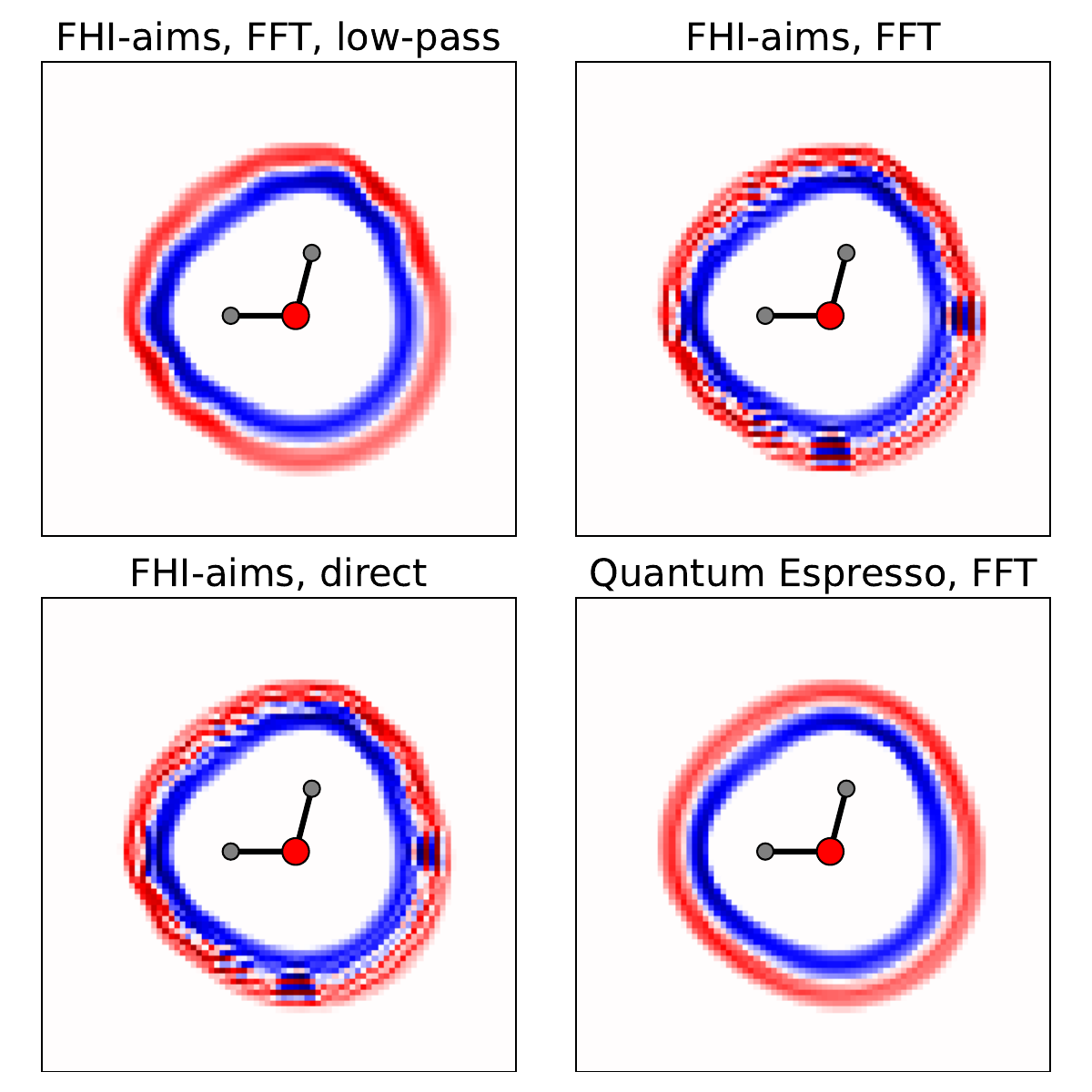}
    \caption{Like \cref*{main-fig:gradients} in the main article but showing $\nabla^2 s$ instead of $\partial s / \partial x$. In the bottom left, $\nabla \Bar{\rho}^\text{el}$ was computed by FHI-aims directly and passed to Environ but $\nabla^2 \Bar{\rho}^\text{el}$ was computed using unfiltered FFT, leading to artifacts.}
    \label{fig:laplacians}
\end{figure}





\FloatBarrier

\newpage

\bibliography{refs}